\documentclass[12pt]{article}
\usepackage{graphicx}
\usepackage{amssymb}
\usepackage{epstopdf}
\newcommand \Pomeron {I\!\!P}
\def\beq{\begin{equation}}
  \def\eeq{\end{equation}}

\begin{document}
%
%
%
\title{DIFFRACTIVE PHENOMENA IN HIGH ENERGY PROCESSES }
\author{L.~Frankfurt,
Tel Aviv University,  Tel Aviv, 69978, Israel \\ 
leonidfrankfurt@gmail.com\\
M.~Strikman\\
The Pennsylvania State University,
 University Park, PA, 16802, U.S.A.\\
strikman@phys.psu.edu}
\maketitle

\begin{abstract} 

We review the evolution of the studies of diffractive processes in the strong interaction over the last   60 years. First, we briefly outline     the  early developments of the theory  based on analyticity and unitarity of the $S$-matrix, including the derivation and exploration  of the Regge trajectories and related moving cuts. Special attention is paid to  the concept of the Pomeron trajectory introduced for description of total, elastic and diffractive cross sections at high energies and to the emergence of the dynamics of multi-Pomeron interactions.
The role of large longitudinal distances  and color coherent phenomena for the understanding of 
inelastic diffraction in hadron--hadron scattering and deep inelastic scattering is emphasized. The 
connection of these phenomena to the cancellation of the contribution of the Glauber approximation 
 in hadron--nucleus collisions  and to the understanding of the Gribov--Glauber approximation 
is explained.  The presence of different scales in perturbative QCD due to masses of heavy quarks  
has led to the emergence of numerous new phenomena including non-universality of  the slopes of 
Regge trajectories made of light and heavy quarks and non-universal energy dependence of elastic 
cross sections.  The application of the perturbative QCD techniques allowed us to calculate from the 
first principles the interaction of small  transverse size   color singlets with hadrons leading to the development of the quantitative theory of hard exclusive reactions and to the successful prediction of 
many regularities in hard large mass diffraction.  It  also led to the prediction of the   phenomenon of  complete transparency of nuclear matter in QCD in special processes. The conflict of perturbative QCD 
with probability conservation for high energy  processes of virtual photon--nucleon scattering is explained.   
Some   properties of the new QCD regime are outlined.
\end{abstract}

\section{Introduction}
\label{section1} 

The aim of this chapter is to demonstrate that the phenomenon of diffraction in quantum chromodynamics (QCD)  is a formative playground  for the fundamental ideas and methods of theoretical physics.  The 
deep disappointment of scientific society in the quantum field theory paradigm  formulated by 
L.~D.~Landau  at the Rochester Kiev conference (1959)~\cite{Landau1} 
was based on the  zero-charge problem  in pre-QCD quantum field theories.    
As a result, the concept of the scattering matrix ($S$-matrix), where all quantities are in principle 
observable,  became popular and replaced studies within the quantum field theory framework.  
The idea was that unitarity of the $S$-matrix, 
its analytic properties and exact symmetries  will allow one 
to avoid dealing with point-like interactions characteristic for a quantum field theory such that 
the need for the ugly procedure of renormalization with all its puzzles will disappear.  
This approach led to the development of such new concepts as single and double dispersion relations, 
Regge trajectories, the Pomeron calculus, string models, etc ~and  to the  prediction of new phenomena. 
The discovery of asymptotic freedom in QCD in the late sixties to early seventies justified   
the space--time description of high energy processes in QCD, which is absent within the $S$-matrix 
concept. The account of the space--time evolution of high energy processes allowed one to predict a 
variety of striking new QCD phenomena such as color fluctuations, complete transparency of nuclear 
matter under special kinematic conditions,  formation of a new QCD regime of the maximally  strong interaction, etc.,  all of   which are absent in the $S$-matrix theory. In the first part of this chapter,   
we consider the  phenomenon of diffraction in the $S$-matrix theory  and then discuss  
new 
diffractive
phenomena that  emerge in QCD studies.

The basic ideas of the $S$-matrix approach are unitarity of the $S$-matrix in all physical channels 
and analyticity of scattering amplitudes in the complex planes of energies and momentum transfers 
that  leads to analyticity of amplitudes in the plane of the angular momentum.  The $S$-matrix   approach  justifies   the concept of Regge trajectories.   The assumption of the  dominance of amplitudes of high energy processes by the Pomeron trajectory exchange predicts an increase with energy in  the  radius of the interacting hadron 
(shrinking with energy of the forward peak in two-body exclusive processes) and, therefore, the dominance 
of peripheral collisions,  the universal dependence on energy of the total and elastic cross sections and 
cross sections of diffractive processes.    
The very existence of Pomeron moving cuts follows from unitarity of the $S$-matrix in the crossed channel. 
The prediction and experimental discovery of the large mass $(M^2 \gg m^2)$  triple Pomeron diffraction 
proves the non-zero value of the effective triple Pomeron interaction.  

The modeling of the contribution of Pomeron moving cuts found  blackening of interactions at central impact  parameters  since the  contribution of the single Pomeron exchange grows with energy.   
However, the fraction of the total cross section due to the elastic scattering slowly grows with energy. 
For the current LHC energy:
\begin{equation}
\sigma_{\rm elastic}(pp)/\sigma_{\rm tot}(pp)\approx 0.25 \,.
\end{equation}
Thus,   the   $pp$ interaction is still very far from the regime of complete absorption where this ratio should 
be close to $0.5$. Note, however, that for  the central $pp$ collisions, almost complete absorption has 
been observed at 
Fermi National Accelerator Laboratory (FNAL) and the LHC, which corresponds to partial amplitudes being   
close to unity. Thus an energetic proton, when interacting with the proton target, behaves as a grey disc 
with a black spot in the center. The size of the black spot rapidly increases with energy.

The assumption that the amplitudes of high energy hadron--hadron collisions
depend on one scale was challenged by the discovery of 
$J/\psi, \Upsilon$ mesons  -- bound states of heavy quarks:
$c \bar c$ and $  b\bar b $.
  The radii of these quarkonia  states  are significantly smaller 
than  for hadrons made of light quarks. The interaction of 
$Q\bar Q$ quarkonia with hadrons made of light 
quarks 
is decreasing with $m_Q$ in the non-perturbative and perturbative QCD domains.

Analyses of the ladder diagrams for cross sections of deep inelastic scattering off a hadron target found that longitudinal distances dominating in the scattering process are linearly increasing with energy in QCD.  As a result, at sufficiently large energies they exceed by far the length of the target, the transitions between different configurations in the projectile slow down and the interaction can be described as a superposition 
of the interaction of instant quark--gluon configurations within the projectile. This feature leads to the fluctuations of strengths of hadron--hadron and hadron--nucleus interactions and to the exact cancellation 
of the Glauber model contribution to hadron (nucleus)--nucleus collisions. 
This cancellation follows directly  from the analytic properties of amplitudes 
and/or
energy--momentum conservation. The Gribov--Glauber model 
replaces the Glauber model in high energy processes where diffraction is a shadow of inelastic processes.

The fluctuations of strengths of the interaction in hadron--hadron collisions
found an explanation in QCD as being due to the color screening phenomenon. 
The fluctuations of strengths of the interaction within the virtual photon wave function 
have been observed directly in the significant cross section of leading twist diffraction in deep inelastic scattering (DIS), which 
is predicted to be negligible in perturbative QCD (pQCD) because of the absence of free quarks and gluons. 
However, 
it was observed in $ep$ DIS at HERA that the cross section of diffractive processes 
constitutes 
$\approx 10\%$ of the total cross section  at $x\approx 10^{-3}$ and the ratio of the 
diffraction cross section to the total one is practically energy, and 
$Q^2$ - independent ($Q^2$ is the photon virtuality).

QCD dynamics  predicts the existence of hard diffractive phenomena that are higher twist effects.
The significant difference in momentum scales characterizing hard and soft (non-perturbative) processes allows one to prove the factorization of hard processes from soft ones and to  calculate cross sections of 
hard diffractive processes. The processes of elastic photoproduction of mesons with hidden heavy flavors 
off the proton target and elastic electroproduction of light mesons observed at HERA revealed an interplay 
of the dependence of the cross sections on energy,  the photon virtuality $Q^2$, and the momentum 
transfer $t$ that is close to that expected in pQCD.   

Diffraction in high energy processes is a shadow of inelastic processes so that,   for sufficiently small $x$, 
an increase with energy of the structure functions of nucleons and nuclei, which is predicted in pQCD approximations, runs into conflict with the probability conservation  at small impact parameters. The range 
of central impact parameters, where the regime of complete absorption dominates, increases with an increase of energy (for fixed $Q^2$). In this regime at ultrahigh energies, $\sigma(pp)  \propto \ln^2(s/s_0)$  and  $\sigma(\gamma^{\ast}p)\propto \ln^3(s/s_0)$. The complete absorption regime is possibly reached for the gluon distribution at the central impact parameters
 and at $Q^2$ of the order of a few GeV$^2$ in electron--proton  collisions at HERA.
Physics   related to the formation of strong gluon fields at sufficiently small $x\approx Q^2/s$ and small impact parameters can be probed at the LHC and 
Large Hadron--Electron Collider (LHeC).  
In this new QCD regime the expansion over powers of $1/Q^2$ (twists) becomes meaningless.  
Also, there arises the question whether continuos symmetries such as conformal and scale
invariances characterizing  pQCD and new QCD regimes are different.

No significant violation of the 
Dokshitzer--Gribov--Lipatov--Altarelli--Parisi (DGLAP) approximation for the structure functions integrated over impact parameters is predicted at achieved $x$. 
This is because the structure functions are dominated by the scattering at the impact parameters
growing with energy where the interaction remains weak even 
though it reaches the black limit at the central impact parameters.  
Moreover, no noticeable slowing down of an increase of the structure functions with energy is expected 
at even smaller $x$ since the black limit
contribution rapidly increases with a decrease in  $x$ 
at the energies achievable in laboratory.

It was suggested to sum the leading $\alpha_s\ln(x_0/x)$  terms in the kinematics of fixed 
$Q^2$ and $x\to 0$ -- leading logarithmic (LL)
approximation. 
There were derived formulas for the collision of two small size ($\approx 1/Q$) wave 
packets, $\gamma^{\ast}(Q^2)+\gamma^{\ast}(Q^2)\to X$, in which case  the diffusion in the $k_t$ space 
is suppressed (which may work for limited range of energies). The same formulas  are often applied to the scattering of a small size ($\approx 1/Q$) wave packet ($\gamma^{\ast}$) off the proton target  that  has 
a size of $\approx  1/(2m_{\pi})$. In this case the neglected 
within the LL approximation
diffusion to small parton momenta within the parton ladder is rather important.   Energy and momentum of the final states calculated within the leading log approaches are significantly different from that for initial state especially within the leading  $\alpha_s\ln(x_0/x)$ approximation. This violation follows from the choice of the kinematical domain characteristic for LL approximation.  Conservation of energy-momentum is guaranteed after resummation over  series of LO, 
NLO, NNLO... approximations. This property of approximation explains negative sign and  huge value of 
next-to-leading order (NLO) ''corrections'' and requires the development of resummation approaches.  However,  diffractive processes were not considered 
yet in the resummation approaches.

The chapter is organized as follows.  In section~\ref{section2} we define the kinematics characteristic 
for diffractive phenomena in hadronic collisions and briefly review the $S$-matrix approach: analytic properties of amplitudes of high energy processes in the energy, momentum transfer and orbital 
momentum planes.  We also explain how the concept of Regge trajectories in the angular 
momentum plane arises in the relativistic theory of the scattering matrix. 

In section~\ref{section3},  we remind basic properties of Regge trajectories and explain that the linearity and 
universality of Regge pole trajectories is confirmed by comparison of Regge trajectories with the data on 
hadron resonances made of light $u$, $d$, and $s$ quarks.  This linearity 
and the assumption that the amplitude is dominated by the  Regge pole trajectory exchange allows one to reproduce the observed  dependence of cross sections of two-body processes with non-vacuum quantum numbers in the crossed channel on energy.  We also explain that the assumption that 
all Regge trajectories have the same
universal slope does not hold for the trajectories with hidden and open heavy flavors. It contradicts the quarkonium models and the data. We also explain how moving cuts accompanying the Regge pole 
follow from the unitarity of the $S$-matrix in the plane of angular momentum.

In section~\ref{section4}, we consider the hypothesis that the
dominant exchange
in the amplitudes of high energy processes, is the one by the Pomeron trajectory  accompanied by Pomeron moving cuts---Pomeron calculus. 
 The dominance of peripheral collisions and  the important role of Gribov diffusion in the 
 impact parameter space are explained. A brief  comparison with data shows that such predicted basic features of 
 high energy processes as the universal energy dependence of all high energy processes and 
 the shrinking with energy of the diffractive peak agree with the data.   We discuss properties of multi-Pomeron interactions, the evidence for triple Pomeron interactions and their role 
in diffraction at the Tevatron and LHC. The impact of multiple rescatterings of Pomerons on 
the elastic differential cross section is also briefly discussed. 

In the framework of the Pomeron calculus, 
we also explain an onset of the regime of the complete absorption at small impact parameters,
some of its properties and compare it briefly with 
the selected FNAL and  LHC  data.  
Implications for the value of the slope of the Pomeron trajectory 
in the regime of complete absorption at small impact parameters are briefly discussed.  

In section~\ref{section5}  the space--time evolution of high energy processes and the linear increase 
with energy in  longitudinal distances in the scattering process are discussed.   As a result, the contribution 
of planar diagrams---known as the Glauber model for hadron--nucleus collisions---is cancelled out. Moreover the contribution of the planar diagrams (the Glauber approximation) violates energy-momentum conservation.

At the same time, the  contribution of non-planar diagrams can be rewritten in the form of the Glauber approximation but with an additional inelastic shadowing term---the Gribov--Glauber  model. 
 
 In section~\ref{section6}  we explain that in QCD, the  increase of longitudinal distances with the collision energy  leads to a variety of coherent phenomena which we refer to as the color fluctuation phenomena.  
 They include the presence of inelastic diffraction at the zero angle and the processes where hadrons 
 fluctuate into small - size configurations and interact with the small strength so that nuclei do not absorb 
 them---color transparency (CT). CT allows one to prove QCD factorization theorems for a number of processes.  At the same time, the concept of color fluctuations allows us to bridge the gap   between the fluctuations to small- and large-size configurations and to reconstruct the distribution over 
  cross sections
  for projectile hadrons and photons.  It also gives us a physically transparent interpretation of the 
  Gribov--Glauber model that could be applied to modeling proton (nucleus)--nucleus collisions.  
  The concept of color fluctuations allows us to build the QCD-improved aligned jet model.  The significant cross section of diffraction in deep inelastic small $x$ processes observed at HERA,  its $Q^2$ and energy dependencies are direct confirmation of the important role of color fluctuations in high energy processes.   
  We also briefly  review the concept of the perturbative Pomeron in pQCD.

In section~\ref{section7}, we  explain the QCD factorization theorem for hard exclusive processes,
derive basic characteristics of hard diffractive processes  and briefly compare the derived formulas 
with the data obtained at FNAL and HERA. 
We point out that complete transparency of nuclear matter for  special 
hard diffractive processes has been predicted and confirmed by the FNAL data.

In section~\ref{section8}, we discuss the onset of the new QCD regime and its basic features
in the limit of fixed $Q^2$ and $x\to 0$. The conflict between pQCD calculations and probability 
conservation in the collisions at central impact parameters, the onset of the black disc regime (BDR), and competition between the soft QCD and pQCD contributions are explained.

Conclusions are presented in section~\ref{section9} .

\section{General properties of the scattering amplitude}
\label{section2}

\subsection{Kinematics}
\label{subsection2.1}
We consider first the scattering of two particles:
\begin{equation}
a+b\to a+b \,.
\label{emconservation}
\end{equation}

The amplitude of this process depends  on the four-momenta of the colliding particles 
$p_i$.   An account of energy--momentum conservation gives the constraint:
\begin{equation}
p_a+p_b= p^{\prime}_a+p^{\prime}_b \,.
\label{EM}
\end{equation}

Lorentz invariance  restricts the number of independent variables. The convenient variables are 
the square of the energy in the $s$-channel  center of mass: $s=(p_a+p_b)^2$,  and the squares of 
the momentum transfer between $a$ and $a^{\prime}$
 $t=(p_a-p^{\prime}_a)^2$, and between $a$ and $b^{\prime}$:  $u=(p_a-p^{\prime}_b)^2$.  
 These three variables are not independent.  
Energy--momentum conservation (Eq.~\ref{EM}) leads to
\begin{equation}
s+t+u=2[m^{2}_a+m^{2}_b] \,.
\end{equation}
If the scattered particles are the  lowest-mass states in the channels with given quantum numbers,  there exist three physical channels where  the scattering process is allowed: 
$s\ge (m_a+m_b)^2$, $(t,u)\le 0$;    $t\ge \max\{4m^2_a,4 m_b^2\}$, $(s,u)\le 0$;
 $u\ge (m_a+m_b)^2$, $(t,s)\le 0$. Thus  the physical meaning of the 
 variables $s$, $t$ and $u$ is that each of them is equal to square of the center-of-mass energy of colliding  particles in the corresponding center of mass of the physical channel: $a$ and $b$ in the $s$-channel, $a$ and $\bar a$ in the $t$-channel 
 and $a$ and $\bar b$ in the $u$-channel. The amplitudes in all three channels  are interrelated by 
 rotation of the four-momenta of the particles.

\subsection{S-matrix approach}

Before the advent of quarks and later QCD the most important ideas of the theory of strong interactions 
were suggested within the concept of the scattering matrix, $S$.  The matrix elements of the $S$-matrix describe amplitudes of the scattering processes where hadrons  in the initial and final states are outside the interaction region.   In the physical region of any physical process, the $S$-matrix is restricted by its unitarity, 
i.e., by probability conservation:
\begin{equation}
 SS^{\dagger}=1 \,.
 \label{unitarity0}
 \end{equation}
 To single out  the contribution  when no interaction occurs, the $T$-matrix is introduced: $S=1+iT$. 
 The unitarity condition for the $T$-matrix has the following form in the $s$-channel:
 \begin{equation}
\mbox{Im} \, T(a+b\to a+b)={1\over 2} \int  \sum_n \left<a+b\right|T\left|n\right> d\tau_n\left<n\right|T\left|a+b\right>^{\dagger} 
 \,,
 \end{equation}
 where $d\tau_n$  is  the phase volume for the state $n$.   The above equation can be rewritten as the optical theorem  which  relates $T$ matrix with the total cross section:
 \begin{equation} 
 \mbox{Im} \,  T(a+b\to a+b)=s\,\sigma_{\rm tot}(a+b) \,,
 \end{equation}
where $s=(p_a+p_b)^2$.
 
The hope was that in a relativistic theory, the conservation of probability, i.e., the $S$-matrix 
unitarity~(Eq.~\ref{unitarity0}), and threshold singularities in the crossed channel
would substitute the non-relativistic concept of the potential. 
  
\subsection{Brief summary of analytic properties of amplitudes in energy, momentum and orbital momentum planes}
\label{subsection2.2}
The aim of this subsection is to briefly remind of basic ideas and results obtained in the $S$-matrix approach.

One starts with imposing causality in the form of the  Lehman, Symanzic and Zimmermann (LSZ) representation of the amplitude as the Fourier transform of the matrix elements of the retarded commutator 
of currents \cite{LSZ}.   This leads to the assumption  that the amplitudes  of the physical processes are 
the boundary values of the same analytic function of the energy and momentum transfers.  The  
singularities of the amplitudes are given by the thresholds for the  physical processes \cite{Landau2},  
see the discussion below.

S.~Mandelstam proposed the double dispersion representation which 
takes into account the singularities both in the momentum transfer  and in the energy planes \cite{Mandelstamrep}:  
\begin{eqnarray}
\label{Mandelstamrep}
A(s, t)={1\over \pi^{2}}\int_{4m^2}^{\infty} ds'dt' {\rho_{st}(s',t')\over (s'-s)(t'-t)}+{1\over \pi^{2}} \int_{4m^2}^{\infty} du'dt' {\rho_{ut}(u',t')\over (u'-u)(t'-t)} \\
\nonumber 
 + {1\over \pi^{2}} \int_{4m^2}^{\infty} du'dt' {\rho_{su}(s',u')\over (s'-s)(u'-u)} \,,  & &
\end{eqnarray}
where we take $m_a=m_b$ for simplicity.
All denominators are understood as having imaginary parts:   
$(s'-s)^{-1}=(s'-s-i\epsilon)^{-1}$.  (This condition selects the outgoing wave.) In the following 
we will  not   need  the exact form of the  spectral densities $\rho_{i, j}$  which 
were  supposed to  follow from unitarity of the $S$-matrix 
and its analytic properties  \cite {CM}.

Dispersion representations over the variables $s$ and $t$ follow directly  from the 
double dispersion representation.
Also, they can be derived directly from the theoretical analysis of the 
LSZ  representation of the scattering amplitude,   
see the discussion and references in \cite{ELO}.   
The dispersion representation of the scattering amplitude in the energy plane is:
\begin{equation}
A(s,t)={1\over \pi}\int_{4m^2}^{\infty} ds' {Im_{s} A(s',t)\over s'-s}+{1\over \pi}\int_{4m^2}^{\infty} du' {Im_{u}A(u',t)\over u'-u} +~{\rm subtractions} \,.
\label{DP}
\end{equation}
By definition subtractions do not have  imaginary part in variables $s$ and $u$.
It follows from unitarity of the $S$-matrix and analyticity that in the physical region, 
the scattering amplitudes at large energies are restricted by the condition: $Im A\le c s\ln^2(s/s_o)$~\cite{Froissart}.  For the amplitude symmetric under  the transformation  $s\to u$, the  subtraction term  
is constant. For the amplitude antisymmetric under the transformation $s\to u$ (negative signature), the  subtraction  term is $\propto s$. 

 The dispersion representation over the momentum transfer $t$ has a similar form:
\begin{equation}
A(s,t)={1\over \pi}\int_{4m^2}^{\infty} du^{\prime} {Im_u A(u^{\prime},t)\over u^{\prime}-u}+{1\over \pi}\int_{4m^2}^{\infty} dt^{\prime} {Im_{t} A(t^{\prime},s)\over t^{\prime}-t} +subtractions \,.
\label{tdispersion}
\end{equation}
Emergence of the concept of Regge poles in the relativistic theory was enabled by the 
combination of Eq.~\ref{Mandelstamrep}  
for the analytic continuation of the decomposition of the scattering amplitudes over partial waves to the crossed channel  with the $S$-matrix unitarity condition in the crossed $t$-channel.  

One starts with the observation that the $S$-matrix unitarity condition becomes diagonal if the 
conservation of the angular momentum is taken into account.
Hence it is convenient to decompose the amplitude over partial waves:
\begin{equation}
A(s,t)=8\pi \sum_{l} f_{l}(s)(2l+1)P_{l}(z) \,,
\label{partialwaverep} 
\end{equation}
where $z=1+2t /(s-4m^2)$; $P_l(z)$ are  the Legendre polynomials.
For simplicity, we consider collisions of hadrons with spin zero. 
(A 
generalization  to  the case of  scattering of 
particles with non zero spin is straightforward but  would 
  make formulae unnecessary lengthy.)
\begin{equation}
f_l(s)={1\over 2}\int_{-1}^{1} dz P_l(z){A(s,t)\over 8\pi} \,.
\end{equation}
The normalization of $f_l$ is chosen so that unitarity of the $S$-matrix has the form:
\begin{equation}
\mbox{Im} m f_l(s)={1\over 2}f_l(s)f^{\ast}_l(s)+{\rm positive~ terms} \,.
\end{equation}

In high energy processes,
orbital momenta essential in the scattering process are large. Hence it is legitimate to substitute 
the sum over the orbital momenta $l$ by the integral over the impact parameters $b$, $l+1/2=pb$,
where $p$ is the center of mass  momentum. 
(The factor of 1/2 follows from the necessity to reproduce the formulae
of the semi-classical approximation in non-relativistic quantum mechanics.)  At large energies, $p\approx \sqrt{s}/2$. Thus we derive the impact parameter representation of the  amplitude:
\begin{equation}
A(s,t)=4\pi s \int db \, b\, f(b,s) J_0(q_t b)=(2s)\int d^2\vec{b} \exp(i\vec{q}_t\cdot \vec{b}) f(b,s) \,,
\label{Impactparameterrepresentation}
\end{equation}
where $q_t=p\sin(\theta)$ and  $\theta$ is the  c.m.~scattering angle.  In the derivation we 
used the asymptotic expression for the Legendre polynomials at large $l$: $P_l(\theta)\approx 
J_0((l+1/2)\theta)$ and the integral representation of  the Bessel function $J_0$: $J_0(q)=(1/2\pi)\int_0^{2\pi} \exp(iq\cos(\phi)) d\phi$.

We  can use the dispersion representation of the amplitude over the momentum transfer and properties 
of the Legendre functions
of the second kind to derive  the representation that can be easily continued into the complex plane of the orbital momentum: 
\begin{equation}
f_l(s)={1\over 2\pi i}\int_{C} {Q_l(z)A(t,z)\over 8\pi} dz \,.
\label{countourl}
\end{equation}
Here the contour of integration 
encircles the $[-1,1]$ interval  on the real axis.  The 
integrand has singularities outside the contour at
\begin{equation} 
z_{1}=1-((4m^2)/(s-4m^2))\,, \quad z_2=-1 +((4m^2)/(s-4m^2)) \,,
\end{equation}
corresponding to the  singularities of the amplitude at $t=4m^2$  and $u=4m^2$.

The Legendre functions of the  second kind satisfy the relation:
\begin{equation} 
Q_l(z)=\frac{1}{2} \int_{-1}^{1}   P_l(z')dz'/(z'-z) \,.
\end {equation}
The advantage of this function is that for $z\gg 1$:
\begin{equation}
Q_l(z)\propto {1\over z^{l+1}} \,,
\end{equation}
and that  for $-1\le z\le 1$,
\begin{equation}
Q_l(z+i\epsilon)-Q_l(z-i\epsilon)=-i\pi P_l(z) \,.
\end{equation}

For the discussion in the next section, it is important to derive the representation of the partial wave in 
$t$-channel.  For sufficiently large $l$, the integration contour  in Eq.~\ref{countourl}  can be deformed 
around  the singularities of the amplitude since the integral over the large circle is equal to zero.  
Thus, another  representation  arises:
\begin{equation} 
f_l(t)={1\over \pi }\int_{z_1}^{\infty}dz {Q_l(z) A_t(z,s)\over 8\pi}+(-1)^{l}{1\over \pi } \int_{z_2}^{\infty} {Q_l(z)A_t(z,u)\over 8\pi}dz \,.
\end{equation}

The presence of the factor of $(-1)^{l}=\exp(i\pi l)$   precludes analytic continuation of the amplitude 
to the  complex plane of the  orbital momentum
since the factor $(-1)^{l}=\exp(i\pi l) $ increases rapidly with $lm\, l$.
To remove the factor of  $(-1)^l$, 
it is convenient to introduce the functions $f^{\pm}_l(t)$ which are  symmetric and antisymmetric under the 
$s\to u$ transposition, respectively---functions with the positive and  negative signature. Thus, 
\begin{equation} 
f_l^{\pm}(t)={1\over \pi}\int_{z_1}^{\infty} {Q_l(z)A_t(z,s)\over 8\pi} \pm {1\over \pi}\int_{z_2}^{\infty} {Q_l(z)A_t(z,u)\over 8\pi} \,, 
\label{GribovFroissart}
\end{equation} 
which provides the analytic continuation of the partial waves to the  complex plane of the angular momentum.  
Above formulae are known as the Gribov--Froissart projection~\cite{GribovFroissart}.

\section {Regge poles in the $S$-matrix theory}
\label{section3}

\subsection{Regge poles and $t$-channel unitarity}
\label{subsection3.1}

T.~Regge  found in the non-relativistic quantum mechanics that the scattering amplitude in the unphysical region corresponding to large imaginary scattering angles, 
$\cos(\theta)\to \infty$,   has the following form:
\begin{equation}
A(s,t)\propto \cos(\theta)^{l(E)} \,,
\label{Regge}
\end{equation}
where  $l(E)$ is  the eigenvalue of  the operator of the orbital momentum $l$  at a given energy $E$. The eigenvalues of the energy $E_n$ follow from the condition that 
$l=n$, where $n$ is an integer number.  Thus  the concept of the   Regge trajectory $l(E)$ allows one 
to describe both the energy eigenstates and the asymptotic behavior of the amplitude 
\cite{Regge}.   In a relativistic theory, 
the $s$-channel $\cos(\theta) \propto s$ and $E $ should be substituted by $t$. 

S.~Mandelstam observed that  in a relativistic theory,  the kinematics of large $s$   
and  fixed $t\ll s$   corresponds to   physical processes with 
$s\gg 4m^2$ which are usually called  the  crossed channel with respect to the $t$-channel processes. 
As cited in \cite{FGZ}, he suggested that the Regge pole behavior, would allow for 
a simple description of bound states. 
The key tool  for the derivation of the basic properties of Regge trajectories and the calculation of the amplitudes of high energy  processes is the partial amplitudes in the $t$-channel analytically continued to the complex plane of orbital momentum---the 
Gribov--Froissart    projection discussed in the previous section.   Regge trajectories describe the sum of poles of these amplitudes which follow from unitarity of the $S$-matrix in the crossed channel.   The concept of the Regge trajectory $l(t)$  is useful for the description of hadron resonances with the same quantum numbers (except for spin) and for the calculation of amplitudes 
of high energy processes.

Important properties of Regge trajectories  follow from the two-particle $t$-channel unitarity condition continued to 
the angular momentum plane:
\begin{equation}
{1\over 2} (f_l^{\pm }(t+i\epsilon))-f^{\pm}_l(t-i\epsilon))=(1/2) f^{\pm}_l(t+i\epsilon) f^{\pm}_l(t-i\epsilon) \,,
\label{Runitarity}
\end{equation}
which can be rewritten as
\begin{equation}
f^{\pm}_l(t)-f^{\pm \ast}_l(t)=f^{\pm}_{l}(t) f^{\pm \ast}(t) \,.
\label{unitaritycom}
\end{equation}
In the proof one uses the observation that for $t\le 4m^2$, $f^{\pm}$ are real and hence  
$f^{\pm}(t-i\epsilon)=[f^{\pm}(t+i\epsilon)]^{\ast}$. The two-particle $t$-channel unitarity condition 
is exact for $4m^2\le t\le 16m^2$ and allows one to prove the existence of Regge poles in relativistic amplitudes
and to establish some of their properties~\cite{Runitarity,G-P62}.

The Gribov--Froissart projection  has pole at $l=l(t)$:
\begin{equation} 
 f_l(t)=c/(l-l(t)) \,.
\end{equation}
Taking into account the real and imaginary parts of the trajectory $l(t)$, it is easy to find out 
that in the vicinity of $l$, the amplitude is described by the    
Breit--Wigner formulas,  see also \cite{CFrautschi2}.
Thus the concept of the Regge trajectory (which is often called the moving trajectory)  
describes  hadronic resonances with the same quantum numbers, except for the spin.

Using these equations and iterating one Regge trajectory, it is easy to show that the Regge trajectory   
generates the moving pole singularities in the complex plane of the angular momentum $l$  
 \cite{G-P62,GPT}.

\subsection{Regge poles and high energy behavior of amplitudes of physical processes}
\label{subsection3.2}

To demonstrate the role of Regge poles  in high energy processes in the relativistic theory, it is convenient to use the method  applied by Sommerfeld   to the problem of diffraction of radio waves around Earth. 
The task is to find an analytic  function of $l$ which coincides with $f_l(t) $ for integer points $l=0$, $1$, $2$, $\dots$.  

The decomposition of the positive signature amplitude over partial waves in the $t$-channel diverges with an 
increase in  $s$ because the Legendre polynomials---being the functions of 
$z=1+2s/(t-4m^2)$---increase as powers of $s$:
\begin{equation}
P_{l}(\cosh(\alpha))_{l\to \infty}\propto \frac{\exp^{(l+1/2)\alpha}} 
{\sqrt{2\pi \sin(\alpha)}} \,.
\end{equation}
The partial amplitudes are restricted by the $S$-matrix unitarity and their imaginary parts are positive.  
Thus one needs to continue partial waves
analytically to the angular momentum plane.
The procedure was explained in the previous section.

The first step is to identically represent the amplitude as a contour integral over the orbital momenta $l\ge 0$ around 
the real axis: 
\begin{equation}
A^{+}(s,t)=(1/i)\int_{C} dl\, \xi_l^+(t) f^{+}_{l}(t)(2l+1) P_{l}(1+2s/(t-4m^2)) \,,
\label{ZW}
\end{equation}
where $\xi_{l}^+(t)$ is called the positive  signature factor. It can be written as
\begin{equation}
\xi_{l}^+(t)=[1+(-1)^{i\pi l}]/\sin(\pi l)=\exp {i(\pi l/2)}/\sin(\pi l/2) \,.
\label{+signature}
\end{equation}
Taking the residues over the poles of $1/\sin(\pi l/2)$ would recover
 Eq.~\ref{partialwaverep}.

For the  amplitude  antisymmetric with respect to the transposition $s\to u$, 
the signature factor is:
\begin{equation}
\xi_{-}(l)=\exp {i(\pi l/2)}/\cos(\pi l/2) \,.
\label{-signature}
\end{equation}

It follows from the location of singularities in the plane of $t$ that
$f_l(t) \propto \exp(-\mu l)$ for $l\to \infty$, where $\mu$ is the minimal mass 
in the singularities over $t$. Hence, for the integration contour 
$C$, one can take  the straight line between the points $l_0 - i\infty$ and  $l_0 + i\infty$. 
The contour can be moved to the left until it encounters singularities of the amplitudes in $l$ that 
do not allow further shifting of the counter to the left.  
In this discussion, $l_0\ge -2$ since the positive signature amplitude   cannot decrease with 
$s$ faster  than $1/s^2$.  This property follows from the fact that imaginary part of this amplitude is always positive.
  In addition, one should take into account poles and cuts in the $l$-plane.
  Thus, the expression for the positive signature amplitude reads: 
 \begin{equation}
A^{+}(s,t)=(1/i)\int_{l_0-i\infty}^{l_0+\infty} dl \xi^+(l)  f^{+}_{l}(t)(2l+1) P_{l}(1+2s/(t-4m^2)) + \Delta \,,
\label{}
\end{equation}
where $\Delta$ is the contribution of the Regge poles and moving cuts.  
A similar analysis can be performed for the negative signature amplitude.

The contribution of the Regge pole $f_l(t)\propto 1/(l -\alpha(t))$  to 
the amplitude $A$ has the form:
\begin{equation}
g_{a,a^{\prime}}(t)g_{b,b^{\prime}}(t)(s/s_0)^{\alpha(t)} \xi(t) \,.
\label{reggepolefact}
\end{equation}
Factorization of the  dependence  of the amplitude on the properties of the particles $a$ and $b$ follows from the unitarity condition for the partial waves in the $t$-channel~\cite{G-P62}.

V.~Gribov has demonstrated \cite{Gribov61}  that the textbook models that assume that the total cross section of hadronic collisions  at large $s$ is energy independent and the $t$ dependence of the elastic amplitude does not depend on $s$, i.e., $A(s,t)=i\,sf(t)$,  are  incompatible with probability conservation 
in the crossed channel as given  by  Eq.~\ref{unitaritycom}.  Such a behavior  corresponds to the fixed 
pole in the orbital momentum plane. However, a single pole $1/(l-1)$ on the left-hand  side of this equation  cannot be equal to a double pole  $1/(l-1)^2$ on the right-hand side of it.

Common wisdom based on the $S$-matrix approach was that any hadron is a bound state of other hadrons but not of some 
elementary constituents.  This  concept was implemented by assuming that hadrons belong to Regge trajectories.   
In quantum chromodynamics, hadrons are bound states of elementary particles---quarks and gluons---and  physical states 
contain no free quarks and gluons (the hypothesis of confinement of quarks and gluons).
This makes  the Regge trajectory description of hadrons even more plausible.

\subsection{Regge trajectories}
\label{subsection3.3}
G.~Chew and S.~Frautchi  suggested to describe the spectrum of hadrons in terms of the 
Regge pole trajectories~\cite{CFrautschi2}. 
The mass of a hadron follows from  the equation:
\begin{equation}
\alpha(t_h)=J_h \,,
\end{equation}
where $J_h$ is the spin of the resonance and $t_h$ its mass. 
The scattering amplitude has poles for these values of $l$ and $t$.

The observed spectrum of hadrons made of light quarks as well as cross 
sections 
of exclusive processes with non-vacuum quantum numbers in the crossed channel are well described 
by the Regge trajectories linear in the momentum transfer $t$~\cite{CFrautschi2}:
\begin{equation}
\alpha(t)=\alpha_0+\alpha^{\prime} t \,.
\end{equation}
The data prefer practically the same slope $\alpha^{\prime}$ for all hadrons made of the light $u$, $d$, and $s$ quarks:
\begin{equation}
\alpha^{\prime} \approx 1 \ {\rm GeV}^{-2} \,.
\label{strings}
\end{equation}  
The values of the intercept $\alpha_0$ for the leading trajectories are the following:
\begin{equation}
\alpha_{\rho}(t=0)\approx 0.5 \,, \quad
\alpha_{A_2}(t=0)\approx 0.3 \,, \quad \alpha_{\pi}(t=0)\approx 0 \,.
\end{equation}

For completeness, we also enumerate  other important results which, however, we will not use in this chapter:

(i)  An analysis of the experimental data on meson resonances and exclusive processes indicates that the 
meson trajectories with positive and negative signatures are close,  see  \cite{Nonvacuum} for  a review and references.

(ii)  Fermion trajectories should also contain the term $\propto \sqrt t$, otherwise the trajectories with 
opposite signatures will be degenerate \cite{GribovComplex}. 
However, the data on two body processes dominated by fermion exchanges
are very limited. 

(iii) There has been proposed a technique
of calculation of  the corresponding QED amplitudes in the angular momentum plane
~\cite{GellMann:1964zz}
which allows  to establish whether constituents of QED belong to moving Regge pole trajectories.   A more powerful technique of identifying  and calculating the leading Feynman diagrams in a quantum field 
theory containing vector particles  has been developed in \cite{GribovComplex}.

(iv)
It has been shown that in non-abelian gauge theories massive  vector 
mesons~\cite{Reggezationvector} and fermions~\cite{Reggezationfermion}  are reggeized in  
the perturbative regime.  In QCD the presence of infrared singularities related to the zero mass of the 
gluon  requires to 
take into account  
addition infrared factor.

(v) It has been observed that the sum of $s$-channel resonances produces linear Regge trajectories in the  
$t$-channel~\cite{Dolen-Horn-Schmidt}. This hypothesis is the basis of string models.

\vspace{0.5cm}

The knowledge of Regge trajectories allows one to predict cross sections of
two-body high energy processes with non-vacuum quantum numbers in the crossed channel.

\subsection{Regge pole theory for non-vacuum exchanges}
\label{subsection3.4}

It has been suggested that reactions with non-vacuum quantum numbers in the crossed channel are dominated at high energies 
by the exchange of
the Regge trajectories allowed by conservation of charge, spatial parity, G-parity and isotopic spin \cite{CFrautschi1}. 

 Assuming the dominance of the Regge pole contribution to the amplitude of diffractive processes, one obtains: 
 \begin{eqnarray}
 \label{Rreggeon} 
\frac{d\sigma(a+b\to a^{\prime}+b^{\prime})}{dt} &=&
{g^2_{aRa^{\prime}}(t) g^2_{bRb^{\prime}}(t)\over \sin^2(\pi\alpha_{R}(t)/2)} 
\left({s\over s_0}\right)^{2(\alpha_{R}(t)-1)} \nonumber \\
&+& {g^2_{aR^{\prime}a^{\prime}}(t) g^2_{bR^{\prime}b^{\prime}}(t)\over \cos^2(\pi\alpha_{R^{\prime}}(t)/2)} \left({s\over s_0}\right)^{2(\alpha_{R^{\prime}}(t)-1)} \nonumber \\ 
&+& 2 \,\frac{g_{aRa^{\prime}}(t) g_{bRb^{\prime}}(t)}{\sin(\pi\alpha_R(t)/2)}
{g_{aR^{\prime}a^{\prime}}(t) g_{bR^{\prime}b^{\prime}}(t)\over 
\cos(\pi\alpha_{R^{\prime}(t)}/2) } 
\left({s\over s_0}\right)^{(\alpha_{R}(t)+\alpha_{R^{\prime}}(t)-2)} 
\nonumber \\
&\times& \cos((\pi/2)(\alpha_{R}-\alpha_{R^{\prime}}))  \,.    
\end{eqnarray}

\begin{figure}[h]  
\centering
 \includegraphics[width=0.8\textwidth]{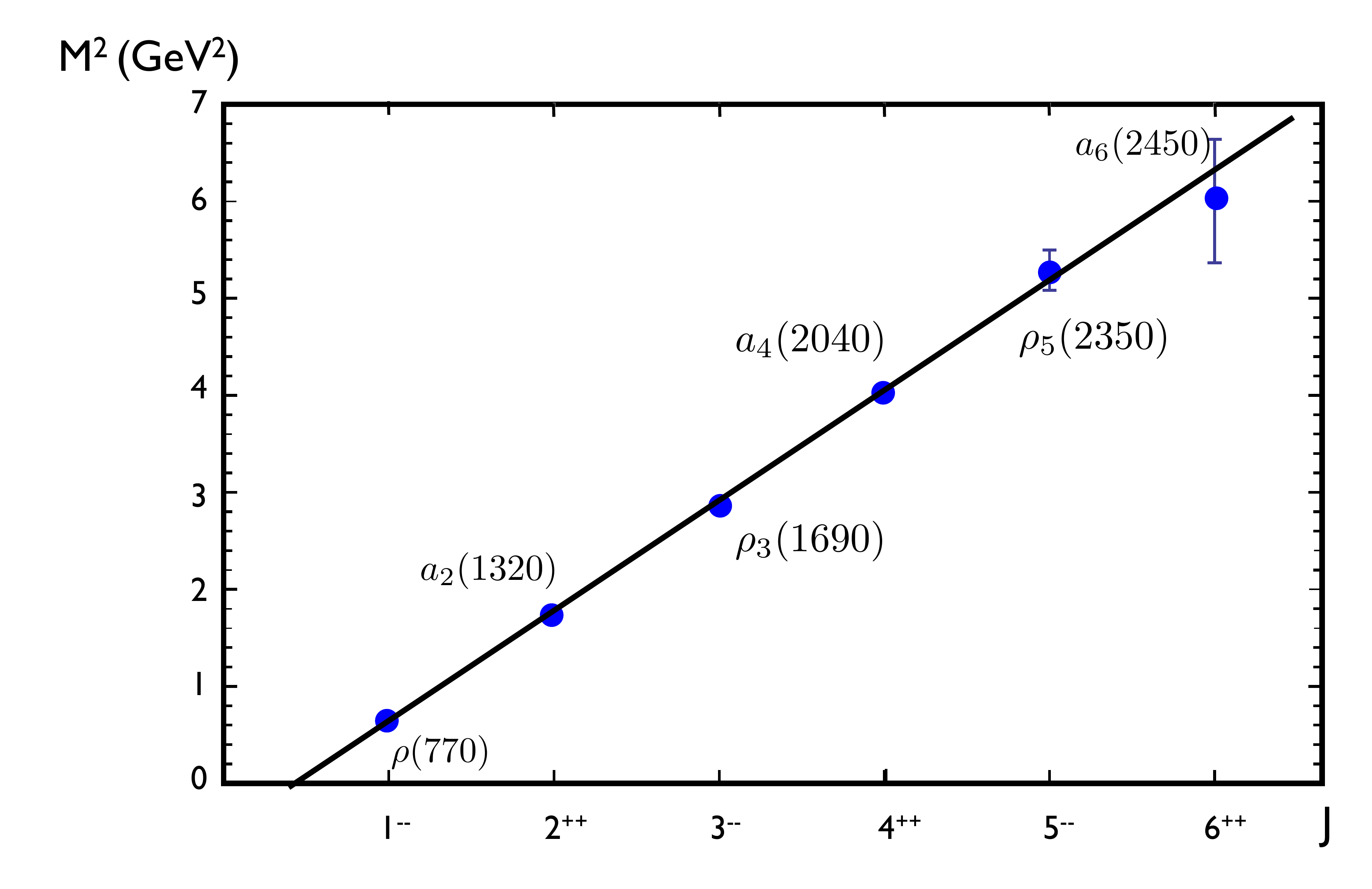}
 \caption{The lowest mass mesons lying on the $\rho$ Regge trajectory and on the nearly degenerate $A_2$ trajectory.}
\label{regge}
\end{figure}

This expression takes into account the exchanges 
of
the trajectories with positive ($R$) and negative ($R^{\prime}$) signatures.  
Fitting Regge trajectories as a linear function of $t$ gives a good description of the mass 
spectrum of the resonances belonging to the corresponding trajectories (see, e.g.  Fig.~\ref{regge}). 
The assumption that the trajectories remain linear for  $t \le 0$ leads  to a reasonable description of 
the data using Eq.~\ref{Rreggeon}.

 The data exist for the cross sections of the following processes (see the review \cite{Nonvacuum}): 
$\pi^{-}+ p\to \pi^0+n$, $\pi^{-}+p\to \rho(A_2)+n$,  $K^-+p\to \phi +n$, $p+p \to \Delta +N$,  etc.
These data are described well at small $t$  by the exchange of 
a few leading Regge meson trajectories.    
In the kinematics of large $s$ and  small $u$, the exchange 
of
the baryon trajectories 
dominate 
such processes as, e.g.,  $\pi^{-}+p\to p+\pi^{-}$.   The data are described well in terms of linear 
meson and baryon trajectories for $-t \le 0.5$ GeV$^2$ (see, e.g., the review \cite{Nonvacuum}). At the 
same time,  for  $-t\ge 1$ GeV$^2$  data can be interpreted as the
evidence for flattening of meson and baryon trajectories to the values 
corresponding to the exchange 
of
reggeized $q\bar q$ ($qqq$) systems \cite{Strikman:2007nz}.  Further experimental studies of high energy two-body reactions in the $-t ={\rm const} \ge 1$ GeV$^2$ limit would be highly desirable.

\subsection{Non-universality of the Regge trajectories 
for the bound states containing heavy quarks}
\label{subsection3.5}

The states containing heavy quarks belong to the Regge trajectories with
the slope different from the one of 
the Regge trajectories for  hadrons made of light quarks.   We will give here 
two examples. 

(i)  In the $M_Q\to \infty$ limit, the masses of $Q\bar q$ states are $M_{Q\bar q}(n) =M_Q+\Delta_n$, 
where $\Delta_n$ are independent of $M_Q$.    This result follows from quarkonium models and is 
probably valid in QCD. Therefore,  for linear trajectories one has:\begin{equation}
\alpha'_{Q\bar q}\approx \frac{1}{2M_Q (\Delta_1-\Delta_0)} \,.
\end{equation}
This  slope is different from the universal slope suggested in the string models 
for a hadron trajectory consisting of light quarks.

(ii) For hadrons with hidden flavor, the slope of the $Q\bar Q$ trajectory decreases  
with the mass of the heavy quark in the  $M_Q\to \infty$ limit as:   
\begin{equation}
\alpha'_{Q\bar Q} \propto {1\over \alpha^2_{s}M^2_Q} \,.
\end{equation}
For an estimate, we used here quarkonium models with the Coulomb interaction between quarks.

To conclude, the large masses of heavy quarks supply extra scales in addition to the 
$\Lambda_{QCD}$ scale, which suggests the existence of a variety of phenomena 
beyond   the framework of the one-scale $S$-matrix theory~\cite{Frankfurt}.   For example,  $J/\psi$ 
and $\Upsilon$  have significantly smaller radii than  the pion:  
\begin{equation}
r_{\pi}=0.5 \ {\rm fm} \,, \quad  r_{J/\psi} \approx 0.2 \ {\rm fm} \,, \quad r_{\Upsilon} \approx 0.1 \ {\rm fm} \,.
 \end{equation}
As a consequence of color screening and asymptotic freedom, heavy quarkonia relatively weakly 
interact with hadrons made of light quarks.
(The probability of the pion field around a heavy quarkonium  in the  ground state is close to zero.)     
The observed total and partial widths  and the cross  section of diffractive photoproduction of $J/\psi$ 
at moderate energies are significantly smaller than those for vector mesons made of light quarks.

\section{Pomeron theory of high energy soft QCD  processes}
\label{section4}

\subsection{Introducing the concept of the Pomeron exchange}
\label{subsection4.1}

We discuss briefly here  the hypothesis of the Pomeron exchange dominance in the amplitudes of high energy processes. 

The Pomeranchuk  theorem, 
\begin{equation}
\sigma(h+T)=\sigma(\bar h+T) \,,
\end{equation}
was proven initially under the assumption that  $\sigma_{\rm tot}(hN) \to \mbox{const}$ for $s\to \infty$. 
The proof uses analytic properties of amplitudes in the energy plane and that  the amplitude is predominantly imaginary at high energies. Indeed, $A^{+}(s,t)=c^{+}[s\ln(-s)+u\ln(-u)] \approx ic^{+}\pi s$ which should be compared with $A^{-}=c^{-} [s\ln(-s)-u\ln(-u)] \approx 2c^{-} s\ln(-s)$. The condition: $A^{+}\gg A^{-}$ requires that $c^{-}=0$---the Pomeranchuk theorem~\cite{Pomeranchuk}.
In the more realistic case of a growing cross section, a weaker form of the theorem for $s\to \infty$ can be 
proven:
\begin{equation}
\sigma_{\rm tot}(h+T)/\sigma_{\rm tot}(\bar h+T)\to 1 \,.
\end{equation}
This theorem is confirmed by the data on $\sigma_{\rm tot}(pp)$ and $\sigma_{\rm tot}(p\bar p)$.

As we explained in section~\ref{subsection3.2}, the behavior of the scattering amplitude
$A(s,t)=isf(t)$, which is typical for quantum mechanical problems with the absorptive interaction, 
contradicts the $S$-matrix unitarity relation for partial waves in the crossed $t$-channel.  
To resolve this contradiction, V.~Gribov~\cite{Gribov61} suggested the behavior of the scattering amplitude 
(for large $s$ and small $t$) of a 
general form that  does not contradict the unitarity of the $S$-matrix in the crossed channel:
\begin{equation}
A(s,t)=i s^{\alpha(t)} F(\ln(s), t) \,,
\end{equation}
where   $F$ is a slow function of $\ln(s)$ rapidly decreasing with an increase of  $-t$. 
The positive value of  $d\alpha(t)/dt$ at $t=0$, which follows from the positive value of the partial waves 
($\mbox{Im} f_l(s)\ge 0$) and properties of the Legendre polynomials~\cite{Gribov61},  
leads to a decrease in  the average $-t$ for elastic scattering  with an increase in  $s$.  

Thus, the  self-consistency of the theory requires  that the radius of the hadron--hadron interaction  
should increase with energy.  

V.~Gribov~\cite{Gribovelastic}  suggested   that amplitudes of high energy processes are dominated 
by the special Regge pole trajectory exchange with the vacuum quantum numbers in the crossed channel 
and calculated elastic and total cross sections.  In parallel,
G.~Chew and S.~Frautschi~\cite{CFrautschi1} 
 also drew attention to the hypothetical Regge pole with $\alpha(0)=1$ and the vacuum quantum numbers
 that  would be responsible 
 for the forward scattering processes.
 This hypothesis reproduces the Pomeranchuk theorem   
 and leads to an increase of the radius of the interaction with energy \cite{GPT,Gribov61,Gribovelastic,GP}:
\begin{eqnarray}
A(hT)=g_{h\Pomeron h}(t)g_{T\Pomeron T}(t)
{\left({s\over s_0}\right)^{\alpha_{\Pomeron}(t)}
+\left({u\over u_0}\right)^{\alpha_{\Pomeron}(t)}
\over \sin(\pi\alpha_{\Pomeron}(t))}
\nonumber \\
=\frac{g_{h\Pomeron h(t)}g_{T\Pomeron T}\left({s\over s_0}\right)^{\alpha_{\Pomeron}(t)}\exp {(i\pi \alpha_{\Pomeron}(t)/2)}}{\sin(\pi\alpha_{\Pomeron}(t)/2)}&& \,,
\label{Pomeron}
\end{eqnarray}
where $g_{h\Pomeron h}(t)$ and $g_{T\Pomeron T}(t)$ are the residues of the Pomeron pole. The signature 
 factor  follows from the symmetry of the amplitude due to the Pomeron exchange  under the $s\leftrightarrow u$ transformation
 as explained in the previous section.

\subsubsection{ Experimental evidence for the Pomeron trajectory}
\label{subsubsection4.1.1}

The Pomeron trajectory  is usually parametrized as a trajectory linear in $t$:
\begin{equation} 
\alpha_{\Pomeron}(t)=\alpha^0_{\Pomeron}+\alpha^{\prime}_{\Pomeron} t \,.
\label{Ptrajectory}
\end{equation}
The observed dependence of the total cross section of  $p p$ collisions on energy, which ranges from the fixed target energies  to the highest energies currently measured at the  LHC ($\sqrt{s}\mbox{=7 TeV}$), is well described  by the Pomeron 
intercept~\cite{Landshoff}:
\begin{equation}
\alpha^0_{\Pomeron}\approx 1.08 - 1.1 \,. 
\end{equation}
The generally accepted value of the  slope of  the Pomeron trajectory is
\begin{equation}
\alpha^{\prime}_{\Pomeron} \approx 0.25 \ {\rm GeV}^{-2} \,.
\label{alphaprime}
\end{equation}

The hypothesis of the dominance of the Pomeron exchange in high energy processes has found a number of experimental confirmations.  Let us briefly outline basic discoveries.

(i) The same Pomeron intercept describes  the energy dependence  of the total and elastic cross sections of $pp$ collisions and of the total  cross section of exclusive photoproduction of 
$\rho$ mesons off the  proton target measured at HERA~\cite{Landshoff}.  
The cross section  of exclusive  photoproduction of $J/\psi$ mesons increases with energy more rapidly, 
see a recent summary in~\cite{Levy:2007fb} (for the explanation, see the following subsections).  

(ii)    The shrinking of the diffractive peak with an increase of the collision energy has been predicted 
in~\cite{CFrautschi1,Gribovelastic}  and observed in $pp$ and $p\bar p$  collisions (see the summary of the data in  \cite{Totem}). The data for $\sqrt{s} \le 1.8$ TeV are consistent with the Pomeron trajectory being linear for $-t \le 0.5$ GeV$^2$.

(iii)  The first LHC data \cite{Totem} report the $t$-slope corresponding to a faster rate of the shrinkage: 
 for $\sqrt{s}$ between $\sqrt{s}=2$ TeV and  $\sqrt{s}=7$ TeV,
$\alpha^{\prime}_{\Pomeron} (-t < 0.15 \ \mbox{GeV}^2)\sim \mbox{0.55  GeV}^{-2}$; this value 
significantly increases with energy  for larger $-t$.  The interference between the single and multi-Pomeron exchanges produces a qualitatively similar behavior which we illustrate below.  
At small $t$, the amplitude for the single Pomeron exchange is conveniently parametrized as:  
$\mbox{Im}  A_1=c_1s^{\alpha_{\Pomeron}(t)-1}\exp(Bt/2)$.
Taking into account interference with the double Pomeron exchange, we obtain for the
square of the ratio of the full amplitude to the single Pomeron exchange amplitude: 
$\left|\mbox{Im}  A/\mbox{Im} A_1\right|^2\approx 1- 2(c_2/c_1) s^{2\alpha_{\Pomeron}(t/4)-\alpha_{\Pomeron}(t)-1} \exp(-Bt/4)$,  where $c_i$ are positive and $c_2/c_1$  can be evaluated in the color fluctuation model.  
This formula produces the minimum which moves to smaller $t$ with an increase of energies.
This phenomenon has been discussed in the eikonal models, see, e.g., \cite{M.Bloc}; 
the problems of this approximation will be discussed in section~\ref{section6}.

\begin{figure}[h]  
   \centering
   \includegraphics[width=0.8\textwidth]{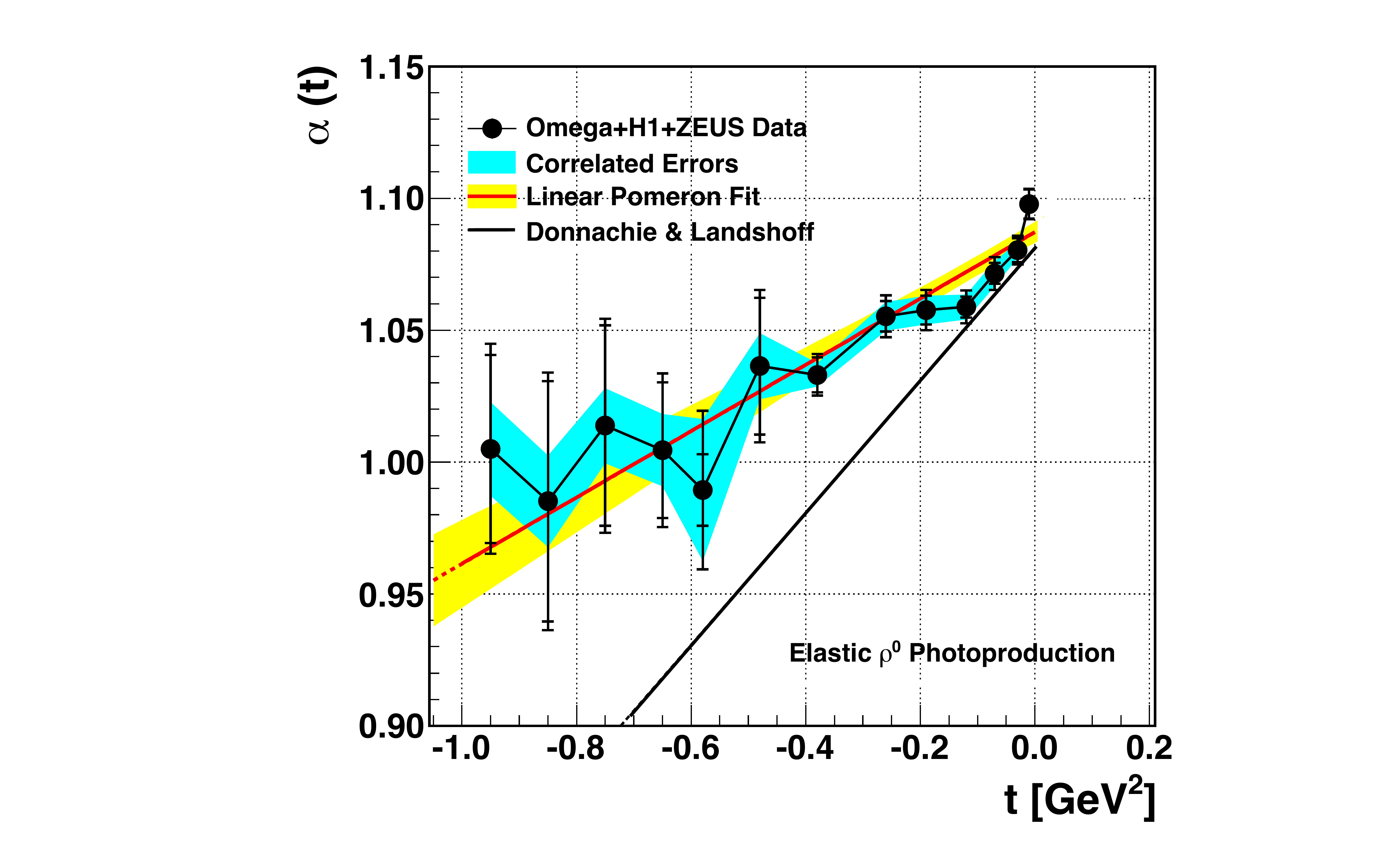}
 \caption{The data on the $t$ dependence of the Pomeron trajectory as extracted from $\rho$-meson photoproduction with a linear fit to the data and the Donnachie--Landshoff parametrization of the $pp$ data. }
\label{alpha}  
\end{figure}

(iv) 
The global analysis of the world data on  exclusive
 $\rho $-meson photoproduction~\cite{List:2009pb} gives $\alpha'_{\Pomeron} = 0.126 \pm 0.013({\rm stat}.) \pm  0.012({\rm syst}.)$ GeV$^{-2}$ indicating non-universality of the $t$ dependence of the effective Pomeron trajectory,  
 see Fig.~\ref{alpha}.  The non-universality can also be seen from the comparison of the $t$ dependence of $\alpha_{\Pomeron}(t)$ with the linear Pomeron trajectory that  describes the $pp$ data for the same energy interval. However the data do not exclude a possibility that $\alpha^{\prime}_{\Pomeron}$ for the $\rho$ case is the same as for the $pp$ case 
 for $-t \le 0.15$ GeV$^2$.  Note here that selection of small $t$   enhances 
the contribution of peripheral collisions and hence suppresses the effects of 
multi-Pomeron exchanges and of
blackening of the interaction at small impact parameters discussed in the next subsections. 
The data also indicate that  
$\alpha_{\Pomeron} \ge 1 $ for the entire studied $t$-range ($-t\le 1$ GeV$^2$).    
In the case of elastic photoproduction of $J/\psi$, $\alpha_{\Pomeron} \approx  1.2$
for forward scattering
and $\alpha_{\Pomeron} \approx  1$  for $-t\approx m^2_{J/\psi}$.
Such a behavior is natural for the pQCD regime where double logarithmic terms are relevant for 
the interrelation between the $t$ and $s$ dependences~\cite{BFS}.   

(v)   
It is well known from the analysis of the ladder 
diagrams  relevant for the total DIS cross section~\cite{BFS94}  that $\alpha_{\Pomeron}^{\prime}$  decreases with the virtuality of the external probe due to the suppression of the Gribov diffusion in 
the impact parameter space.  The observed  value of  $\alpha_{\Pomeron}^{\prime}$ for $J/\psi$ photoproduction is definitely much smaller than that 
for $pp$ scattering, see Eq.~\ref{bg_param_last}.  However, the errors in the $J/\psi$ case (see  Eq.~\ref{bg_param_last})  do not allow one  to establish whether  $\alpha_{\Pomeron}^{\prime} $
 extracted from the global analysis of the data for the $J/\psi$ case  is  smaller than for the $\rho$ case.

(vi)   One of important confirmations of the dominance of the Pomeron exchange is the observation of 
the triple Pomeron diffraction~\cite{KancheliTPL,MuellerTPL}  at FNAL~\cite{DFNAL} and at the LHC~\cite{DLHC}, see  the discussion in the next subsection.

The dominance of the Pomeron exchange in the total cross section allows for a relationship between 
the cross sections of diffractive and inelastic processes~\cite{AGK}. Thus, the total cross section of diffraction 
is unambiguously calculable in terms of the cross section of inelastic processes, i.e., diffraction in high energy processes is a shadow of inelastic processes.

\subsection{Pomeron calculus}
\label{subsection4.2}
In this section we will discuss the physics of interacting Pomerons.  The ideas and methods discussed 
in this section are now widely used in the evaluation of amplitudes of high energy processes, 
for a detailed discussion of the subject and proper references, see~\cite{GribovComplex}.

A single Pomeron trajectory generates multi-Pomeron branch points and related cuts in the angular momentum plane.  This result follows from  unitarity of the $S$-matrix for partial amplitudes continued 
into the angular momentum plane  in 
the crossed $t$-channel~\cite{G-P62,GPT2}. An exchange 
of $n$ Pomerons  leads to a branch point in the angular momentum plane located 
at~\cite{S.Mandelstam63, G-P62,GPT2}:
\begin{equation}
j(t)\alpha_{\Pomeron}(t/n^2)-n+1\approx n [\alpha_{\Pomeron}(0)-1] +(\alpha^{\prime}_{\Pomeron}/n) t \,.
\label{branch}
\end{equation}

Realization of the Pomeron in the form of ladder diagrams helped to develop  the diagram technique~\cite{PomeronCalculus},   which allows one to calculate the contribution of any number of Pomeron cuts 
to the cross section and, hence, helped to develop the Pomeron calculus.  

\begin{figure}[h]  
   \centering
   \includegraphics[width=0.9\textwidth]{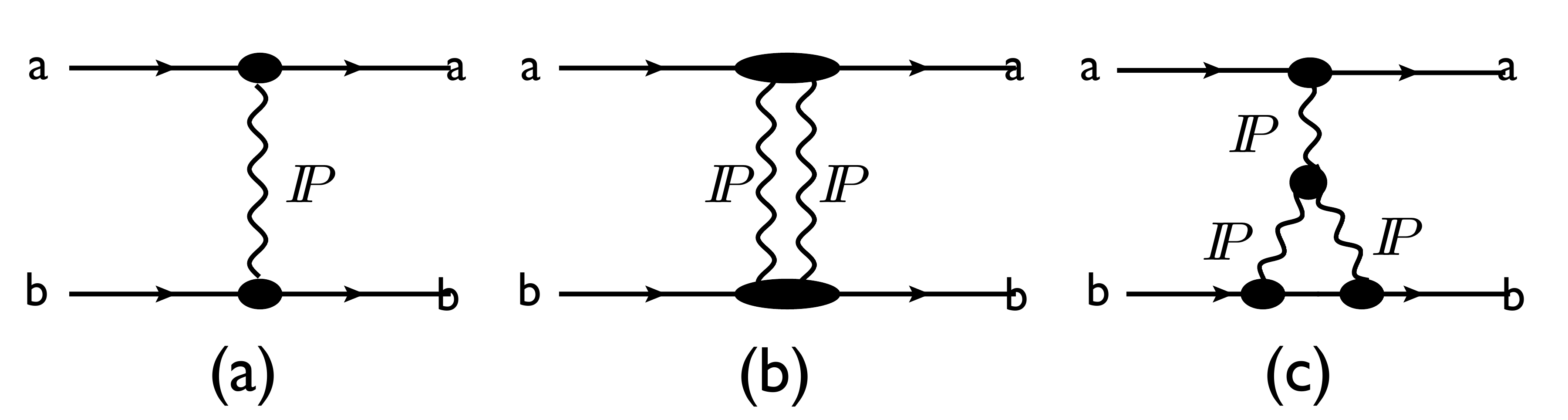}
 \caption{The single Pomeron exchange (a), the two-Pomeron exchange (b)  and the triple Pomeron (c) diagrams.}
\label{pomeron}
\end{figure}

The simplest  diagrams of the Pomeron calculus, which include a single Pomeron exchange, a 
two-Pomeron exchange, and a triple Pomeron exchange, are shown in Fig.~\ref{pomeron}. The 
coupling of two Pomerons to a hadron is expressed through the diffractive $a+b \to X +b$ cross 
section  that  includes both the elastic and inelastic contributions. For $t$ away from zero,
the latter is enhanced as compared to the elastic contribution since the $t$ dependence of 
diffraction is weaker than that of the elastic scattering. This is because in quantum mechanics form factor of bound state decreases significantly more 
rapidly with an increase in -t than does the  form factor describing sum of inelastic transitions. 

 If the  intercept  of the Pomeron were   equal to unity ($\alpha_{\Pomeron}(0)=1$), all branch points 
 would be located at  $j=1$ for $t=0$. In this case, the calculation of the energy dependence 
 of amplitudes of high energy processes leads to the scaling behavior of the Green's 
 functions in the angular momentum plane in the vacuum channel.  This technique developed in~\cite{GribovMigdal} helped to build  the theory of second-order phase transitions and to analyze 
 long-range fluctuations near the  critical point.

There exists a variety of experimental evidences
for the important role of multi-Pomeron interactions: 

\noindent 
(i)  The triple Pomeron diffraction gives the pattern 
of how multi-Pomeron interactions arise, see the discussion in subsection~\ref{subsection4.4};

\noindent (ii)  The shift to smaller $-t$ of the position of the minimum of the elastic cross section with increase of  energy indicates that 
the increase of the role of the multi-Pomeron exchanges with increase of $s$ and $t$.

(iii) The phenomenon of nuclear shadowing in hadron--nucleus collisions, 
where the incident hadron interacts with several nucleons of the nucleus, arises mostly due 
to multi-Pomeron exchanges.

 \begin{figure}[h]  
   \centering
   \includegraphics[width=0.7\textwidth]{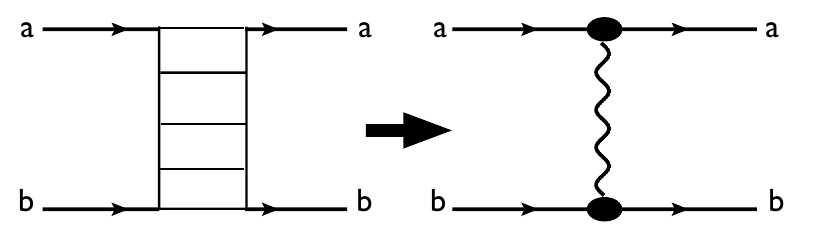}
 \caption{Ladder diagrams building the Pomeron exchange.}
\label{ladder}  
\end{figure}  

\subsection{Gribov diffusion in the impact parameter space within the Pomeron ladder  }
\label{subsection4.2-3}

The dynamical interpretation of the Pomeron exchange in the multiperipheral model or, equivalently, in the parton model, allows us to visualize the Pomeron in the phase space and to evaluate diffusion to large 
impact parameters.   The Pomeron is modeled as the parton ladder (Fig.~\ref{ladder}) where transverse momenta of produced partons are  
independent of energy, $\left<k^{2}_t\right> = k^{2}_0$~\cite{Gribovladder, Feynman}. 
Experimental analyses of the transverse momenta distributions of produced pions and kaons show 
that $\sqrt{k^2_0} \approx 0.3-0.4$ GeV/c. However, the majority of observed mesons originate  
from  decays of heavier hadrons (direct hadrons) which have  significantly  larger average transverse momenta  $\sqrt{k^2_0} \sim 0.5-0.6$ GeV/c already at moderate $\sqrt{s} =10 - 20$ GeV. 
The distances in rapidity $\Delta y$ between partons adjacent in the ladder   are independent of the  
collision energy. (The data on inelastic hadron production in high energy scattering find significant  
rapidity correlations between hadrons  only for $\Delta y\le 1$.)   In a naive picture of the Pomeron 
described by a single ladder, each decay corresponds to a step of a random walk in the $b$ space of 
the length $\propto 1/k_{0}$ and the number 
of steps $\propto \ln (s/\mu^2)/\Delta y$. This leads to an increase of $b^2$ with energy, 
$b^2 \propto \ln(s/\mu^2)$, and hence to finite $\alpha^{\prime}$~\cite{Gribovladder}.

Diffusion to large impact parameters manifests itself directly in elastic hadron--hadron collisions. In the  
impact parameter space, the imaginary part of the 
partial amplitude due to the Pomeron exchange has the following form: 
\begin{equation}
\mbox{Im}  f(s,b^2)=\int {d^2 \vec{q}_t\over (2\pi)^2} \exp(i \vec{q}_t \cdot \vec{b}) {A(s,q^2_t)\over 2s}=\nonumber \\ 
\frac{cs^{\alpha_P(0)-1}\exp(-b^2/2B)}{B} \,,
\label{Pomeronimpact}
\end{equation}
where $B=B_0+2\alpha^{\prime}_{\Pomeron}\ln(s/s_0)$ is the $t$-slope of the elastic cross section.
We parametrize the product of the residues as $c\exp(B_{0}t/2)$ to take into account that  
Pomeron  dynamics is 
dominated by peripheral, that is,  small $t$ processes.

It follows from Eq.~\ref{Pomeronimpact} that the impact parameters essential for elastic $pp$ collisions 
increase with energy as:
\begin{equation}
\left<b^2\right>_{\rm el} \approx B_{0}+2\alpha'_{\Pomeron}\ln(s/s_0) \,.
\label{diffusion}
\end{equation}
In the case of the total cross section, 
\begin{equation}
\left<b^2\right>_{\rm tot} =2 \left<b^2\right>_{\rm el} \,.  
\label{diffusion1}
\end{equation}

 Thus the assumption of the dominance of a single Pomeron exchange implies the dominance of 
 peripheral collisions in hadron--hadron  interactions at ultra high energies.
 
If spin-flip effects are neglected,  the amplitude $A(s,t)$ can be extracted from the differential 
cross section within the forward peak: 
\begin{equation}
d\sigma_{\rm el}/dt=\frac{1}{16\pi}|A(s,t)/s|^{2} \,.
\label{elamp}
\end{equation}
In elastic $pp$ collisions, spin-flip effects are small for $-t\le 1$ GeV$^2$~\cite{asymmetry}.
There exists an insignificant correction due to the possible ambiguity in the sign of the amplitude beyond 
the cross section minimum, which 
for $\sqrt{s}=7$ TeV occurs at $-t \approx 0.5$ GeV$^2$.
   Thus the impact parameter distribution can be measured using the elastic cross section data.

The parton ladder model 
of the Pomeron
allows us to evaluate properties of multi-Pomeron exchanges and Pomeron loops. To explain the 
role of multi-Pomeron exchanges, let us consider a simple model where the product  of the 
Pomeron residues is parametrized  
as above by the factor of $g_{h_1\Pomeron h_1} g_{h_2\Pomeron h_2}\exp B/2t$. In this case 
the impact parameters corresponding to an exchange 
of $n$ Pomerons rapidly decrease with $n$:
\begin{equation}
\left<b^2_{n}\right>\approx \left<b^2_{\Pomeron}\right>/n \,,
\end{equation}
where $\left<b^2_{\Pomeron}\right>=B$  are the impact parameters characteristic for the single Pomeron exchange; 
$B$ is the slope of $t$ dependence of the elastic cross section.  
(The inclusion of inelastic intermediate states for multi-Pomeron exchanges would result in a slower $t$ dependence 
and, hence, in the further reduction of the average $b^2$ for multi-Pomeron exchanges.)

The relative contribution of the 
multi-Pomeron exchanges as well as  the effects of the  
Pomeron self-interactions  grow with an increase of energy.   As  a result, the  amplitude of 
the elastic $pp$ collision becomes completely absorptive, $\mbox{Im}  f(b,s)\approx 1$ for small impact parameters, see the discussion in the next subsection. The blackness of the interaction at small impact parameters somewhat
suppresses diffusion in the impact parameter space  because the trajectories 
entering the absorption region disappear from Gribov diffusion. 

Another effect leading to the suppression of Gribov diffusion is an increase with collision energy
of the probability of hard processes with the corresponding tendency of the disappearance of significant 
differences between soft and hard QCD processes. It is rather difficult to observe this phenomenon 
directly since the  majority of observed pions results from the decays of heavier resonances 
(section~\ref{subsection4.2-3}) 
which have average transverse momenta  as high as $0.6$ GeV/c  already at fixed target energies.

\subsection{Observation of multi-Pomeron interactions: soft diffraction in the triple Pomeron limit}
\label{subsection4.4}

Probability of the processes with a large gap in rapidity  can be evaluated 
using   the Pomeron exchange. The Pomeron calculus predicts diffraction of an incident hadron into 
hadronic states whose invariant masses $M_X$  are large and the rapidity gap is large and
increasing with energy,    
\begin{equation}
h+T\to M_{X} +{\rm rapidity \ gap} +T^{\prime} \,,
\end{equation}
in the kinematics where 
$M^2_{X}/s=x_{\Pomeron}$
 is small and constant and $s\to \infty$. Here $1- x_{\Pomeron}$ is the fraction of the target momentum carried by  the final state hadron $T^{\prime}$.  Since $M_X$ is large and increases with $s$,  the sum over diffractively produced states can be substituted by the Pomeron exchange  for 
 $x_{\Pomeron}  \ll  1$.  Thus the  process of diffraction of an incident hadron into a  
 large mass state probes the triple Pomeron vertex 
 (the process corresponding to the diagram in Fig.~3c).
 
 The observation of this process at FNAL~\cite{DFNAL} and the LHC~\cite{DLHC} 
 is a direct demonstration of how the Pomeron branch points arise and of 
 their important role in high energy processes.

The consideration of the diagram $c$ in Fig.~\ref{pomeron} allows one to predict the dependence of the cross section on $x_{\Pomeron}$~\cite{KancheliTPL,MuellerTPL}:
\begin{eqnarray}
\lefteqn{d\sigma(h+T\to "M_X"+{\rm rapidity \ gap} +T')/dx_P  d^2q_t=} \\ \nonumber
&&s^{\alpha_{\Pomeron}(0)-1}g_{h{\Pomeron} h}(t=0)g^2_{T{\Pomeron} T'}(q^2_t) (1/x_{\Pomeron})^{2\alpha_{\Pomeron}(t)-1} x_{\Pomeron}^{\alpha_{\Pomeron}(0)-1}=\\ \nonumber
&&s^{\alpha_{\Pomeron}(0)-1}g_{h\Pomeron h}(t=0)g^2_{T\Pomeron T^{\prime}}(q^2_t) (1/x_{\Pomeron})^{\alpha_{\Pomeron}(0)-2\alpha^{\prime}_{\Pomeron} q^2_t} \,,
\end{eqnarray}
where $q_t$ is the transverse momentum transferred to the target. 
A distinctive  feature of this formula is the singularity of the cross section at $q_t=0$ and 
$x_{\Pomeron}\to 0$,  if the triple Pomeron vertex $g_{3P}(q_t=0)$ is different from zero. 
If $\alpha_{\Pomeron}(0)>1$, this singularity will be present at non-zero $t$ as well.  
This  singularity lies in the unphysical region since $x_{\Pomeron}=0$ requires infinite energies.   
With an increase of $s$, multi-Pomeron exchanges and Pomeron loops, which were effectively
forbidden at lower energies by energy--momentum conservation,
become progressively important.

Inelastic diffraction can occur only at large impact parameters since 
inelastic processes fill the rapidity gap at small impact parameters, where the interaction is practically 
completely absorptive.  The Pomeron calculus takes this effect into account by including multi-Pomeron exchanges that  strongly screen the triple Pomeron contribution, especially at small impact parameters.  
The overall effects are the reduction in the large-mass diffraction, a faster decrease in the cross section with $q^2_t$, 
and a reduction of 
$\alpha_{\Pomeron}^{\rm eff}(t\sim 0)$.  Screening of the triple Pomeron vertex was evaluated  
in the generalized eikonal approximation in~\cite{AbrBat} and was found to be very large.  
The behavior of inelastic diffraction in $pp$ collisions that we described above  
has been observed at FNAL~\cite{DFNAL}. Most of the FNAL data~\cite{DFNAL} correspond to relatively 
large $x_{\Pomeron} >  0.01$ so that the contributions of secondary trajectories 
play an important role. The recent LHC data~\cite{DLHC} observed 
similar regularities but at smaller $x_{\Pomeron}$, where the contribution of secondary 
trajectories can be neglected, while the screening effects may play a role;
$\alpha_{\Pomeron}^{\rm eff} \approx 1.05$ was reported~\cite{DLHC}.

\subsection{Blackening of hadron--hadron interactions at central impact parameters}
\label{subsection4.5}
\

Probability conservation, i.e., unitarity of  the $S$ matrix in the $s$-channel, restricts the  high energy behavior 
of the total cross sections of hadronic collisions and cross sections of diffractive processes:  
\begin{equation}
\mbox{Im} \, f(b,s)=\frac{1}{2}\left|f(b,s)\right|^2\,\mbox{+ positive  terms} \,.  
\label{unitarity}
\end{equation}
High energy processes are predominantly inelastic so that the partial waves are predominantly imaginary.  Indeed, it   follows 
from Eq.~\ref{unitarity}  together with the expression for the cross section of inelastic processes~\cite{LandauN} that:
\begin{equation}
\sigma_{\rm inel}(s,b)=1 - \left|S_l-1\right|^2 \,,
\label{sigmain}
\end{equation}
where $S_l=1+if_l$  is the matrix element of the $S$-matrix corresponding to the orbital momentum $l$.
If $\mbox{Im} \, f(b,s)$ exceeds unity, $\sigma_{\rm inel}(s,b)$ would start decreasing with an increase of 
$\mbox{Im} \, f(b,s)$ in contradiction with the dominance of inelastic processes. As a result, one concludes 
that at high energies the partial waves 
for elastic collisions
cannot exceed unity:
\begin{equation}  
\mbox{Im} \, f(b,s) \le 1 \,.
\label{unitaritybound}
\end{equation}
The upper boundary for the total cross section follows from the $S$-matrix theory~\cite{Froissart}.
 Indeed,   if in the kinematical region of $t$ restricted by the  singularities of the  amplitude in $t$ plane
 the scattering amplitude increases with energy not faster than a polynomial,  i.e., if  $\mbox{Im}  A(s,t)\le s ^N$, then 
\begin{equation}
\sigma_{\rm tot}\le c\ln^2(s/s_0) \,.
\label{eq:Fr}
\end{equation}
The formal derivation \cite{AMarten} uses analytic properties of the amplitude in the $t$-plane   
and unitarity of the $S$-matrix to derive the polynomial boundary on the amplitude that we mentioned above.   
This derivation also allows one to evaluate the coefficient $c$ in Eq.~\ref{eq:Fr} whose  
value turns out to be unrealistically large, being significantly larger than that 
arising from fits to the $pp$ data.

As we explain above, the amplitude $A(s,t)$ can be unambiguously extracted from the data on elastic 
$pp$ collisions using Eq.~\ref{elamp}.   In the kinematics achieved at FNAL for $pp$ scattering, 
$\sigma_{\rm tot}(pp) \approx 16\pi B$.This means that  for the $pp$ interaction  at the zero impact parameters, the  partial amplitude is close to unity.  If we define (for the illustration purposes) proximity 
to the black regime as a condition that probability 
 of the inelastic interaction is $\ge 0.75$ (
 see Eq.~\ref{sigmain}), 
 the interaction will be close to being black for $b\le 1.0$ fm at $\sqrt{s}=7$ TeV  and 
 for $\sim 15\%$ smaller values of $b$ at $\sqrt{s}=2$ TeV.  For larger $b$, which dominate 
 in the inelastic cross section (the median $b\sim 1.4$ fm at $\sqrt{s}=7$ TeV), 
 the interaction is grey and rather far from the black regime.

It is well known from textbooks 
that the assumption that the interaction is black at all impact parameters leads to the following two predictions:
\begin{equation}
\sigma(hT)_{\rm el}/\sigma(hT)_{\rm tot}=\frac{1}{2}
\label{blacklimit}
\end{equation}  
and
\begin{equation}
d\sigma/dt(pp\to {\rm diffractive \ state}+{\rm rap. \ gap} +p)_{t=0}=0 \,.  
\label{blackort}
\end{equation}
The second property of the black disc regime (BDR)  follows from orthogonality of the wave functions
of the eigenstates of the Hamiltonian corresponding to different eigenvalues. 

Eq.~\ref{blacklimit} is in the evident disagreement with the data as the  
$\sigma_{\rm el}/\sigma_{\rm tot}$ ratio at the LHC, while slowly increasing with energy, still reaches only 
the value of about 25\%.  Also, at the LHC the ratio of inelastic and elastic diffraction is close 
to unity in the apparent contradiction with Eq.~\ref{blackort}.  Taken together, 
the discussed features of the data  point out  that, effectively, the $pp$ scattering at FNAL and the LHC 
has a small black spot at small $b$  surrounded by a large grey area. 
The grey part mostly represents the contribution of peripheral collisions to the total cross section 
and it is dominated by exchanges 
of 
the  Pomeron and Pomeron cuts.

\section{Space--time evolution of  high energy processes}
\label{section5}
\subsection{Introduction}

 The phenomenon of the linear increase of longitudinal distances with an increase of the collision energy 
 was first understood in QED for the propagation of charged particles through the medium    
  and led to the so-called Landau--Pomeranchuk--Migdal effect  \cite{LPTM}.  
  In QCD, the linear  increase of longitudinal distances with the collision energy~\cite{GPI} is a  
  fundamental property of high energy processes. 
  This and following 
  sections discuss new coherent phenomena that  follow from taking into account 
  the space--time evolution of high energy processes  in QCD. Note that the concept of the space--time evolution of high energy processes~\cite{GribovST} is beyond the framework of  the $S$-matrix approach.   The aim of this section is to explain some properties of the space--time evolution of the scattering processes and their implications.

\subsection{Linear increase of longitudinal distances in high energy processes with energy}
\label{subsection5.1}
It follows from the application of the energy--time uncertainty principle to scattering processes in  QCD that 
when the energy of the projectile in the target rest frame, $E$, is large enough,  
the quark--gluon configurations satisfying the condition: 
\begin{equation}
z\approx (1/(\Delta E)) \approx  2E/(M^2-m^2_h)\gg R_T
\label{ld}
\end{equation}   
are  formed before the target and these configurations are frozen before the collision.  
In Eq.~\ref{ld}, $(M^2-m^2_{hadron})=\sum_{i}  (m^2_q +k^2_{t,i})/z_i -m^2_{hadron})$  and $z_i$ is the fraction of the incident particle momentum carried by the constituent $i$.  Thus, the space--time evolution of the scattering processes  
differs from that in non-relativistic physics.  Eq.~\ref{ld} directly follows from the Lorentz transformation 
applied to projectiles with an energy-independent number of constituents.  One can easily check the 
validity of the above estimate of longitudinal distances  by analyzing the dominant ladder diagrams using popular approaches to high energy processes.

Here we will present reasonings applicable also in the non-perturbative QCD regime. 
Let us consider the virtual photon--target scattering in the deep inelastic (Bjorken) limit:
$-q^2\to \infty, -q^2/2(q\cdot p_T)={\rm const}$, where $q$ and $p_T$ are the four-momenta of the photon and the target, respectively.  It follows from the optical theorem that  the  total $\gamma^{\ast} $--target cross section
 has the following form: 
\begin{equation}
\sigma={1\over s}\mbox{Im}  A(\gamma^{\ast}+T\to \gamma^{\ast} +T)=\frac{1}{s} \int \exp(iq\cdot y) 
\left<T\right|\left[J_{\mu}(y),J_{\mu}(0)\right]\left|T\right>d^4 y \,.
\label{commutator}
\end{equation}
As a consequence of causality,
only the $y^2=t^2-z^2-y^2_t\ge 0$ region contributes 
to Eq.~\ref{commutator}.  At large energies,  $q_0=\sqrt (q^2_z-Q^2)\approx q_z-Q^2/2q_z$ and, thus, 
$i(q \cdot y)\approx i(q_0(t-z)-zQ^2/2q_0)$.   Since the $Q^2$ dependence
 is contained only in the second term and the cross section decreases with an increase of $Q^2$, 
 the  direct analysis of the representation of the cross section in the form of Eq.~\ref{commutator} 
 shows~\cite{GPI} that  essential distances in the integral are  $t\approx z$  and 
\begin{equation}
z \approx 2q_0/Q^2 \,.
\end{equation}

\subsection{Cancellation of the contribution of planar/Glauber-approximation diagrams} 
\label{subsection5.2}
It has been understood 
long ago that the Glauber (eikonal) approximation---being a very popular  
method of modeling of high energy processes in nuclear and particle  physics---is actually inapplicable beyond the non-relativistic domain. The dominance of large longitudinal distances changes qualitatively 
the pattern of multiple interactions. 

\begin{figure}[h]  
   \centering
   \includegraphics[width=0.9\textwidth]{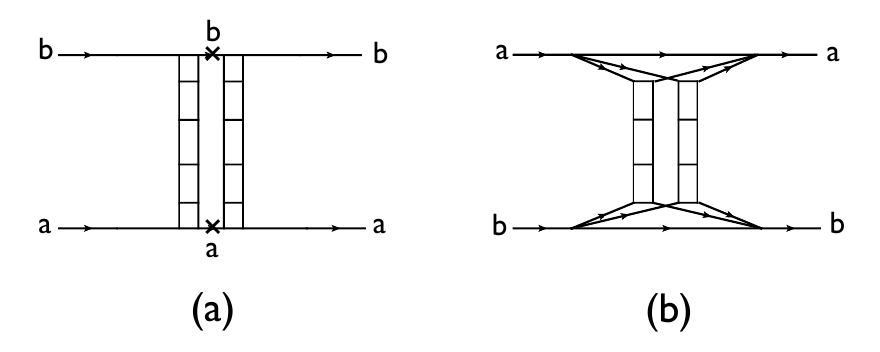}
 \caption{(a) The planar diagram for double scattering. (b) A non-planar diagram for double scattering. }
\label{cross}  
\end{figure}

In the non-relativistic quantum mechanics, the eikonal approximation follows from the Schrodinger
equation when the kinetic energy of the incident particle significantly exceeds the potential of the 
interaction~\cite {LandauN}.  
In  the Glauber approximation, high energy interactions of the projectile with a target occur via    
consecutive rescatterings of the projectile off the constituents of the target. 
The projectile is on its mass shell between the interactions---one takes 
the residue in the propagator of the projectile (Fig.~\ref{cross}a).  However, the Glauber approximation contradicts the QCD-based space--time evolution of high energy processes 
dominated by particle production.  Indeed, as the essential distances  become significantly larger than the distances (time intervals) between consequent  rescatterings~\cite{GPI}, there is no time for a frozen configuration in the projectile to recombine into the projectile during the time of the order of $R_T $ since
it is much shorter than the lifetime of the configuration (Eq.~\ref{ld}).
 
There are two independent theoretical  proofs that the contribution of planar diagrams to the 
double  Pomeron scattering amplitude is actually zero.  It was found that in the case of high energy 
scattering in a quantum field theory, the contribution of the planar diagrams with intermediate states corresponding to the projectile on its mass shell
drops with the incident energy as $1/s$.  Indeed, the integral over the square of the mass of hadrons produced  in the $\Pomeron$--hadron  collision, $M^2$,   is zero in the case when the Feynman diagrams have only $s$ or $u$ cuts because the contour of integration can be moved  in the  direction where there 
are no singularities in $M^2$~\cite{S.Mandelstam63,PomeronCalculus}. The integral over the large circle is zero as a consequence of a decrease of the amplitude with $M^2$.  
 The eikonal diagram 
 (Fig.~\ref{cross}a) belongs to  the class of Feynman diagrams where the cancellation occurs.  
 Crossed (non-planar)
 diagrams (Fig.~\ref{cross}b), which have cuts both  in $s$ and $u$, give a non-zero contribution.  
 
 Taking in account  energy--momentum conservation leads to the same conclusion~\cite{Blok:2006ns}.   Indeed, the eikonal diagrams correspond to an inelastic intermediate  state described by the Pomeron exchange at the double energy $2s$.  On the contrary,  in the crossed diagram  {\it the energy is divided between constituents before the collision}.  If  one parton carries the fraction $z$ of the incident hadron momentum and   another parton carries the fraction  $z^{\prime} \le 1-z$, the total energy of the produced hadronic state is $sz+sz^{\prime} \le s$. Both arguments can be easily generalized to the case when the  wave function of the initial hadron contains many constituents.  

Using the technique of the Pomeron calculus, V.~Gribov showed that in a quantum field theory, the contribution of the non-planar diagrams relevant for the multiple Pomeron exchanges
to the total cross section can be rewritten as a sum of the eikonal term and the inelastic diffraction contribution.  The resulting Gribov--Glauber model~\cite{Gribov-Glauber} is in 
agreement 
with the data on nuclear shadowing in hadron--nucleus interactions~\cite{Karmanov-Kondratuk}. 
 We will explain in section~\ref{subsection6.3.2} how taking into account
 color fluctuations allows one to evaluate the relative contributions of multiple scatterings 
 in the Gribov--Glauber model.

\section{Fluctuations of color in  diffractive phenomena }
\label{section6}
\subsection{Introduction}
QCD predicts  new types of diffractive phenomena as compared to the $S$-matrix approach since the 
wave function of an energetic incident hadron is formed long before the target  and the transitions between different configurations in the wave function occur at  distances comparable with 
the characteristic longitudinal distance called the coherence length.  Therefore, the cross section is 
calculable in terms of the instant quark--gluon configurations in the projectile (see the discussion in the previous section).  In the  exclusive processes, where the incident hadron is squeezed in the 
transverse direction by the choice of the specific final state,  the spatially small wave packet of quarks and gluons weakly interacts  with a target in a rather large interval of  collision energies. This phenomenon is calculable for hard   diffractive processes in the form of the special  QCD factorization theorem~\cite{BFS94,CFS96} and has been observed in a variety of experiments (see section~\ref{subsection7.7} 
and the review and references in~\cite{CT}).  Thus QCD predicts fluctuations 
 in the strength of the interaction since the interaction differs for different 
configurations 
 of constituents in the wave function of the incident hadron.   
This is an intuitive justification of the necessity to use the concept of the distribution over cross sections instead of the average cross section.

\subsection{Suppression of  the strong interaction due to screening  of color }
\label{subsection6.2}
At large energies the wave function of the incident hadron is
formed before the target and is frozen, if $(2E/\Delta M^2) \gg R_T$. Different  configurations of constituents in the wave function of the incident hadron interact with 
the target  at  different strengths.  This is an important property of QCD where the interaction is proportional 
to  the area occupied by color since   the color charge of a hadron is zero.  For illustration purposes
we  begin with quark models of hadrons and then derive formulas in QCD.

A popular model was suggested by Low and Nussinov~\cite{F.Low, S.Nussinov}   in which 
the total hadron--hadron cross section
is described by the exchange of two gluons. Low further argued that the cross section is  proportional  to  the region occupied by color in the hadrons:
\begin{equation}
\sigma(hT)= c r^2_t \,,
\label{F.Low}
\end{equation}
where $r_t$ is the transverse radius of the smaller hadron. The derivation  involves taking into account
 the gauge invariance, the zero color charge of hadrons as well as an implicit assumption that the 
 momenta of constituents within the hadrons are significantly larger than the transverse momenta of the exchanged gluons. Eq.~\ref{F.Low} was elaborated on in the constituent quark model with a two-gluon exchange between $h$ and $T$~\cite{GS:1976}.  
   Eq.~\ref{F.Low} can be questioned since  in the model,  the average size of configurations involved in the scattering is comparable  to the scale of non-perturbative QCD phenomena   and the restriction by a two-gluon exchange cannot be justified. 
   
Eq.~\ref{F.Low}   can be  reformulated  to include the full QCD.  If  the incident meson  is in a quark-gluon configuration whose transverse size is significantly smaller than the scale of non-perturbative QCD phenomena,  $r_t^2\Lambda^2_{QCD}\ll 1$,    the application  of the technology of the QCD factorization theorem~\cite{CFS96} and the QCD evolution equation 
for parton densities allows one to calculate the cross section of the hadron interaction with a target $T$.  
The derived expression~\cite{Blaettel:1993rd,Miller93,Radushkin97} also contains the factor 
of $r^2_t$ as in Eq.~\ref{F.Low}  which represents the  coordinate space equivalent of 
approximate Bjorken scaling for DIS processes. However, in addition, the final expression 
 contains the factor  $xG_T$---the gluon distribution in the  target $T$
 absent in  Eq.~\ref{F.Low}: 
\begin{eqnarray}
&&\sigma_{hT}(r_t\to 0)_{\left| 4r_t^2s/m_N \gg R_T\right.} \nonumber\\
&&=\psi^2_{h}(r_t=0, r_z=0)  {F^2 \pi^2\over 4}r^2_t\alpha_s(Q^2)xG_T(x,Q^2= {1\over 4r^2_t}) \,.
\label{sigmah}
\end{eqnarray}
Here $F^2$ is the Casimir operator for the quark (gluon) dipole; $r_t$ is the transverse distance between the quark and 
the antiquark;  $\psi_h(r_t)$ is the small transverse size component of the incident hadron wave function;  
$x\sim 1/(4sr^2_t)$.   Eq.~\ref{sigmah} can be obtained 
also
from the 
formulae
 derived in~\cite{Muellereikonal}  in the leading 
$\alpha_s\ln(x_0/x)$ approximation.
 In the  case  of nucleon one needs to take into account that  the  square of nucleon  wave function  at small $r_{it}$ is proportional to $r_{it}^2$ and to make the substitution 
 \begin{equation}
 r_t^2 \to (r_{t1}-(r_{t2}+r_{t3})/2)^2 +(r_{t2}-(r_{t1}+r_{t3})/2)^2 +(r_{t3}-(r_{t1}+r_{t2})/2)^2 \,.
 \end{equation}
In this case,  the three quarks act as a symmetrized superposition of three dipoles stretched between one quark and  the transverse center of mass of the other two quarks.

\subsection{Perturbative Pomeron }
\label{subsection6.3} 

The asymptotic behavior of the amplitudes 
of high energy processes 
in the vacuum channel in a quantum field theory containing vector particles 
was first investigated 
 in QED.   In \cite {FGL} the sum of the  leading $\alpha_{em}^{2}\ln(s/s_o)$ terms in the cross section of 
the photon--photon scattering was calculated.  (Note that in QED, in the lowest--order over coupling constant, this cross section is independent of energy.) 
This idea was applied to QCD in \cite{BFKL, MuellerBFKL} to study amplitudes of high energy processes 
in the kinematics where $\ln(x_0/x)\gg \ln(Q^2/Q^2_0)$ by summing  leading $\alpha_s\ln(x_0/x)$ 
terms---the so-called perturbative Pomeron.   A priori one can try to justify  this approximation 
in the case of scattering of two small dipoles of the transverse size $\propto 1/Q$ 
within the restricted kinematical domain  of rapidities $y\le y_0(Q^2,x)$,  where  the coupling to the colliding dipoles is perturbative. The kinematical boundary for the  applicability of this approximation---$y_0(Q^2,x)$---as well as 
for the decomposition over powers of $1/Q^2$  arises due to diffusion in the space of transverse momenta 
to the non-perturbative domain \cite{Muellerrestriction}.  

In the leading log approximation, the  cross section  grows as:
\begin{equation}
\sigma_{\rm dipole - dipole} \propto (1/x)^{\beta} \,,
\label{dd}
\end{equation}
where 
\begin{equation}
\beta= {N_c\alpha_s 4\ln 2\over \pi } \,.
\label{LO}
\end{equation}
For $Q^2 \sim 2$ GeV$^2$, $\alpha_s\sim 0.25$ leading to $\beta \approx 0.7$.  
Note that the  actual formula derived in \cite{BFKL,MuellerBFKL} is significantly more complicated than Eq.~\ref{LO}, which  
is just a  popular fit to this formula.  The derived expression \cite{BFKL} 
corresponds to a cut  in the angular momentum plain 
so that it involves a mathematical object that is 
different from the Pomeron Regge trajectory discussed in section~\ref{section4}.   
It becomes a sum of the poles in the angular momentum plane   in the large $N_c$ approximation.

 Assuming that the number of radiated gluons is sufficiently large, the diffusion equation was derived  
 for the motion in the plane of $\ln(p^2_t/p^2_{t 0})$. It was observed that diffusion both to large and  small $p_t$  is present \cite{BFKL,CTM}.  Significant diffusion to small $p_t$, i.e., into the  non-perturbative 
 domain raises questions  about  the validity of pQCD approach because the answer depends on the treatment of the  badly understood infrared region. The main  difference  between the perturbative 
 Pomeron and the Pomeron trajectory discussed in section~\ref{section4} is 
 significant  diffusion to large parton momenta  (in addition to diffusion to small parton momenta). 
 Such diffusion is absent in the non-perturbative  Pomeron which is modeled by the non-perturbative ladder 
 discussed in  section~\ref{section4}.

 The NLO correction to Eq.~\ref{LO} was found to be so large~\cite{Ciafaloni,Lipatov-Fadin} that it dominates 
 the LO expression for a wide range of $\alpha_s$, which leads to $\beta \sim 0$. 
 This is primarily because the LO as  well as NLO approximations ignore energy--momentum conservation.   
 The poor convergence of the series for the total cross section in terms of powers of 
 $\ln(s/\mu^2)$ was first demonstrated in QED~\cite{KL} by the direct calculation of the lowest 
 order diagrams for the $e^+e^-$ pair production in electron--electron scattering. 
 For the contribution of the dominant two-photon mechanism, it was found that 
\begin{equation}
 \sigma=\alpha_{\rm em}^4 c(1.04\ln^3(s/\mu^2) -6.59\ln^2(s/\mu^2)-11.8\ln (s/\mu^2)  +104 +O(\mu^2/s)) \,,
 \end{equation}
 where $\mu$ is the electron mass.
 It was explained in  \cite{KL} that a fast growth of the coefficients in front of the powers of $\ln(s/s_0)$
 reflects the highly restricted  phase space for obtaining logarithmic contributions.
 
 The resummation models~\cite{Altarelli,Ciafaloni:2003rd} more smoothly match with the formulas 
 of  the DGLAP approximation where the conservation of the longitudinal component of the momentum is exact.

The same approach is often applied to scattering of a small dipole of the diameter $d\approx 1/Q$ off the nucleon whose diameter is $\approx 1/(2m_{\pi})$, which is relevant for such processes as
inclusive DIS, 
exclusive production of vector mesons, etc. 
Although the energy behavior given by Eq.~\ref{dd}  obviously contradicts the data, resummation approaches 
can fit the data since they lead to the results close to those obtained in the DGLAP approximation.  
Up to now the resummation  approaches have not been applied to the description of hard diffractive processes.

One should note that the probability conservation in the form of Eq.~\ref{inequality} 
is also violated in the resummation models at sufficiently small $x$ and small impact parameters (see 
the discussion in section~\ref{section8}).

\subsection {Distributions  over strengths of the interaction for hadron and photon projectiles}
\label{subsection6.3.2}

Before QCD was recognized as the theory of the strong interactions,
in the framework of the parton model in which the strength of interactions 
is proportional to a number of wee partons in the projectile configuration,
Pumplin and Miettinen suggested 
the description  of high energy diffractive processes  in terms of the probability 
distribution over cross sections, $P_h(\sigma)$~\cite{PM}. 
It was understood later on that such a distribution originates from the 
dependence of the cross section on the instant transverse radius of the color distribution 
in the incident hadron.

Constructively, $P_h(\sigma)$  is defined in terms of its moments, 
\begin{equation}
\langle \sigma^{k} \rangle=\int P_h(\sigma)\sigma^{k} d\sigma \,,
\end{equation} 
with additional general QCD restrictions on the form of $P_h(\sigma)$.
The case of $k=0$ 
corresponds to the normalization condition for $P_h(\sigma)$. By definition, the first moment of 
$P_h(\sigma)$ ($k=1$)  is the total cross section of $hN$ scattering.  The dispersion of the distribution 
over $\sigma$  is given by the ratio of inelastic and elastic diffraction at $t=0$~\cite{PM}:
\begin{equation}
\frac{d\sigma(h+N\to h^{\prime}+N)/dt_{\left|t=0\right.}} {d\sigma(h+N\to h+N)_{\left|t=0\right.}}=\frac{\left<\sigma^2\right>-\left<\sigma\right>^2}{\left<\sigma\right>^2} \equiv \omega_{\sigma} \,.
\end{equation}
where the state $h^{\prime}$ differs from the state $h$.

\begin{figure}[h]  
   \centering
   \includegraphics[width=0.6\textwidth]{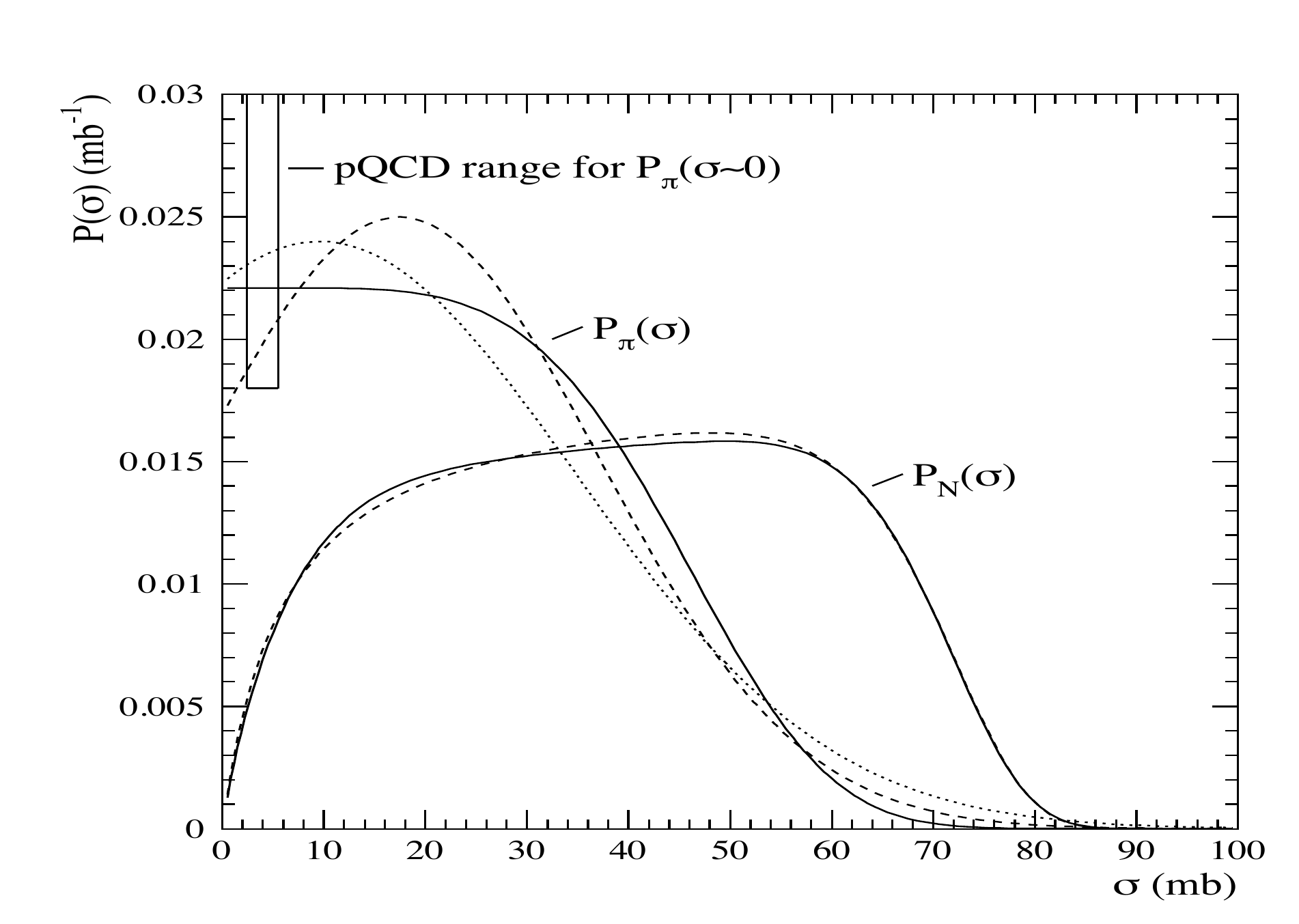}
 \caption{The distribution over $\sigma$ for fixed target energies extracted in~\cite{Baym93}. 
 The set of curves reflects uncertainties in the extraction procedure.  The rectangular area in the upper left corner is the pQCD  evaluation of $P_{\pi}(\sigma\to 0)$~\cite{Radushkin97}.}
\label{psigma}  
\end{figure}

 The behavior of 
 $P_h(\sigma)$  at small $\sigma$ follows from the 
 kind of quark counting rules.  
 Taking into account  the number of valence quarks in a hadron  $h$  
 and  using 
 approximate Bjorken scaling   in the form explained above 
 ($\sigma\propto r_t^2$),     it was found that   \cite{Blaettel:1993rd,Baym93}:
 \begin{equation} 
P_h(\sigma\to 0) \propto \sigma ^{(n_q+n_{\bar q}+n_g-2)} \,,
\label{qcr}
\end{equation}
where $n_i$ is the number of valence constituents in the incident hadron  in the configuration participating in the scattering process\footnote{In the 
analysis of~\cite{PM} it was assumed that $P_N(\sigma)$ contains the term $\propto \delta(\sigma)$.}.  
The data on diffractive $p+^2H\to X+ ^2H$ scattering 
provides an additional constraint, $\left<(\sigma\right. -\left<\sigma\right> )^3\left.\right> \approx 0$ 
at $\sqrt{s_{NN}}=30$ GeV~\cite{Baym93}.  The information on the first three moments of $P_h(\sigma)$
and its behavior at $\sigma\to 0$ allows one to reconstruct the form of $P_h(\sigma)$ for the
pion and nucleon projectiles (Fig.~\ref{psigma}).

At present, the distributions are  reconstructed for the energies of incident hadrons in the range of a few hundred GeV in the target rest frame.  With an increase of the energy, the edge of the distribution moves to the right diminishing probability of  weakly interacting configurations. The dispersion of the distribution grows with $\sqrt{s}$ up to $\sqrt{s}=50 - 100$ GeV, 
where $\omega_{\sigma}\sim 0.3$ is reached,
and it starts  dropping for larger $\sqrt{s}$. Preliminary LHC data indicate that for $pp$ scattering at $\sqrt{s}$=7 TeV, $\omega_{\sigma}\sim 0.2$.    Note that  $\omega_{\sigma}\sim  0.2 - 0.3$ corresponds to very large fluctuations 
of the strength of the interaction. For example, if one models $P_N(\sigma)$ as a superposition of two scattering states, $\sigma$'s for these states would be $\sigma_{\rm tot}(1\pm \sqrt{\omega_{\sigma}})$ corresponding to $\sigma_1\sim 55$ mb and 
$\sigma_2 \sim 145$ mb at $\sqrt{s}=7$ TeV.

In the case of photon, 
\begin{equation}
P_{\gamma}(\sigma\to 0)\propto 1/\sigma \,,
\end{equation}
which follows from the presence of point-like 
$q\bar q$ configurations in the photon the wave function.

The Gribov--Glauber model allows for a transparent interpretation in the formalism
of cross section eigenstates of Good and Walker~\cite{Good:1960ba}. 
Indeed, different configurations  are absorbed with the strength given by 
the Glauber model for a given $\sigma$ and 
incoherently contribute to the total cross section~\cite{Kopeliovich:1978qz}.

The concept of $P_h(\sigma)$ allows us to  build a compact implementation  of the Gribov--Glauber  series for the  total cross section of $hA$ scattering:
\begin{equation}
\sigma_{\rm tot}(h A)=\int d\sigma P_{h}(\sigma) \int d^2b\, 2\,[1-e^{-\sigma T_A(b)/2)}] \,.
\label{CF}
\end{equation}
It allows us also to calculate the total  cross section of inelastic coherent diffraction off nuclei ($hA\to h^{\prime} A$) in a good agreement with the data~\cite{Frankfurt:1993qi,SG,LFGS}. It also allows us 
to model deviations from the Glauber model in inelastic proton (nucleus)--nucleus collisions~\cite{Baym:1995cz}.  Note here that deviations from the eikonal approximation for the interactions with $j$ nucleons  
is given by the  $\left<\sigma^j\right>/\left<\sigma\right>^j$ ratio, which rapidly grows with $j$.

Knowledge of the  first three moments of the distribution over $\sigma$  is sufficient to describe many 
nuclear phenomena, 
in particular, the total and coherent inelastic diffraction cross sections.  
The evaluation of more complicated phenomena such as,  the tail of the hadron multiplicity, at 
present is model-dependent since the behavior of $P_h(\sigma)$ at large $\sigma$ is far from being understood.

\subsection {Diffraction in deep inelastic collisions as a pattern for the fluctuations of color}
\label{subsection6.4}
In this subsection we will consider diffraction in  deep inelastic $ep$ scattering: 
\begin{equation}
\gamma^*+p\to \mbox{X +rapidity~ gap +} p
\end{equation}
in the Bjorken limit.

The standard picture of DIS  is that of the absorption of the virtual photon by a parton (quark or antiquark) that  carries a fraction $x$ of  the light-cone  momentum of  the nucleon with radiation of gluons in the 
initial and final states. In such a  picture,  which was quite popular before the first measurements at HERA,  the hard gluon radiation should fill the whole available rapidity interval (in addition,  partons should be 
emitted to screen the delocalized color in the final state)   and lead to  the disappearance of  diffractive processes.   However, such processes were observed with a significant probability even at very large $Q^2$.

 We will focus our attention on  the limit when $M_{X}^2 / Q^2={\rm const}$. In this limit,  it is 
 convenient to introduce the variable $\beta$:
 \begin{equation}
 \beta={Q^2\over Q^2 +M_X^2} \,.
 \label{beta}
 \end{equation}
 The variable $\beta$ is related to the fraction of the momentum lost by the nucleon, $x_{\Pomeron}$, as  
\begin{equation}
 \beta=x/x_{\Pomeron} \,.
 \label{beta1}
 \end{equation}

 It is convenient to introduce ''conditional'' or fracture structure functions for the processes where one hadron is fixed in the fragmentation region. For diffractive processes, one usually uses the notation $F^{D(4)}_p(\beta, Q^2,x_{\Pomeron},t)$ and one can also introduce quark and gluon  diffractive parton distribution functions (PDFs) that depend on the same variables.   
 Since hard processes occur locally in transverse momenta and rapidity,  the increase in  the resolution should not affect the properties of the nucleon fragmentation   region. Hence one should expect that the diffractive PDFs  should satisfy the same DGLAP evolution  equations as 
 usual PDFs.

Extensive data on various large-mass diffractive processes have been obtained at HERA (for the recent results, see~\cite{HERA}). The principal findings are the following:

\noindent 
$\bullet$ The leading twist approximation with the same diffractive PDFs consistently 
describes the $Q^2$ evolution of the inclusive  diffractive cross section and the diffractive  cross sections of dijet 
(X= jet$_1$ + jet$_2$ +X$^{\prime}$) and heavy flavor production for fixed $x_{\Pomeron}$. 
Factorization was  formally proven in \cite{Collins}.

\noindent 
$\bullet$ The data are also consistent with the Pomeron
factorization:
\begin{equation}
f_j^{4D}(\beta, Q^2,x_{\Pomeron},t) = r(x_{\Pomeron},t)f_j(x,Q^2) \,.
\label{softfact}
\end{equation}
The $x_{\Pomeron}$ dependence of $r(x_{\Pomeron},t)$ is given by the same expression as that for soft diffraction, 
which employes $\alpha_{\Pomeron}(0)=1.11$, which is  close to $\alpha_{\Pomeron}(0)$ extracted from the analysis of soft diffractive processes,  total $pp$ cross sections, and exclusive light vector meson photoproduction   (subsection \ref{subsection5.1}).
The  observed value of the intercept is significantly smaller   than $\alpha_{\Pomeron}(0)$ for such hard 
exclusive diffractive processes as $J/\psi$ photo(electro)production. It was also found 
that gluons play a very  important role in diffractive dynamic: 
\begin{equation}
{\int_0^1 d\beta \beta f_g(\beta,Q^2)  \over 
\sum_{q_i, \bar q_i} \int_0^1 d\beta \beta f_{q_i}(\beta,Q^2)  } \sim 4
\end{equation}
for $Q^2 \sim$  a few GeV$^2$.

\noindent 
$\bullet$  The overall probability of diffraction in DIS
\begin{equation}
R(x,Q^2)=\sigma_{\rm diff}(x, Q^2)/\sigma_{\rm DIS}(x,Q^2) \,,
\label{diffratio}
\end{equation}
is of the order of 10\% and grows with a decrease of $x$ for fixed $Q^2$.

As we mentioned above, hard processes cannot screen the quark (antiquark) emitted by a highly virtual photon.   
Therefore, pQCD states should contain no rapidity gaps and, thus, diffraction should be 
part of the non-perturbative initial condition for the QCD evolution equation,  
which is far from trivial to implement in the infinite momentum frame.{\it The presence of the leading twist diffraction imposes constraints on sea quark and gluon nucleon 
PDFs at the starting point of the evolution---they should exhibit the small $x$ behavior 
consistent with the soft Pomeron limit, i.e., 
they are allowed to grow only slowly with a decrease of $x$.}

The significant value of the cross section of diffraction can be understood in the formulation 
of the parton model in the target rest frame suggested by Bjorken in 1970  twenty years before diffraction in DIS was observed. 
 At small $x$ in this reference frame, the virtual photon transforms into $q\bar q $ pairs well before the target (section \ref{section6}).   
 To    satisfy Bjorken scaling,  it is necessary  to assume that only the  $q\bar q $ pairs with $k_t \le k_{t0}$  and the 
 light-cone fractions satisfying the condition $k_{t0}^2/z(1-z) \sim Q^2$ should interact with the target at small 
 $x$ with a strength comparable to that of the pion--nucleon interaction, while the contribution of the pairs with 
 $k_t \gg  k_{t0}$ should be strongly suppressed. The low $k_t$ pairs  are aligned along the photon  direction---hence they are referred to as the aligned jet model (AJM)~\cite{Bj1}. 
The $1/Q^2$ behavior of this contribution is due to the small phase space allowed for these configurations. 
In the coordinate space this corresponds to production of a $q\bar q $ pair at a distance $2q_0/Q^2$ from the target and 
the expansion of the pair to the hadronic-scale size of $1/k_{t0}$ by the time it reaches the target~\cite{Bj2}.

 In  QCD,
 the parton picture is modified by the following two effects. First, the emission of a large-size  $q\bar q $ pair 
 without the associated gluon  emission is suppressed by the Sudakov form factor. Inclusion of this emission leads to the scaling violation, but it does not change the size of the quark--gluon configuration---the QCD AJM~\cite{Frankfurt:1988nt}.
  Second, while the interaction of $q\bar q $ pairs with the large transverse momenta up to $k_t \propto Q$
  is suppressed by the $1/Q^2$ factor due to color transparency, it also contains the factor $\alpha_s(Q^2)xG_N(x,Q^2)$.   
  As a result, there is conspiracy between  the hard and soft  contributions---both of them are $\propto 1/Q^2$,   
  with the hard contribution being numerically suppressed at moderate $x \sim 10^{-2}$, but gradually growing in importance with a decrease in  $x$  due to the corresponding increase in  $xG_N$.
  The contribution of large masses $\gg Q^2$, i.e., $\beta\ll 1$ (the triple Pomeron  processes) 
  requires sufficiently small $x$ to reveal itself. Otherwise,  it is suppressed by energy--momentum conservation.

Thus, the probability of diffraction in the aligned jet model (DGLAP approximation)   
is comparable to that in hadron--nucleon scattering. At the same time,  
the contribution of small-size configurations
to the cross section of diffraction (integrated over $\beta$) 
is suppressed relatively to the inclusive cross section by the factor
\begin{equation}
{\sigma_{\rm diff}\over \sigma_{\rm tot}}_{\left |{\rm hard} \right.} =c \frac{\alpha^2_s(Q^2) (xG_T(x,Q^2))^2/BQ^4}{\alpha_s xG_T(x,Q^2)/Q^2)} \propto \alpha_s(Q^2)xG_T(x,Q^2)/BQ^2 \,,
\label{DGLAPratio}
\end{equation}
where $B$ is the slope in  the $t$ dependence of the diffractive cross section.  This ratio rapidly decreases with 
an increase  in $Q^2$ and increases (for fixed $Q^2$) with a decrease in  $x$.   In the fast frame, AJM configurations are equivalent to the presence of local (in rapidity) color screening of $q\bar q$ pairs in 
the small $x$ nucleon wave function~\cite{Abramowicz}.

The observed soft $x_{\Pomeron}$ dependence of diffraction is natural in the QCD AJM in the kinematic range of the validity of the  DGLAP approximation.

The approximate soft factorization (Eq.~\ref{softfact})  is natural in the ladder models of the Pomeron  since in these models, the structure of the ladder does not depend on the rapidity for $\alpha_{\Pomeron}(0)\approx 1$. 
An increase of the probability of hard small $x$ processes  would result in breaking of the soft factorization at very small $x_{\Pomeron}$.

 At sufficiently small $x$ and moderate $Q^2$, the hard contribution may become significant.  
 Attempts to incorporate these higher twist contributions were taken in a number of the dipole models 
 of the $\gamma^{\ast} N$ interactions 
 (see, e.g., \cite{Kowalski:2008sa} and references therein).  However, most of these models ignore the $Q^2$ evolution of the AJM component.

To conclude, the significant cross section of  diffraction in DIS is  another  demonstration of the important role of color fluctuations in the virtual photon wave function and of the dominance of soft Pomeron physics even in seemingly hard processes.
In such processes soft dynamics 
together with $Q^2$ evolution gives a significant contribution to the total cross section.

\section{Hard Exclusive Processes}
\label{section7}
\subsection{Introduction}  
\label{subsection7.1}

It has been  understood since nearly two decades ago that a number of two-body and 
quasi  two-body processes off nucleons, nuclei, photons, etc.~can be  legitimately calculated in 
QCD  in the kinematics of fixed $x$ and $Q^2\to \infty$ as a consequence of the QCD factorization 
theorem:   $\pi + T \to 2 \ {\rm jets} +T^{\prime}$~\cite{Miller93}; 
$\gamma_L^{\ast}+ N \to V(\rho,J/\Psi,\rho^{\prime},\dots)+N^{\prime}$~\cite{BFS94},  
(where the excitation energy of the state $T^{\prime}$  $\ll Q$);
 $\gamma_L^{\ast} N \to {\rm Meson}(\pi,K,\eta,) + {\rm Baryon}$~\cite{CFS96};
 $\gamma_L^{\ast} \to [\rm Few~meson~ system]+{\rm Baryon}$~\cite{CFS96};
 $\gamma^{\ast} +N \to \gamma + N$~\cite{Bartels, BB, Dittes, Abramowicz, Ji1, Rad, Jireview, Freund};
and $\gamma^{\ast}  + \gamma \to {\rm Meson} +{\rm Meson}^{\prime}$~\cite{Diehl}.  These processes provide new ways to 
investigate  the three-dimensional partonic  structure of nucleons (transverse distribution of partons with 
a given light-cone fraction)  and to compare it to that of $\Delta$-isobars, hyperons, and $N\pi$.
A theoretical 
analysis of these processes allows one also to address such novel questions of short-range parton correlations in nucleons
as:  What is the probability to find a small color singlet cluster in the nucleon made of  a 
quark-antiquark pair, three quarks or even three antiquarks?  These processes also probe the minimal light-cone    $q\bar q$   components of   various mesons and few meson systems.   In addition,  these processes 
provide an effective probe of high energy dynamics of QCD and test whether/at what energies 
the strength of the interaction of small dipoles with nucleons/nuclei reaches the maximal strength allowed 
by unitarity, which leads to breakdown of the DGLAP QCD evolution equations.

An investigations of the same processes 
off nuclear targets  reveals another distinctive property of QCD: at 
fixed $x$ and $Q^2\to \infty$, nuclear matter is completely transparent to the propagation of spatially 
small colorless clusters of quarks and gluons---this regime is usually referred to as color transparency. 
In this limit, the complete transparency for hard diffractive processes unambiguously  follows from 
basic properties of the QCD evolution equations~\cite{LFGS}.
 The observation of color transparency of nuclear matter is the striking confirmation that the interaction  
in QCD is due to the color charge that  is screened within such clusters. Such a phenomenon would be absent, if 
high energy processes were dominated  by exchanges of usual mesons, as was assumed before
the emergence of QCD as the theory of hadronic interactions.

A characteristic feature of these processes is that the final state contains a particle (few particles)
that has  small momentum in the target rest frame.  Hence to study these processes in the fixed target 
mode, one needs to design a detector  that  would be able to 
(1) detect slow particles
(including neutrons) over a large range of laboratory angles,   (2)  measure momenta of the leading 
hadrons with high resolution,  and (3) operate at high luminosity to reach high enough $Q^2$---a challenging, 
though not impossible, task.

Detection of these reactions in the collider kinematics is somewhat   easier 
since the particles that are  slow in the target rest frame fly along the beam direction. 
Also, it is much easier to select coherent interactions with nuclei. The challenge in the case of these reactions  
is  to reach high enough luminosities---so far only 
channels with vacuum quantum numbers in the $t$-channel were investigated at HERA.

The prediction and discovery of the quarks--gluon configurations in hadrons that weakly interact 
with a target  requires also the presence of  configurations in the hadron wave function whose interaction 
with the target is larger than average (see the discussion in section \ref{subsection6.3.2}).

\subsection{QCD factorization theorem}  
\label{subsection7.2}
\subsubsection{The statement of the theorem \protect\cite{CFS96}}
\label{subsubsection7.2.1}

The starting point for the analysis is the  factorization theorem  for the process
\begin{equation}    
\gamma_L ^{\ast}(q) + p \to  {\rm "Meson"}(q+\Delta ) + {\rm "Baryon"}(p-\Delta ) 
\label{process}    
\end{equation}   
at large $Q^{2}$, with $t$ and $x=Q^{2}/(2p\cdot q)$ fixed.  It asserts that    
the amplitude has the form
 of convolution of the three blocks depicted in Fig.~\ref{fig:fact}:
 
\begin{figure}[h]  
   \centering
   \includegraphics[width=0.75\textwidth]{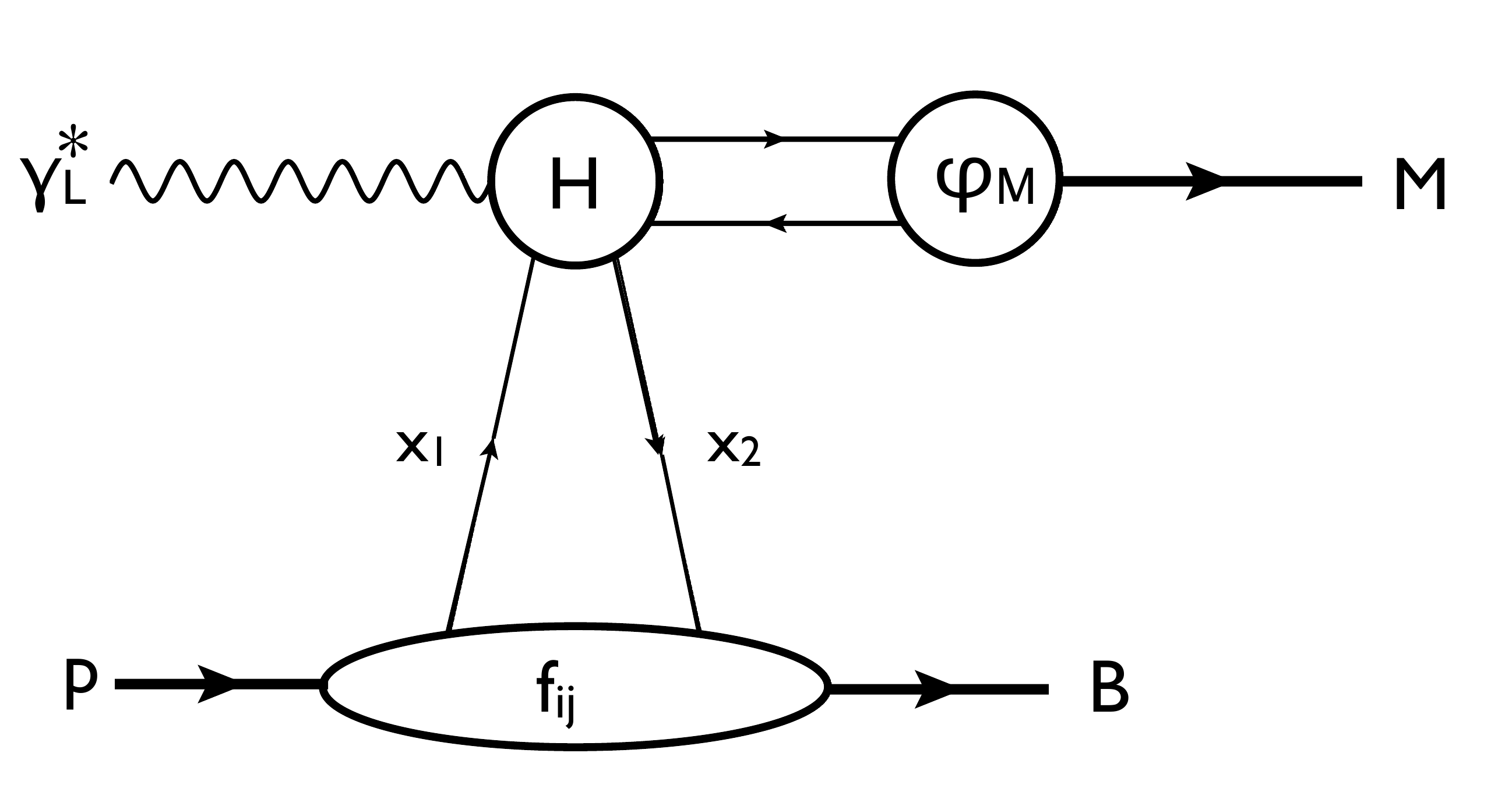}
 \caption{The block structure of the DIS  exclusive  process $\gamma_L^{\ast}+p \to \mbox{"meson" + "baryon"}$.}
\label{fig:fact}  
\end{figure}

\begin{eqnarray}
   {\cal M} &=&
   \sum _{i,j} \int _{0}^{1}dz  \int dx_{1}
   f_{i/p}(x_{1},x_{2},t,\mu ) \,
   H_{ij}(x_{1}/x,Q^{2},z,\mu )
   \, \phi_{j}(z,\mu )
\nonumber\\
&&
   + \mbox{power-suppressed corrections} \,,
\label{factorization}
\end{eqnarray}
 where $f_{/p}$ is the ``generalized  parton density'' (GPD);   $x_1-x_2=x$;
 $\phi$  is the light-front wave function of the    
meson; $H_{ij}$ is the hard-scattering coefficient usefully    
computable in terms of the powers of $\alpha_{s}(Q)$.  The contribution of the  diagrams, where  an extra gluon is exchanged between the hard blocks, is suppressed by an additional factor of $1/Q^2$.  The formal proof~\cite{CFS96}  is very lengthy  so we restrict ourselves in the further discussion by the qualitative explanation only.   Qualitatively,  the factorization in these processes is   due to the color screening/transparency:  the small  transverse size of $\gamma^{\ast}_L$   selects small-size (point-like)
 configurations ($b\sim 1/Q$)  in the meson and the interaction with such "white" configurations 
 is suppressed by the factor of $~1/Q^2$.
The relation of the color screening to factorization  is  best seen in the Breit frame.  
Before the interaction, $\gamma_L^{\ast}$  is static, while 
after the photon is absorbed,  the quark-gluon
 system (which would form the meson)  moves with a  large velocity in the direction of
the photon keeping the small transverse
size, while the baryon system rapidly moves in the opposite direction. No soft 
interactions between the left and right movers is possible, provided that
the meson system has a small transverse size. The same argument is likely to work for the processes where a forward (anti)baryon is produced:  $\gamma_L^{\ast}+ p \to {\rm forward} \ N+ \pi$,
 $\gamma_L^{\ast}+ p \to {\rm  forward} \ \Lambda +K^+$, and
 $ \gamma_L^{\ast} +p \to {\rm forward} \ \bar{p} +NN$~\cite{FPPS}, though no formal proof has been  given so far.

In the case of the transverse polarization of $\gamma^{\ast}$, the non-perturbative QCD contribution is only suppressed by 
the power of $1/\ln Q^2$ (similar to the case of $F_{2N}(x,Q^2)$). It originates from the contribution of
highly asymmetric $q\bar q$ pairs in the $\gamma^{\ast}_T$ wave function
which have the transverse size similar to that of hadrons.

\subsubsection{Definitions of light-cone distributions and 
            amplitudes: longitudinally polarized  vector meson}

{\it   The wave function of longitudinally polarized vector meson.}\\

\noindent
The light-cone wave function of a longitudinally
polarized vector meson is
\begin{eqnarray}
   \phi ^{V}_{j}(z,\mu ^{2})  &=&
   \frac {1}{\sqrt {2N_{c}}}
   \int _{-\infty }^{\infty } \frac {dy^{+}}{4\pi }
   \;
   e^{-izp^{-}y^{+}} \langle 0|\; {\bar \psi }(y^{+}, 0, {\bf 0}_{T})\gamma
^{-}
  {\cal P} \psi (0) \; |V\rangle  \,,
\label{wf.def}
\end{eqnarray}
where $\cal P$ is a path-ordered exponential of the gluon field 
along the light-like line joining the quark operators entering the matrix element.

 \vspace{0.5cm}
  
 {\it Quark density of the nucleon:}\\

\noindent
  For a quark of  flavor $i$, its density in the nucleon ($q_i$) reads:
\begin{eqnarray}
   f_{i/p}(x_{1},x_{2},t,\mu )  &=&
   \int _{-\infty }^{\infty } \frac {dy^{-}}{4\pi }
   \;
   e^{-ix_{2}p^{+}y^{-}}
   \langle p'|\; T {\bar \psi }(0,y^{-},{\bf 0}_{T})\gamma ^{+}
     {\cal P} \psi (0)\; |p\rangle  \,,
\label{pdf.q.def}
\end{eqnarray}
Note that in the case of charged mesons, $i$ stands for the flavor indices of the initial and final quarks.

\vspace{0.5cm}

{ \it   Gluon density of the nucleon:}\\

\noindent
For the gluon density in the nucleon, one can give the definition 
symmetric with respect to the $x_1\to x_2$ transposition:
\begin{eqnarray}
   f_{g/p}(x_{1},x_{2},t,\mu )  &=&
   - \int _{-\infty }^{\infty } \frac {dy^{-}}{2\pi }
   \, \frac {1}{x_{1} x_{2} p^{+}}
   \;
   e^{-ix_{2}p^{+}y^{-}} \nonumber \\
   &\times & \langle p'|\; T G_{\nu }{}^{+}(0,y^{-},{\bf 0}_{T}) \,
     {\cal P} \, G^{\nu +}(0)\; |p\rangle \,.
    \nonumber \\
    & \ & 
\label{pdf.g.def}
\end{eqnarray}
Note that the factor of $1/(x_{1}x_{2})$ cancels the inverse factor that appears in the
derivative part of the product of the two gluon field strength tensors 
 $G_{\nu }{}^{+}(0,y^{-},{\bf 0}_{T}) G^{\nu +}(0)$.
The normalization condition is \begin{equation}
 x f_{g/p}(x,x,t=0,\mu)=  f_{g/p}(x,\mu) \,,
   \end{equation}   
   where $f_{g/p}(x,\mu)$ is the usual (diagonal) gluon PDFs.
An additional factor of $x$ reflects the difference of the symmetric definition of the gluon correlation function
   from that in the diagonal case.

The $t$ dependence of the gluon GPDs in the ``diagonal'' case of $x_1=x_2$ is of special importance for the 
interpretation of various hard $pp$ processes (see the discussion in section  7.5). It is described by the normalized 
two-gluon form factor 
$F_g (x, t, Q^2)$, where $t = -{\bf \Delta}_\perp^2$ 
is the transverse momentum transfer to the target.  
Its Fourier transform describes the transverse spatial distribution of gluons with 
given $x$:
\begin{equation}
F_g (x, \rho | Q^2) \; \equiv \; \int\!\frac{d^2 \Delta_\perp}{(2 \pi)^2}
\; e^{i ({\bf \Delta}_\perp {\bf \rho})}
\; F_g (x, t = -{\bf \Delta}_\perp^2 | Q^2) \,,
\label{rhoprof_def}
\end{equation}
where $\rho \equiv |{\bf \rho}|$ measures the distance from the transverse
center of momentum of the nucleon.  The distribution is normalized
such that $\int d^2\rho \, F_g (x, \rho | Q^2) = 1$. The information on 
 $F_g (x, t,  Q^2)$ that  can be extracted from the hard exclusive processes like $\gamma + p \to J/\psi + p$:
  \begin{equation}
 F_g(x,q_{t}^2,Q^2)=1 /(1- t /m_{g}^2)^2
 \end{equation}
Here  $m_g$ is $\approx 1 GeV$ and slowly decreases with $x$ decrease.
 \vspace{0.5cm}

Modifications necessary for the case of pseudo-scalar meson 
electroproduction are given in \cite{CFS96}.

The GPDs are different from zero  for 
\begin{equation}
-1\le x_1 \le 1 \,, \quad -1\le x_2\le 1 \,.
\end{equation}
There are two physically  different regions. In region I,
$x_1 \ge 0$ and $ x_2 \ge 0$. It corresponds to the 
  knockout of a parton with the light-cone fraction
 $x_1$ of the initial target momentum
and its absorption in the final state with the light-cone fraction $x_2$. The 
  $Q^2$ evolution is described by  the DGLAP-type  evolution equations.
In the $x\rightarrow 0$ limit, a simple connection with the diagonal distributions holds.
In   region II,   $x_1 \ge 0$ and $ x_2 \le 0$. This  corresponds to 
scattering off a small-size color singlet ($\bar{q} q$, $gg$)
emitted by the  target. The $\bar qq$ case is loosely analogous to  
scattering off  the meson cloud of the target,  provided the meson is 
collapsed into a small-size configuration. In this case,
the  $Q^2$   evolution is similar to the one for the meson wave function and  is governed by 
the Brodsky--Lepage--Efremov--Radyushkin evolution equation~\cite{Radyushkinrho}.

The imaginary part of the scattering amplitude originates from region I. 
Using a dispersion representation in energy, it is possible  to calculate 
the real part of the amplitude  for small  $x_{bj}$ and to avoid the consideration 
of region II. Secondly, at small  $x_{bj}$, the space--time evolution of the processes 
allows for a simpler  visualization of the interaction process. Also, most of 
the currently available data at large $Q^2$ were obtained at HERA for the 
small-$x$ kinematics. Thus, as a next step, we summarize the small  $x$-theory and compare it with the data.
 
 \subsection{Hard diffractive production of vector mesons}
\label{subsection7.3}

\subsubsection{Space--time evolution of high energy processes }

Vector meson production   at small $x$ in the target rest frame can be described as a three-stage process~\cite{BFS94}:

(i) The longitudinally polarized virtual photon $\gamma^{\ast}_L$ with the four-momentum
$q=(zq_0,k_t)$
breaks up into $\bar q q$ with the lifetime 
(which follows from the energy--time uncertainty principle):
 \begin{equation}
\tau_i=l_{\rm coh}/c= 
 {2q_0\over Q^2 + {k_\perp^2+ m_q^2 \over z(1-z)}}
\approx {1\over m_Nx} \,. 
\end{equation}
The coherence length is $l_{coh} \ge$ 100 fm at HERA. 

(ii)  The $\bar q q$ pair then scatters off the target proton.

(iii) The $q\bar q$ pair then lives for the time 
\begin{equation}
\tau_f=l_f/c= {2q_0 \over {k_\perp^2+m^2\over z(1-z)}} 
\end{equation}
before the final state vector meson is formed.  We note that
$\tau_f \geq \tau_i$.

 As a result, the production amplitude $A(\gamma^{\ast}_L+p\to V+p)$
 can thus be written as convolution of the light-cone   wave function of 
 the photon,   $\Psi_{\gamma^{\ast} \rightarrow |q\bar q\rangle}$,
the scattering amplitude of the hadron state,  $A(nT)$,
and the wave function of the vector meson,  $\psi_{V}$:
\begin{equation}
A= \Psi^\dagger_{\gamma^*_L  \rightarrow |n\rangle}  \otimes
A(nT) \otimes \Psi(q\bar q \to V) \,.
\label{conv}
\end{equation}

In the  impact parameter space:
\begin{equation}
A= \int d^2b \,\psi_{\gamma^{\ast}_{L}}(b)\sigma(b,s)\psi_V(b) \,,
\end{equation}
where $b$ is the  transverse separation of $q$ and $\bar q$.

 The leading twist expression is~\cite{BFS94}:
\begin{equation}
\left. {d\sigma^L_{\gamma^*N\rightarrow VN}\over dt}\right|_{t=0} =
{12\pi^3\Gamma_{V \rightarrow e^{+}e^-} M_{V}\alpha_s^2(Q)\eta^2_V
\left|\left(1 + i{\pi\over2}{d \over d\ln x}\right)xG_T(x,Q^2)\right|^2
\over \alpha_{EM}Q^6N_c^2} \,.
\label{master}
\end{equation}
Here,
$\Gamma_{V \rightarrow e^{+}e^{-}}$ is the decay width of
 $V\to e^+e^-$ and 
\begin{equation}
\eta_V\equiv {1\over 2}{\int{dz\,d^2k_t\over z(1-z)}\,\Phi_V(z,k_t)\over
\int dz\,d^2k_t\,\Phi_V(z,k_t)}\rightarrow 3
\end{equation}
 for $Q^2\rightarrow \infty$. The rapid onset of the leading twist formulas  for $\sigma(e\bar e\to hadrons)$
suggests that for $\rho$ and $\phi$ mesons, $\Phi_V(z,k_t)$ and hence 
$\eta$ are already close to the asymptotic value  at $Q^2 \sim $ a few GeV$^2$. 

Note here that in this expression, the difference between the light-cone fractions $x_1$ and $x_2$ was neglected. For large $Q^2$, the non-diagonal GPD is calculable~\cite{Frankfurt:1997ha,Shuvaev:1999ce}  through the diagonal one since the DGLAP evolution for GPDs  conserves $x=x_1-x_2$, while the
light-cone fractions  essential  at the starting point of the evolution grow with an increase of $Q^2$.

In \cite{Ryskin}  elastic photoproduction of $J/\psi$  was evaluated in the leading $g^2\ln(x_0/x)$ approximation. 
As we discussed above this approximation ignores huge NLO effects.

At extremely small values of $x$, that  are significantly smaller than those characteristic for the 
applicability of the DGLAP approximation, the $\ln(x_0/x)$ terms  
not enhanced by $\ln(Q^2/Q_{0}^2)$ and thus
neglected in the DGLAP approximation, become important.  
The restriction on the region of applicability leading $\log(x_0/x)$ approximation follows from the necessity to take into account energy--momentum conservation. Indeed, in 
multi-Regge kinematics, the interval in rapidity  between adjacent radiations within the ladder is 
$\Delta y\gg 2$. This number is comparable with the interval in rapidities achieved (to be achieved) in DIS: 
\begin{equation}
\Delta y=\ln(1/x)+2 \ln(Q/m_N) \,.
\end{equation}
For the edge of the kinematics achieved at HERA,  $\Delta y\approx 10$. Since four units of rapidity are occupied by the two fragmentation regions, two-to-three gluons are allowed to be radiated in this kinematics.  This is insufficient  for the dominance of multi-Regge kinematics 
characteristic for LL approximation. 
So far there have been no attempts to describe hard diffractive processes in the resummation approach. 
Note that formulas obtained in the double logarithmic approximation,  $\alpha_s\ll 1$ and
 $\alpha_s\ln(x_0/x)\ln(Q^2/Q^2_0)\sim 1$,  should coincide for the BFKL and 
 DGLAP approximations.

\subsubsection{Modeling finite-$Q^2$  effects}

\begin{figure}[h]  
   \centering
   \includegraphics[width=0.9\textwidth]{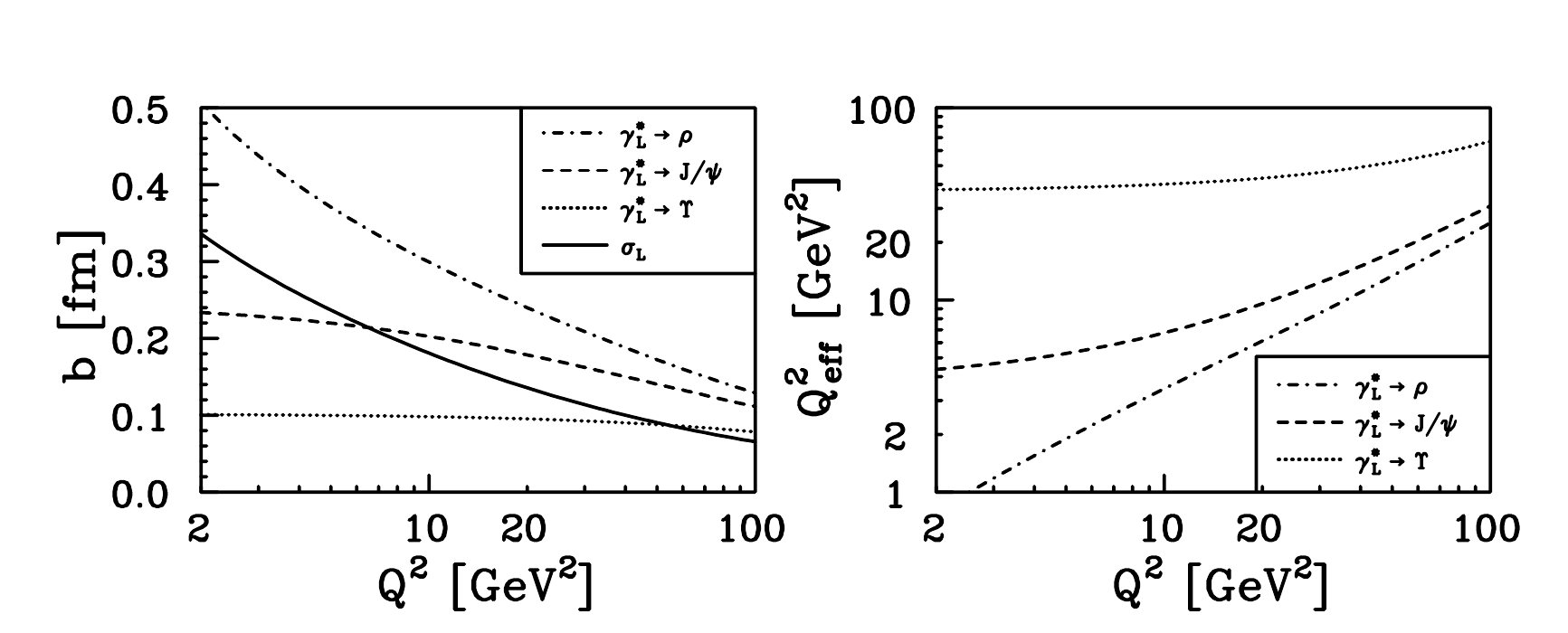} 
 \caption{The dependence of the average $b$ (left) and 
the effective $Q^2$ (right) on $Q^2$
for production of vector mesons \protect\cite{Koepf}.} 
\label{bsize}
 \end{figure}

In the convolution integral (Eq.~\ref{conv}), $\Psi_{\gamma^*}^L(b)$  with $b\propto {1 \over Q}$ is 
convoluted  with the broad wave function  of a light vector meson\footnote{To obtain this expression, 
one needs first to apply conservation of the electromagnetic current to express the bad component of the current through the good one.}. 
Hence, the  average distances  contributing to  $\sigma_L$  are significantly 
 smaller than those contributing to $\sigma_T$. 
 As a result,  the effective $ Q^2$ is smaller for  vector meson production than for 
$\sigma_L$ \cite{Koepf} (Fig.~\ref{bsize} ). 
This effect is taken into account by evaluating $\sigma(q\bar q - N)$ using the dipole model. 
One also has to include the difference between $x_1$ and $x_2$, 
which is absent when the dipole model is applied to inclusive DIS.

A  related effect is that, at  pre-asymptotic energies, one cannot  substitute 
$\Psi_V(b)$ by $\Psi_V(0)$.  This higher twist correction leads to the 
suppression of the amplitude by the factor:
\begin{equation}
T(Q^2)=
{\left|\int d^2b\,dz\, \Psi_{\gamma^{\ast}_L}(z,b)\sigma(q\bar q-N)\phi_V(z,b)\right|^2
\over 
\left|\int d^2b\,dz \, \Psi_{\gamma^{\ast}_L}(z,b)\sigma(q\bar q-N)
\phi_V(z,0)\right|^2} \,.
\label{tfac}
\end{equation}
  
\begin{figure}
   \centering
\includegraphics[width=.6\textwidth]{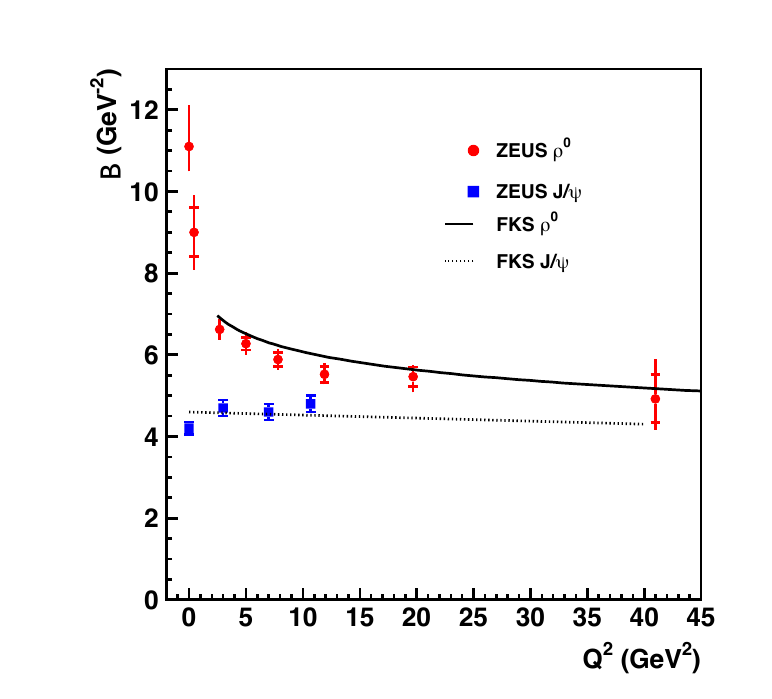}
\caption[]{The convergence of the $t$-slopes, $B$,  of $\rho$ and $J/\psi$
 electroproduction  at high $Q^2$. The data are from~\cite{Chekanov:2004mw,Chekanov:2007zr}; 
 the curves are the predictions 
 of~\cite{Koepf}.}\label{slope}
\end{figure}

The HERA data (for a  recent summary, see~\cite{Levy:2007fb})
 have  confirmed the following  basic predictions of pQCD~\cite{BFS94}:

\noindent 
\begin{itemize}
\item 
The rapid increase with energy---$\left|xG_N(x,Q^2_{eff})\right|^2 \propto W^{0.8}$ for $Q^2_{\rm eff} \sim 4$ GeV$^2$---of
  $\rho$ production for $Q^2=10-20$ GeV$^2$ and  of $J/\psi$ production for $Q^2\le 10$ GeV$^2$. (Note that $\sigma(W) \propto W^{0.32}$ for soft physics at $t=0$ and is even slower for the cross section integrated over $t$.)
For $\Upsilon$ production, $Q^2_{\rm eff} \approx 40$ GeV$^2$ which leads to 
$\sigma(W) \propto W^{1.7}$. This prediction maybe tested in the 
 ultraperipheral collisions at the LHC~\cite{Baltz:2007kq}. 
 
 \item The absolute values of the cross sections of vector meson production are well  reproduced,  provided 
that  the factor $T$ (Eq.~\ref{tfac}) is taken into account. In the case   of  $\Upsilon$ photoproduction, the
skewedness effects due to large difference between $x_1 $ 
 and $x_2$ as well as the large value of the real part of the amplitude  are important.  
 Together they  increase the predicted cross section by a factor of about four~\cite{Frankfurt:1998yf,Martin:1999rn}.
 \item 
The  decrease of $\sigma_L$ with $Q^2$ is slower than ${1/Q^6}$ because of the 
$\left|\alpha_SG_N\right|^2$ and $T(Q^2)$ factors. 

\item
The ratio $\sigma_L/\sigma_T \gg 1$ for $Q^2\gg m_V^2$.

\item There is a universal $t$ dependence for large $Q^2$ originating solely from the two-gluon--nucleon form factor. 
The model, which takes into account squeezing of $\gamma_L$ with $Q^2$, provides a reasonable description 
of 
the convergence of the $t$-slopes of light mesons and $J/\psi$ production 
and makes the observation that the slope of $J/\psi$  production is practically $Q^2$ independent (Fig.~\ref{slope}).

\end{itemize}

\subsubsection{Lessons and open problems}

$\bullet$ Transition from soft to hard regime.

We can estimate the effective size of a $q\bar q$ dipole as 
\begin{equation}
{B(Q^2)-B_{2g} \over  B(Q^2=0)-B_{2g}} \sim {R^2({\rm dipole}) \over R^2_{\rho}} \,,
\end{equation}
where $B_{2g} $ is the slope of the square of the two-gluon form factor.
Based on the HERA data \cite{Chekanov:2007zr},  we  conclude that
\begin{equation}
 {R^2({\rm dipole}) (Q^2\ge 3 \ {\rm GeV}^2)/ R^2_{\rho}} \le {1/2  - 1/3} 
\end{equation}
for collider energies.  
Accordingly, it appears that the soft  energy dependence of the 
cross section persists over a significant range of the dipole sizes. (This is consistent with the observed similarity of the energy dependence of $\rho$ and $\phi$ photoproduction.)

 $\bullet$  In the pQCD regime, the $t$-slope the dipole--nucleon amplitude should be a weak  function 
 of $s=W^2$, $B(s)=B(s_0)+2\alpha^{\prime}_{\rm eff}\ln(s/s_0)$, since the Gribov diffusion in the hard regime is small 
 (see the discussion in section 4).  Hence, a significant contribution to $\alpha^{\prime}$  comes 
 from the variation of the $t$ dependence of the gluon GPD with an decrease in  $x $ at $Q_0^2$.  

$\bullet$ 
The contribution of soft QCD physics in the overlapping integral between the wave functions of the virtual photon and 
the transversely polarized vector meson is suppressed by the Sudakov form factor (see, e.g., the discussion 
in~\cite{Mankiewicz:1999tt}), which is absent in the case of  the processes initiated by longitudinally polarized photons. 
This is probably relevant for 
the understanding of 
 similar dependence of $\sigma_{L,T}$ on $x$ and on $t$ that were observed at HERA.

\subsection{Meson production at intermediate $x$}
\label{subsection7.4}

An analysis of the HERA data on vector meson production indicates  that,  up to rather large  $Q^2$, cross sections of hard exclusive processes
are suppressed significantly as compared to the leading twist QCD  predictions. 
The origin of this is
higher twist effects  originating from the  contribution of the transverse sizes that are  
comparable in the longitudinal photon and the meson wave functions.

At the same time,  the overall transverse size of the produced meson  is  quite small ($\le$ 0.4 fm) 
already for $Q^2 \ge$~5 GeV$^2$.    Due to the color transparency phenomenon,  this leads to  a strong suppression of the final state interaction of the  $q \bar q$ pair,  which  in the end  will fragment into
the meson  and the residual baryon system. For $W \le 20$~GeV, this cross section   is  of the order of few mb. Besides, the expansion of the  $q \bar q $ system  to a normal hadron size in the nucleon rest frame
takes a significant distance
\begin{equation}
 l_{coh} \sim 2p_M/(\Delta m_M^2)\sim
\frac{1}{x M_N} Q^2/(\Delta m_M^2) \,
\end{equation}
where  $\Delta m_M^2\le  1$ GeV$^2$, and 
$(\Delta m_{M}^2)/2p_M$
is the characteristic light-cone energy denominator  for a meson $M$.
The  condition  $l_{coh} \gg r_N$  is  satisfied for  $x \le 0.2$ already for 
 $Q^2 \ge$ 5 GeV$^2$.  Hence, it seems likely that   
 the precocious factorization into the three blocks (Fig.~\ref{fig:fact})---the overlap integral between the photon 
 and  the meson wave functions, the  hard blob, and the  skewed distribution---could 
 be valid already at moderately high  $Q^2$,  leading to 
 precocious scaling of the spin asymmetries  
and of the cross section ratios as a function of  $Q^2$. 

The discussion of numerous promising channels such as production of charged and neutral vector 
mesons ($\rho^{\pm,0}$, $K^{*+}$, $\dots$), pseudoscalar mesons $\pi^0,\eta,\eta'$ that are sensitive to the QCD  axial anomaly, and  
$\Delta$-isobars is beyond the scope of this review (for a detailed discussion, 
see~\cite{Goeke:2001tz}).

\subsection{Transverse structure of the nucleon at small $x$ from GPDs}
\label{subsection7.5}

 At small $x$  many processes are dominated by the two-gluon ladder
 with no contribution from quark GPDs---like production of $J/\psi$ and $\Upsilon$.
	In the case of GPDs linked to sea quarks, the   situation is more complicated. In the case of the 
	deeply virtual Compton scattering (DVCS) amplitude  in the NLO
approximation,  quark and gluon contributions enter with opposite signs and are of a comparable magnitude~\cite{Freund:2001hm}.  As   a result,  a relatively small difference of the transverse sizes of the sea quarks and gluons, which is expected due to pionic cloud  effects~\cite{Strikman:2003gz}, is amplified in the $t$-slope of DVCS. Thus, 
precision measurements of the quark GPDs at 
small $x$ require an accurate measurement of the gluon GPD.

Higher twist (HT) effects modify  the $t$-distribution of light mesons  up to $Q^2 \sim 15$ GeV$^2$. 
Therefore,  it appears that the only practical chance to perform  a precision measurement of the gluon GPDs 
is the production of onium states. 

The effects of non-diagonality in the 
gluon GPD appear to be small for the $J/\psi$ case, this is due to the large transverse momenta in 
the wave function, 
which lead to the comparable light-cone fractions of the gluons attached to $c\bar c $: $x_1/x_2 \sim 2$, 
$x_1-x_2=x=(Q^2+m_{J/\psi}^2) /W^2$. In the $\Upsilon$ case, the effect of non-diagonality can be taken into account via  the DGLAP evolution.  As a result, 
\begin{equation}{d\sigma^{\gamma + p \to J/\psi + p}
\over dt}\propto F_g^2(x,t) \exp(\Delta B t) \,,
\label{twogl}
\end{equation} 
where $F_g(x,t) $ is the two-gluon form factor of the nucleon;
the second factor takes into account a small but finite correction due to the finite size of $J/\psi$ that 
was estimated in~\cite{Koepf} to be
 $\Delta B \approx 0.3$ GeV$^{-2}$.
 
 The $t$--dependence of the measured differential cross sections of exclusive processes for $|t| < 1$ GeV$^2$ 
 is commonly  described either by an exponential or by a dipole form inspired 
by analogy with the nucleon elastic form factors. The data are not precise enough to distinguish  between the
two forms since they mostly differ at small $t$, where the resolution of the measurements is moderate, and at 
large $-t \ge 0.8$ GeV$^2$, where the measurements rather strongly  depend  on the procedure of subtraction of the inelastic background contribution.

The data can be fitted in the following form~\cite{Frankfurt:2010ea}:
\begin{equation}
B_g (x) = B_{g0} \; + \; 2 \alpha^{\prime}_g \; \ln (x_0/x)  \,,
\end{equation}
where
\begin{equation}
x_0 = 0.001 \,, \quad B_{g0} = 4.1 \; ({}^{+0.3}_{-0.5}) \; \mbox{GeV}^{-2} \,, \quad
\alpha'_g = 0.140 \; ({}^{+0.08}_{-0.08}) \; \mbox{GeV}^{-2} \,. 
\label{bg_param_last}
\end{equation}

Fits of similar quality are produced with the dipole form: 
\begin{equation}
F_g (x, t|Q^2)=(1 - t/m_g^2)^{-2} \,, \quad B_g= 3.2 /m_g^2\,\, ({\rm for} \, m_g^2 \sim \mbox{1 GeV}^2) \,.
\end{equation}
The  spatial distributions of gluons in the transverse plane corresponding to the two fits are: 
\begin{eqnarray}
F_g (x, \rho | Q^2) \;\; = \;\; 
\left\{ \begin{array}{l}
\displaystyle 
(2 \pi B_g)^{-1} \, \exp [-\rho^2 / (2 B_g)] \,,
\\[2ex]
\displaystyle 
[m_g^2/(2\pi)] \; (m_g \rho/2) \; K_1 (m_{g} \rho ) \,,
\end{array}
\right.
\label{f_rho_param}
\end{eqnarray}
These transverse distributions are  similar for the average $\rho$, leading, for example, to the 
nearly identical distributions over the impact parameter for  production of dijets in $pp$ collisions at 
 the LHC~\cite{Frankfurt:2010ea}.   At the same time, the dipole fit gives a significantly larger $F_g (x, \rho | Q^2)$ for 
small $b$. As a result,  analyses of  the proximity to the black disc regime for the interaction of a
small dipole with the nucleon for small $b$ are sensitive to the choice of the model for $F_g (x, \rho | Q^2)$.
 A related effect is a factor of 1.6  difference of $\left<b^2_g\right>$  in the two fits.  The conclusion \cite{Frankfurt:2002ka}    that gluons are localized in a smaller transverse area than 
 that   given by the e.m.~form factor  is based on the use of the same shape for the e.m.~and  two-gluon form factors.
 
 The current knowledge of  $F_g (x, \rho | Q^2)$ allows us to study the impact parameter dependence of 
 dijet production at collider energies. One finds  that the median  impact parameters for the 
 inelastic $pp$ collisions with jet production are a factor of two smaller than those for the minimal bias 
 inelastic events, and weakly depend on the rapidity and $p_t$ of the jets. This may explain the 
 regularities in the multiplicity of the underlying events on $p_t$ of the trigger~\cite{Frankfurt:2010ea}.
   It also provides an important constraint  for the models of inelastic $pp$ collisions  at the  LHC and, 
   in particular, of the dynamics of multiparton interactions.

 \subsection{Break-up processes with gaps: from small to large~$t$}
 \label{subsection7.6}
 
\subsubsection{Probing fluctuations of the gluon field}

In the high energy $\gamma^{\ast}+ p \to M +{\rm rapidity}~gap+ X$ process, at $t\approx 0$ 
the two-gluon ladder couples only to one parton in the target in the leading twist approximation.
 If the strength of the coupling  to all configurations containing partons with a given $x$ were the same,
  it would be  impossible to produce an inelastic final system $X$. As a result, similarly to the case 
  of inelastic diffraction of hadrons off hadrons, the discussed process measures the variance of the gluon field. 
  It is given by the ratio of the diffraction dissociation and elastic cross sections for 
  vector meson production at $t=0$~\cite{Frankfurt:2008vi}: 
\begin{equation}
\omega_{\rm hard}={d\sigma^{\gamma^{\ast} + p \to V +{\rm rapidity \ gap} +M_X}(x,Q^2)/dt \over d\sigma^{\gamma^{\ast} + p \to V +p}(x,Q^2)/dt}={\langle G^4 \rangle - \langle G^2 \rangle^2 \over   \langle G^2 \rangle^2 } \,.
\end{equation}
	Our estimates of the strength of the fluctuations due to the fluctuations of the overall size of the 	
	nucleon, as seen in soft inelastic diffraction, find 
	that
	  $\omega_{hard} \approx 0.15$ at small $x$, which is consistent with the current data. 
	  
	\begin{figure}[h]
   \centering
\includegraphics[width=.35\textwidth]{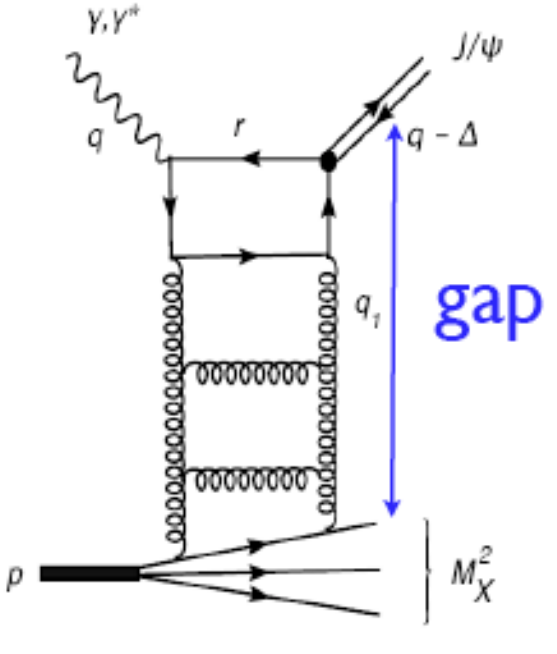}
\caption[]{A typical QCD diagram for the rapidity gap process (\ref{gap}).} 
\label{fig:gap}
\end{figure}

\subsubsection{Onset of new regime at large $t$  }

In order to determine the spacial distribution of gluons at $b\sim 0$, one needs to study exclusive processes 
at large $t$ since in this case the integral over $-t$ in the Fourier transform  converges very slowly.
For example, in the case of the dipole fit, half of the contribution to $F_g(x,b=0)$ originates 
from $-t\ge m_g^2$ and one quarter, from 
$-t\ge 3 m_g^2$. This implies that one needs a detector which will be able to separate exclusive processes from diffractive dissociation in a very broad range of $t$, where the dissociation dominates  by far.

A further complication is that the QCD factorization theorem \cite{CFS96} was derived in the limit of 
$t={\rm const}$, $x={\rm const}$, and $Q^2\to \infty$.   Recent studies~\cite{BFS} found
that the pattern of the QCD evolution changes in the kinematical domain
when $-t$ becomes comparable to the intrinsic  hardness scale of the process 
($Q^2_{\rm eff} \sim 3$ GeV$^2$ for $J/\psi$ photoproduction). 
The  DGLAP equations require modifications in this case. Let us, for example, consider the process 
	\begin{equation}
	\gamma^{\ast} + p\to J/\psi + \mbox{rapidity gap} + M_X 
	\label{gap}
	\end{equation}
	for  large $-t$, which is still smaller than $Q^2 + M_{J/\psi}^2$. The typical  leading QCD diagrams (Fig.~\ref{fig:gap}) correspond to the attachment of the two-gluon ladder to a parton with $x_J=-t/(-t+M_X^2 -m_N^2)$.
The cross section has the factorized form similar to~\cite{Frankfurt:1990nc}:
\begin{equation}
\frac
{d\sigma_{\gamma+p\to V+X}} {dt dx_J}=
\frac {d\sigma_{\gamma+{\rm quark} \to V+{\rm quark}}} {dt} 
 \left [{81\over 16} g_{p}(x_J,t) 
+\sum_i (q_{p}^{i}(x_J,t)+{\bar q}_{p}^{i}(x_J,t)) \right ] \,. 
\end{equation}

At fixed $x_j$, the energy dependence is determined by the evolution 
of the dipole--parton elastic scattering amplitude with $x/x_J$. In difference from the small $t$ limit,
 the DGLAP evolution is strongly suppressed and completely disappears for $-t$ close to the intrinsic scale. 
 The HERA data~\cite{HERAgap} on the energy dependence of the process (\ref{gap}) are consistent 
 with the behavior expected in QCD.

		Consequently, the  effective $\alpha_{\Pomeron}(t)$ in this limit stays close to unity until very 
small $x/x_J$ (not available at HERA) where the  Pomeron-type behavior may reveal itself.

The same mechanism  may be responsible for part of the drop of $\alpha_{\Pomeron}(t)$  
with an increase in $-t$ observed for elastic $J/\psi$ production. This may indicate that precision measurements 
of the $t$ dependence of the two-gluon form factor at high energies would require 
using of
electroproduction rather than of photoproduction.  The discussed phenomenon may also be 
 relevant for the explanation of the pattern observed in photoproduction
 of  $\rho$  mesons, where $\alpha_{\Pomeron}(t)$  appears to flatten out around 
 $\alpha_{\Pomeron} (t) =1 $ for large $|t|$   (Fig.~\ref{pomeron}).

\subsubsection{Probing minimal quark component in the pion}
\label{subsubsection7.3}
 
\begin{figure}[h]  
   \centering
   \includegraphics[width=0.6\textwidth]{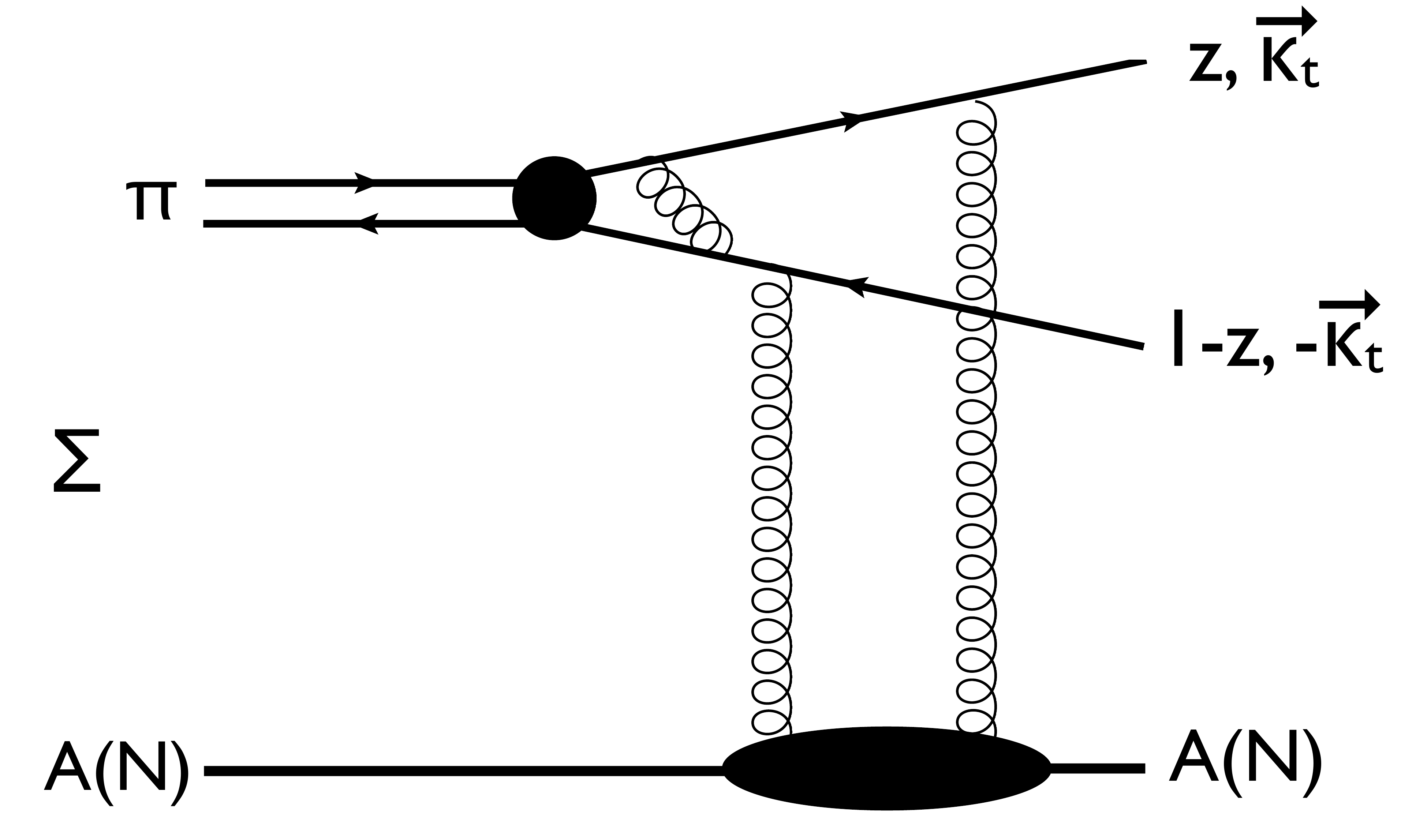}
 \caption{A typical  two-gluon ladder exchange diagram contributing to the process of pion coherent diffractive dissociation.}
\label{dijet1}
\end{figure}

The  QCD factorization theorem discussed in the previous 
subsections allows us to calculate another  group of hard processes where the selection of the final state 
dictates squeezing of the initial state. This is a particular case of the pre-selection phenomenon familiar from 
non-relativistic quantum mechanics. 
The most straightforward process is  
\begin{equation}
\pi+T\to {\rm two \ jets}+ {\rm rapidity \ gap} +T \,.
\end{equation}
This process is in a sense a mirror image of vector meson production in DIS. The pion in the initial 
state collapses into a small-size configuration due to the hard interaction; this $q\bar q$ pair interacts coherently with the target and transforms into two jets~\cite{Miller93} (Fig.~\ref{dijet1}).

In the limit of large transverse momenta of the jets, one can 
 justify applicability of the QCD factorization theorem and 
 obtain~\cite{Frankfurt:2000jm}:
 \begin{equation}
{d\sigma(\pi + A\to 2jet +A)(q_t=0)\over dt\, dz\, d^2 \kappa_{t}}=
\frac{(1 +\eta^2)}{16\pi (2\pi)^3}
{\left[\Delta\left({\chi_{\pi}(z,\kappa_t)}\right)
 {\alpha_s \pi^2\over 3} x_1 G_A(x_1,x_2,Q^2) \right]^2} \,,
\label{dijet}
\end{equation}
where $\chi_{\pi}(z,\kappa_{t})\equiv{4\pi
C_{F}}{\alpha_{s}(\kappa_{t}^2)\over \kappa_{t}^2} \sqrt{3} f_{\pi}z(1-z)$;
$\Delta$ is the Laplacian in the $\kappa_t$ space; $z$ is the fraction of the
light-cone plus-momentum  carried by the quark in the final state;
$x_{1}G_{A}(x_{1},x_{2},\kappa_{t}^2)$ is the generalized
gluon density of  the nucleus, where $x_{1}$ and $x_{2}$ are the fractions of the target momentum
carried by exchanged gluons 1 and 2 respectively, $x_1-x_2=M^2_{\rm 2jet}/s$ and $x_2\leq x_1$  and the integral over $x_2$ is not written explicitly;
$\eta=\mbox{Re}  F/\mbox{Im}  F$, where $F$ as the dipole--nucleon scattering amplitude).
Note that the resulting $\kappa_t^{-8}$ dependence is  a consequence
of the comparatively well-understood wave function of the pion.  This wave function determines the 
asymptotic behavior of the pion electromagnetic  form factor.

For $x\ge 0.03$, $G_A(x,Q^2)=AG_N(x,Q^2)$. Hence, for this kinematics there are no absorptive effects---the 
amplitude at $t=0$ should be $\propto A$. This prediction has been confirmed at 
FNAL with $p_{\pi}$=500 GeV/c: the experiment~\cite{Aitala:2000hc} observed a strong coherent peak for dijet production
from carbon and platinum targets and measured the $A$ dependence for this interval of $A$ that was found  to be $A^{1.54}$. 
The ratio of the cross sections for the two targets is a factor of seven larger than that for soft coherent diffraction.  Furthermore, the observed
dependence of the cross section on the pion momentum fraction and the 
jet transverse momentum is well-consistent with the perturbative 
QCD prediction of~\cite{Miller93,Frankfurt:2000jm} for $k_t \ge 1.5$ GeV/c.
The relatively early onset of scaling for this process---as compared to 
diffractive electroproduction of vector mesons 
discussed above---maybe due to the presence of the plane wave in the convolution formula for the 
dijet production cross section. This should be compared to the case of the virtual photon wave function that 
restricts the phase space much stronger.

We would like  to note that the derivation of the QCD factorization theorem for this process~\cite{Frankfurt:2000jm} 
 heavily used the fact that the trigger on
 two high-$\kappa_t$ jets along with the Ward identities enforce the $q\bar q$ pair and, therefore, the color  
 to be concentrated in the interaction volume $\propto 1/\kappa_t^2$. 
 The use of the asymptotic freedom and the restriction by the leading twist contribution guarantees the 
 dominance of the $q\bar q$-component in the pion wave function at  sufficiently large $\kappa$.  
 The generalized gluon distribution of the  target depends on variables $x_1\gg x_2$  and $x_1-x_2=M^2_{\rm 2jet}/s$. 
 The kinematics     follows  from energy--momentum conservation.  
 The mass of the two jets, $M_{\rm 2jet}$,   is significantly larger than the  pion mass; that allowed one 
 to justify the applicability of pQCD and the  Ward identities. The calculation  also explores  the 
 asymptotic solution of the QCD evolution equation for the  pion wave function.

It was assumed in~\cite{Braun}  that for the calculation, one can substitute the pion by a system 
of the non-interacting quark and antiquark with the mass equal that of the two-jet system $M_{\rm 2jet}\gg m_{\pi}$. 
 This assumption violates the conditions of the applicability of the QCD factorization theorem discussed above 
 and, therefore, produces a different amplitude.      There is no requirement for the 
 $q\bar q$ pair (color) to be within a small volume. Therefore, there is no way to justify the applicability of pQCD 
 and to neglect other quark--gluon configurations.  
 The calculation of leading Feynman diagrams due to the two-gluon exchange produces the 
 factor  $1/(M^2_{\rm 2jet}-M^2_{\pi})$, while in the kinematics of~\cite{Braun}, $M_i=M_{\rm 2jet}$.  
 A prescription is needed to remove this artificial singularity.  Violation of the Ward identities 
 requires an additional prescription of how to derive the condition that  
 the leading amplitude for  $q\bar q$ scattering off the two-gluon ladder exchange  is  $\propto r^2_t$,  
 where $r_t$ is the momentum of the exchanged gluon.    In this kinematics, $x_1=x_2\approx 0$.   
Besides, in the framework of the approximations discussed  above,
the 
authors~\cite{Braun} found 
that the scattering amplitude contains 
singularities that  contradict  the  Landau rules  for the amplitudes of physical processes.

For a proton projectile, the related process would be proton diffraction into three jets:
\begin{equation}
p+A\to {\rm three \ jets}+{\rm rapidity \ gap}+A \,.
\end{equation}
So far, the observation of this process at the LHC looks very difficult---current 
detectors do not have acceptance for realistic $p_t$ of jets of this process.

\subsection{Nuclear effects in hard exclusive reactions}
 \label{subsection7.7}

 The use of  nuclear targets provides complementary probes of the QCD dynamics of diffractive processes.   QCD predicts that a sufficiently energetic,  spatially small color-neutral wave packet of quarks and gluons   should traverse  hadron medium without absorption. This prediction is equivalent to the QCD factorization theorem  of~\cite{CFS96}.  
 The complete transparency  of nuclear matter [color transparency (CT)] has been unambiguously observed at several 
 experiments at high energies (see discussion below). Such processes are becoming  a promising tool for the 
 detailed investigation of the quark and gluon structure of nuclei.  
 
We give here just  a few examples. The factorization theorem \cite{CFS96} predicts 
for coherent electroproduction of vector mesons~\cite{Miller93,BFS94}:
\begin{eqnarray}
\frac{{d\sigma\over dt}(\gamma^{\ast} A \to V A)\big\vert_{t=0}} {{d\sigma\over dt}(\gamma^{\ast} N \to V N)\big\vert_{t=0}} =
\left [{F^L_A(x,Q) \over F^L_N(x,Q)}\right ]^2
= {G^2_A(x,Q) \over G^2_N(x,Q)} = A^{2 \alpha_g(x,Q)} \,,
\end{eqnarray}
with  $ \alpha_g(x,Q)\approx 1$  for $x\ge 0.02$. 
Thus  final-state interaction in this process is 
a higher twist effect.

$\bullet$   Almost complete transparency has been observed at FNAL~\cite{Sokoloff:1986prl} 
in both the coherent  and incoherent  processes: 
$\gamma A \to J/\Psi+A$, $\gamma A \to J/\Psi+A^{\ast}$. 

$\bullet$ Complete transparency of nuclear matter, which follows from the QCD factorization theorem for the sufficiently large transverse momenta of jets  $k_t\ge 2$ GeV, has been observed in the processes 
$\pi + A \to {\rm 2\ jets} +A^{\prime}$~\cite{Aitala:2000hc}.

 $\bullet$  For quasi-elastic scattering off nuclei and for large enough $\left|t\right|\ge 0.1$ GeV$^2$,
  one can use closure over the processes of nuclear disintegration    $A^{\prime}$. 
  For example,   for production of neutral mesons, one obtains~\cite{FPPS}:
\begin{eqnarray}
\sum_{A'} {d\sigma(\gamma^{\ast}_L+A \to M +A')\over dt} &=&
Z\,{d\sigma(\gamma^{\ast}_L+p \to M +p^{\prime})\over dt} \nonumber \\
&+&N\,{d\sigma(\gamma^{\ast}_L+n \to M +n^{\prime})\over dt} \,.
\end{eqnarray}

 $\bullet$ The color  transparency phenomenon leads to a strong suppression of the cross 
 section of coherent scattering off the lightest nuclei  for  the $t$ range, 
 where double scattering dominates~\cite{FPSS}.

At lower energies, where the Lorentz factor is not large enough  to guarantee the sufficiently large 
lifetime of a spatially small wave packet of quarks and gluons traversing a nuclear target,  
CT is masked to large  extent by the following quantum mechanical phenomenon:   
the small-size wave packet is not an eigenstate of the QCD Hamiltonian and, hence, it rapidly expands~\cite {FFS}. 

\begin{itemize} 
\item   The onset of complete transparency has been observed~\cite{Adams:1995} in the incoherent process 
$\gamma^{\ast}+A \to \rho +X$ at $\nu \sim 200$ GeV in the kinematics where hadron production 
in the nuclear fragmentation region was allowed.  A smaller effect was observed at $\nu \sim 15$ GeV under 
similar conditions~\cite{Airape:2003}. 

\item  A gradual increase of transparency in exclusive processes with an increase of $Q^2$ was observed in 
high precision experiments at JLab at $\nu \sim $ a few GeV that measured production of $\pi^+$~\cite{Clasie:2007} and 
$\rho^0$~\cite{ElFassi:2012nr}. The observed $A$ and $Q^2$ dependencies are consistent with 
the familiar quantum mechanical effect of the expansion of a 
small $q\bar q$ wave packet with distance~\cite{Frankfurt:1988nt}. 

\end{itemize}

 In  collider kinematics,  it is much easier to check 
whether the  nucleus remained intact
and to also select certain exclusive break-up channels. Hence, 
it may be easier to study color transparency effects for
 these processes at  colliders.

  Overall,  the use of nuclei would add substantially to the program
of studies on hard exclusive processes for  the $x\ge 0.05$ range.
Since nuclear 
parton densities 
for a wide range of   $x\ge 0.05$
are $ \propto A$ up to a small correction due to the EMC effect,
one would not have to deal simultaneously
with the effect of leading twist shadowing.

  \subsection{Summary}
  Hard exclusive meson  (few meson) production is calculable in QCD
in the same sense as the leading twist DIS processes.
  Hard exclusive processes provide unique ways to
 study minimal Fock state components 
of mesons and structure functions of mesons,
to compare skewed parton distributions
 in a multitude of baryons, and to investigate the onset of color transparency.

Colliders have a number of advantages for observing many of these processes.

To have a successful program of studies on the hadron and nuclear structure
in the next decade, one needs a 
 triad of electron accelerator facilities optimized to study
the following three $x$ ranges: (a) the $x\ge 0.3 $ range relevant for the study of
short-range correlations in nucleons and nuclei, (b) the $0.05\le x\le 0.3$
range relevant for the study of multiparton correlations in nucleons and the 
origin of the nuclear forces, (c) the $x \le 0.05$ range relevant for the nuclear
shadowing phenomena, high parton density physics, etc.  This 
corresponds well to the $W \le 8$ GeV range discussed for the upgrade of the Jefferson Lab facility 
and for the COMPASS experiment with the recoil detector as well as to the 
$W \sim 30 - 150$ GeV range discussed for the electron--nucleon/nucleus colliders at 
JLab and RHIC and for the LHeC collider at CERN.

\section{ A new regime of high energy QCD}
\label{section8}
\subsection{Introduction}
\label{subsection8.1}
At large energies, hard QCD interactions become  strong at  central impact parameters   in spite of the small running coupling constant and a larger scale of momenta. 
The necessity to take into account an entire series in $(1/Q^2)^n$ (which resembles the situation in the second-order phase transitions)
shows that this new phase of QCD has a different continuous symmetry (conformal symmetry), thereby  
distinguishing it from  the perturbative phase of QCD.

With a further increase of energy,   predictions based on the QCD factorization theorems  start to contradict probability conservation. For the interaction of a spatially small dipole with a target,
this can be be formulated as:
\begin{equation} 
\sigma_{\rm el}(dipole+T)\le{1\over 2}\sigma_{\rm tot}(dipole+T) \,. 
\label{unitarity2}
\end{equation}
The upper limit on the $\sigma_{\rm el}/\sigma_{\rm tot}$ ratio is reached in the regime of complete absorption for all essential impact parameters.  The restriction due to probability 
conservation is stronger for the scattering at central impact parameters (see the discussion in section~\ref{subsection8.2}).

In the leading twist approximation, the inequality~(\ref{unitarity2}) is violated at sufficiently small $x$ 
since the   QCD factorization theorem for the interaction of a small-size colorless dipole leads to 
$\sigma_{\rm el}\propto (xG_T(x, Q^2)/Q^2)^2/B$,  which becomes larger than   
$\sigma_{\rm tot}\propto xG_T(x, Q^2)/Q^2$ for sufficiently small $x$ ($B$ is the $t$-slope of the 
differential cross section of the dipole--target elastic scattering).  
Thus, in the limit of fixed and large $Q^2$, $x\to 0$
 and fixed impact parameters, the decomposition over high powers of $(1/Q^2)^{n}$ (over twists)  becomes ineffective. 
 Indeed, the dependence of the higher twist term  $T_{n+2}$  on $x$ and $Q^2$ can be easily evaluated at large 
 energies:  $T_{n+2} \propto (1/Q^2)^{n} (x_0/x)^{ (1+n)(\lambda-1)}$. 
 Therefore the ranking over twists disappears at  sufficiently small $x$.  As a result, the concept of a 
 spatially small dipole becomes ineffective as well since the contribution  of various higher twist effects 
 (for example, the splitting of a small dipole into two small dipoles each interacting with the target)   
 is not suppressed at these energies. In other words, the effective number of dipoles
continuously increases with energy.

The regime of complete absorption at fixed impact parameters and conservation of probability does not preclude a 
rapid increase in  cross sections of hard processes in $pp$ collisions at central impact parameters 
as well as of the $\gamma^{\ast}N$ cross section with a decrease in  $x$ at fixed $Q^2$.   
At collider energies,  the hard contribution to the nucleon structure functions increases with energy 
as  $\sigma(\gamma^{\ast}N) \propto \ln^2(x_0/x)$, which is  faster  than the $x^{-0.2}$ behavior   
observed at HERA  at $Q^2\sim$ a few GeV$^2$.   At very high energies, an 
even faster increase is expected, 
$\sigma(\gamma^{\ast}N)\propto \ln^3(x_0/x)$~\cite{MGFS,BF1}.  The additional $\ln(x_0/x)$  is 
a consequence of the singular behavior of the light-cone wave function of the virtual photon in the 
coordinate space, which in the momentum space corresponds to the contribution of the quarks with momenta $\gg Q$
in the box diagram.   Numerical studies on the energy dependence of  $F_{2N}(x,Q^2)$ that took into account
the taming of partial waves at small impact parameters, were performed in a number of papers using 
the dipole model (see, for example, \cite{Watt:2007nr}).

 Note also that  QCD predicts taming of structure functions of nuclei  which competes  with the nuclear shadowing phenomenon.   
 However, it is hard to observe the violation of the DGLAP approximation 
by an analysis of the experimental data in the current energy range 
 since the evolution equation is linear and has the flexible initial condition.

Deep inelastic interactions studied so far at HERA are far from the strong absorption regime for the QCD interactions 
with a small coupling constant. The only possible exception is the the gluon dipole--nucleon interaction at central impact parameters. However, the situation may change in the case of scattering off heavy nuclei and also in the case of 
central $pp/pA/AA$ collisions at the LHC.

\subsection{Conflict of  pQCD  with probability conservation.}
\label{subsection8.2}
Asymptotic freedom in pQCD does not guarantee probability conservation since pQCD amplitudes  are rapidly increasing with energy. The conflict with probability conservation reveals itself in the scattering of a color singlet wave packet of quarks and gluons 
of the small diameter $d$ off a hadron target, see Eq.~\ref{unitarity2}.   The cross section of 
elastic scattering of a dipole off the nucleon target can be parametrized as:
\begin{equation}
\label{eq:dipscat}
\frac{d\sigma_{\rm el}(dipole+N\to dipole +N)}{dt} = \frac{\sigma_{\rm tot}(dipole+N)^2}{16 \pi} \, F_{2g}^2(t) \,,
\label{2gl}
\end{equation}
where $F_{2g}(t)$ is the two-gluon form factor of the nucleon extracted 
in~\cite{Frankfurt:2002ka} from the HERA data on hard diffractive electroproduction of $J/\psi$.
At $t=0$, Eq.~\ref{2gl} is  the optical theorem. 
In the expression above we neglected a small contribution of the real part of the elastic dipole scattering amplitude.
Using Eqs.~\ref{2gl} and \ref{unitarity2}, we obtain the following inequality:
\begin{equation}
\frac{\sigma_{\rm tot}(dipole+N)}{16 \pi} \int dt\, F_{2g}^2(t)  \le 1/2 \,.
\end{equation}
In the case of scattering at the zero impact parameter, the inequality is 
a factor of $\sim 2$  stronger:
\begin{equation}
\frac{\sigma_{\rm tot}(dipole+N)}{8\pi} \int dt\, F_{2g}(t)  \le 1 \,.
\label{inequality}
\end{equation}
One can also study the unitarity condition as a function of an impact parameter using 
Eq.~\ref{unitaritybound}~\cite{MSM,Rogers:2003vi}.

The cross section for a spatially small color singlet dipole scattering off the nucleon target follows directly from the pQCD factorization theorem  in the leading 
$\alpha_s\ln(Q^2/Q^2_0)$  approximation~\cite{Blaettel:1993rd,Miller93,Radushkin97}.
It    can also be derived from the Born term
obtained in the leading $\alpha_{s}\ln(x_0/x)$ approximation in~\cite{Muellereikonal}. 
Using the machinery of pQCD calculations, it should not be too   difficult to calculate 
a series of LO, NLO, and NNLO approximations including corrections to the cross section and to the dipole  
wave function itself.

Since the expressions in the leading log approximation  are too bulky, for 
illustration purposes, we use the fit $xG \propto (x_0/x)^{\lambda(Q^2)}$. Due to an increase 
in  $\lambda(Q^2)$ with $Q^2$, the drop of the amplitude due to a decrease in  the dipole size  
is compensated to some extent by a faster increase in the gluon density. 
Thus, $x_{cr}$ at which unitarity breaks down, increases with an increase 
in  $Q^2$  rather slowly.

 A similar conflict   with probability conservation exists for the perturbative "Pomeron". The energy at which 
 this conflict becomes acute obviously  depends on the  impact parameter, see~\cite{Watt:2007nr} and references therein. 
 Note also that the conflict with unitary exists not only for scattering of small dipoles off hadrons, but also in the case of scattering of two small objects~\cite{UBFKL}.

\subsection{Gluon dipole--proton scattering  at the upper edge of HERA kinematics} 
\label{subsection8.3}
To demonstrate  that the new QCD regime for a gluon cloud is not far from the kinematical domain probed at the  LHC,
 we consider here scattering of a colorless small gluon dipole off the nucleon
 (for a detailed discussion, see~\cite{Ted}). One can evaluate  $\Gamma_{gg}(x, b)$ based on the information on the total cross section of the $q\bar q $ dipole--nucleon interaction extracted from DIS inclusive data~(Eq.~\ref{sigmah}),  the $t$ dependence of the two-gluon form factor (section~\ref{section7}), and the relation  
\begin{equation}
\Gamma_{gg}^{\rm inel}(x, b)={9 \over 4} \Gamma_{q\bar q}^{\rm inel}(x, b) \,.
\end{equation}
The knowledge of $\Gamma_{gg}(x, b)$ allows us to calculate the total and elastic cross sections 
of the colorless ''gluon--gluon'' dipole--nucleon scattering.  (By definition, the imaginary part of the  partial wave 
for the dipole--nucleon scattering is 
$\Gamma_{gg}(x, b)$.)   The ratio of the elastic and total cross sections for the scattering of a 
color-singlet dipole off the nucleon is:
\begin{equation}
\label{eq:R}
R_g(x,Q^2)={\int d^2b \left|\Gamma_{gg}(b)\right|^2  \over 2 \int d^2b \,\Gamma_{gg}(b)} \,.
\end{equation}

Numerical estimates~\cite{Ted}  indicate that for $d=0.1$~fm ($Q^2\sim 40$ GeV$^2$) corresponding to 
$\Upsilon$ photoproduction,  the value of $R_g(x,Q^2)$ is small ($\sim 0.14$) for $x\sim 10^{-4}$.
However, $R_g(x,Q^2)$ reaches 
the value of $\sim 0.4$, when $x$ goes down to $10^{-7}$ indicating that for such $x$, most of the 
cross section maybe due to the interaction in the BDR. At the same time, for $Q^2\sim 4$ GeV$^2$ one finds 
that the interaction is nearly black over a rather large range of impact parameters already for $x\sim 10^{-4}$, 
leading to $R_g\sim 0.3 - 0.4$.  At the same time, in the $q\bar q$--nucleon case, significantly  smaller $x$ 
are necessary  to reach the BDR. So the interaction in this channel is rather far 
from the regime of complete absorption at  $x$ achieved at HERA.

The predictions discussed above can be compared with the experimental data on diffraction in DIS obtained at HERA.  
The sum of the cross sections of color-singlet gluon dipole elastic scattering and
of inelastic diffraction scattering is calculable in terms of diffractive gluon PDFs (when an external hard probe couples directly to the gluon) integrated over the momentum of  the diffracted proton with the cut $x_{\Pomeron} \le 0.03$. 
The total cross section of the color singlet dipole--proton scattering
 is calculable in terms of the gluon density  of the  nucleon.  Thus, to compare  
$R_g$ with the data, one should  measure the  ratio of the diffractive and inclusive cross sections 
induced by a hard probe coupled to gluons:
\begin{equation}
R_g(x,Q^2)={\int_x^{0.03} dx_{\Pomeron} \int_0^{-\infty}dt f_g^{D(4)}(x/x_{\Pomeron}, Q^2 , x_{\Pomeron}, t)\over g_N(x,Q^2)} \,,
\end{equation}
where $f_g^{D(4)}$ is the diffractive gluon PDF. It can be evaluated using the most recent analyses of hard diffraction at HERA.
One finds that $R_g(x\sim 10^{-4}, Q^2 = 4 \ {\rm GeV}^2)  \sim 0.3$ and
$R_q(x\sim 10^{-4}, Q^2 = 4 \ {\rm GeV}^2)  \sim 0.2$     
which confirms proximity of the gluon interaction to the BDR at small impact parameters
(for the recent update, see Fig.~69 in~\cite{GFS}).

A word of caution is necessary here. These calculations are based on the leading log approximation where such higher order Fock states as $q\bar q g$ are neglected.  Hence, these calculation should only be considered as semi-quantitative ones.

\subsection{The regime of complete absorption}
\label{subsection8.4}
Bjorken scaling, i.e., the dependence of cross sections of hard processes only on the hard scale, completely 
disappears in the black-disc regime only for a specific range of $Q^2$ that depends on $x$. 
Basic formulas of the black-disc regime for the total cross section  of photo(electro)production of hadrons off a 
heavy nucleus at high energies were derived by V.~Gribov~\cite{Gribovphoton}  by analyzing the contribution  of 
diffraction   to the structure functions of heavy nuclei and an increase in  absorption  with an increase in  the atomic number. 
The theoretical  observation that enabled the calculation was that 
in the BDR, non-diagonal transitions between diffractively produced hadronic states are absent at zero angles. 

Interesting features of this regime include the dominance of diffraction into dijets which constitutes 50\% of the total cross section, a gross change of the $Q^2$ dependence of exclusive meson production (from $1/Q^6$ to $1/Q^2$ for the case of the longitudinal photon), and gross suppression of leading hadron production in the current fragmentation  region 
(suppression of the effective fractional energy losses)~\cite{MGFS}.

A  big challenge is to establish the mechanism of the onset of the BDR
and the basic features of the BDR.  
A popular approach is to tame the rapid increase in  the perturbative LO "Pomeron" with energy by taking into account 
the simplest  non-linear effects~\cite{BK}.  This approach has even more severe problems as compared to the perturbative "Pomeron". Indeed,  energy--momentum conservation  restricts the number of allowed  branchings of the "Pomerons" due to the 
triple "Pomeron" and the number of multi-"Pomeron" exchanges. The restriction  follows from the fact that an exchange of 
the perturbative
 "Pomeron" is dominated by inelastic processes which carry energy and momentum.   This restriction is especially severe 
 for the multi-Regge kinematics explored in~\cite{BK}.  
 Thus,  the contribution of fan diagrams relevant in this approach  for the screening of the 
 perturbative  "Pomeron"   requires  collision energies that cannot be 
treated in a realistic way at the collider energies.

The appearance of new scales in addition to $Q^2$  follows from the fact that the decomposition over 
twists becomes meaningless at sufficiently small $x$ since the decomposition parameter starts to exceed 
the radius of convergence. As a result, an    analytic continuation becomes necessary.

 The new scale  depends on the incident energy as a result of probability conservation.  This property of the 
 new QCD regime distinguishes it from the pQCD regime.  It may indicate that in this regime,  violation of the 
 two-dimensional translation symmetry, or even conformal invariance, takes place.
This option has been suggested for soft processes  in~\cite{amati1,amati2}   
and for hard QCD in~\cite{BF1}. In this case, the quasi-Goldstone mode is the motion along the 
impact parameter in the region of complete absorption.  Moreover it is unclear whether pQCD  is applicable even qualitatively for the description of the new QCD   regime since in the kinematics where the interaction becomes strong, 
one should 
take into account 
the Gribov copies  found in~\cite{Gribovcopies}.

\section{Conclusions}
\label{section9}
The theory of the $S$-matrix  allowed one to understand many regularities of diffractive processes in the regime that is now referred to as the soft QCD regime. Further progress became possible with an advent of QCD and the focus on the processes involving a hard scale.

Investigations of diffractive phenomena led to the ideas, concepts, and technologies
of the calculations   that form the foundation of modern particle physics and are the basis for the future developments aiming at bridging the gap between soft and hard phenomena.  
The exploration of the color transparency phenomenon supplies a new method of investigations of hadrons and nuclei.    
Investigation of the new QCD regime of strong interaction will allow one to find new phenomena and to develop new methods of treating  phase transitions in the relativistic kinematics. The long-standing challenges include confinement of quarks and gluons and spontaneously broken chiral symmetry.

Also, QCD is the only quantum field theory leading  to non-linear phenomena that  can be probed in laboratory.  
Hence, it may  
provide new tools for developing a theory beyond the Standard Model of strong and electroweak interactions.

 \section{Acknowledgments}

We thank  J.~Bjorken, V.~N.~Gribov and A.~H.~Mueller for the illuminating discussions of diffractive phenomena in high energy processes.  We thank  V.~Guzey for reading manuscript and valuable comments.
The research was supported by DOE and BSF.

\


\begin{thebibliography}{0}
\bibitem{Landau1} L.D. Landau,Talk at Kiev conference 1959,
published in, Theoretical physics in the 20th century: A memorial volume to W Pauli, p. 245 (Eds M Fierz, V F Weisskopf) (New York: Interscience Publishers, 1960).

\bibitem{LSZ}H.~Lehmann, K.~Symanzik and W.~Zimmermann,
  Nuovo Cim.\  {\bf 6} (1957) 319.


\bibitem{Landau2} 
 L.~D.~Landau,
  Nucl.\ Phys.\  {\bf 13} (1959) 181.



\bibitem{Mandelstamrep}     S.~Mandelstam,
  Phys.\ Rev.\  {\bf 112} (1958) 1344.
  
  
\bibitem{CM}   	
 G.~F.~Chew, S.~C.~Frautschi and S.~Mandelstam,
  Phys.\ Rev.\  {\bf 126} (1961) 1202.

\bibitem{ELO} R.~ J.~Eden, P. ~ V.~ Landshoff ,D.~ I.~ Olive,  "Analytic S- matrix" {\it  Cambridge Univ. Pr. (2002) 296}

\bibitem{Froissart}  M.~Froissart,
  Phys.\ Rev.\  {\bf 123} (1961) 1053. 
  
  \bibitem{GribovFroissart}V.~N.~Gribov,
  Sov.\ Phys.\ JETP {\bf 14} (1962) 1395
   [Zh.\ Eksp.\ Teor.\ Fiz.\  {\bf 41} (1961) 1962].

M.Froissart,   Report to the La Jolla Conference on the Theory of Weak and Strong Interactions, La Jolla,1961(unpublished)

\bibitem{Regge}  T.~Regge,
  Nuovo Cim.\  {\bf 14} (1959) 951; {\bf 18} (1960) 947.

\bibitem{FGZ}  Cited in S.~C.~Frautschi, M.~Gell-Mann, and F.~Zahariasen,  Phys.Rev.
\  {\bf 126} (1962) 2204.

\bibitem{Runitarity}   V.~N.~Gribov,
  Sov.\ Phys.\ JETP {\bf 15} (1962) 873
   [Zh.\ Eksp.\ Teor.\ Fiz.\  {\bf 42} (1962) 1260]
   [Nucl.\ Phys.\  {\bf 40} (1963) 107]. 
 
  \bibitem{G-P62} V.~N.~Gribov and I.~Y.~.Pomeranchuk,
  Sov.\ Phys.\ JETP {\bf 15} (1962) 788L
   [Zh.\ Eksp.\ Teor.\ Fiz.\  {\bf 42} (1962) 1141]
   [Phys.\ Rev.\ Lett.\  {\bf 8} (1962) 343]. 
   \bibitem{CFrautschi2}   G.~F.~Chew and S.~C.~Frautschi,
  Phys.\ Rev.\ Lett.\  {\bf 8} (1962) 41.
  \bibitem{GPT} V.~N.~Gribov, I.~Ya.~.Pomeranchuk and K.~A.~Ter-Martirosian,
  Phys.\ Lett.\  {\bf 9} (1964) 269.

\bibitem{Gribov61}	V.~N.~Gribov,
  Nucl.\ Phys.\  {\bf 22} (1961) 249;
 Zh.\ Eksp.\ Teor.\ Fiz.\  {\bf 41} (1961) 667 
[Sov.\ Phys.\ JETP {\bf 14} (1962) 478 ].

\bibitem{Nonvacuum}    A.~C.~Irving and R.~P.~Worden,
  Phys.\ Rept.\  {\bf 34} (1977) 117.
  
\bibitem{GribovComplex} V.~N.~Gribov,
  ``The theory of complex angular momenta: Gribov lectures on theoretical
  physics,''  {\it  Cambridge Univ. Pr. (2003) 297 p}

\bibitem{GellMann:1964zz} 
  M.~Gell-Mann, M.~L.~Goldberger, F.~E.~Low, V.~Singh and F.~Zachariasen,
  Phys.\ Rev.\  {\bf 133}, B161 (1964).
\bibitem{Reggezationvector}  L.~L.~Frankfurt and V.~E.~Sherman,
  Sov.\ J.\ Nucl.\ Phys.\  {\bf 23} (1976) 581.
                                                	
L.N. Lipatov 
Sov.\ J.\ Nucl.\ Phys.\  {\bf 23} (1976) 338.
       
\bibitem{Reggezationfermion}    	
 V.~S.~Fadin and V.~E.~Sherman,
  Pisma Zh.\ Eksp.\ Teor.\ Fiz.\  {\bf 23} (1976) 599.

\bibitem{Dolen-Horn-Schmidt}  R.~Dolen, D.~Horn and C.~Schmid,
  Phys.\ Rev.\  {\bf 166} (1968) 1768.

\bibitem{CFrautschi1} G.~F.~Chew and S.~C.~Frautschi,
  Phys.\ Rev.\ Lett.\  {\bf 7} (1961) 394.


\bibitem{Strikman:2007nz}
  M.~Strikman,
  Nucl.\ Phys.\ A {\bf 805} (2008) 369
  [arXiv:0711.1634 [hep-ph]].
 
\bibitem{Frankfurt} L.L.Frankfurt and V. A.Khose 
Materials of 10 Winter School on nuclear and particle physics,v. || Leningrad 1975, pp
196-408.


\bibitem{Pomeranchuk}   I. ~Ya.~ Pomeranchuk, JETP, {\bf 34} (1958) 725.



\bibitem{Gribovelastic}
V.~N.~Gribov,  JETP\ Lett. {\bf 41} (1961) 667-669;
 \bibitem{GP}V.~N.~Gribov and I.~Y.~.Pomeranchuk,
  Sov.\ Phys.\ JETP {\bf 16} (1963) 220
   [Zh.\ Eksp.\ Teor.\ Fiz.\  {\bf 43} (1962) 308]
   [Nucl.\ Phys.\  {\bf 38} (1962) 516].
\bibitem{Landshoff}P.~ V.~Landshoff,
  Acta Phys.\ Polon.\ B {\bf 40} (2009) 1967
  [arXiv:0903.1523 [hep-ph]].
  
  \bibitem{Levy:2007fb}
  A.~Levy,
   arXiv:0711.0737 [hep-ex].
  
 


\bibitem{List:2009pb}B.~List [H1 Collaboration],
  arXiv:0906.4945 [hep-ex].
 \bibitem{Totem}G.~Antcheva et al [The TOTEM collaboration], TOTEM 2012-002.
 
 
\bibitem{M.Bloc}M.~ Block and R.~ N.~ Cahn, \ Rev.\  Mod. \ Phys. \
{\bf 57} (1985) 563. 

\bibitem{BFS} B.~Blok, L.~Frankfurt and M.~Strikman,
  Eur.\ Phys.\ J.\ C {\bf 67} (2010) 99
  [arXiv:1001.2469 [hep-ph]].


\bibitem{BFS94} 
ibitem{Brodsky:1994kf}
  S.~J.~Brodsky, L.~Frankfurt, J.~F.~Gunion, A.~H.~Mueller and M.~Strikman,
  Phys.\ Rev.\ D {\bf 50} (1994) 3134
  [hep-ph/9402283].



\bibitem{KancheliTPL} O.~V.~Kancheli,
  JETP Lett.\  {\bf 11} (1970) 267
   [Pisma Zh.\ Eksp.\ Teor.\ Fiz.\  {\bf 11} (1970) 397].
   
   \bibitem{MuellerTPL}    A.~H.~Mueller,
  Phys.\ Rev.\ D {\bf 4} (1971) 150.
 
 
\bibitem{DFNAL}  CDF Collaboration, Phys. Rev. {\bf D50} (1994) 5535.

\bibitem{DLHC} 
G.~Aad {\it et al.}  [ATLAS Collaboration],
  Eur.\ Phys.\ J.\ C {\bf 72} (2012) 1926
  [arXiv:1201.2808 [hep-ex]].

B.~Abelev {\it et al.}  [The ALICE Collaboration],
  arXiv:1208.4968 [hep-ex].

\bibitem{AGK} 	 V.~A.~Abramovsky, V.~N.~Gribov and O.~V.~Kancheli,
  Yad.\ Fiz.\  {\bf 18} (1973) 595
   [Sov.\ J.\ Nucl.\ Phys.\  {\bf 18} (1974) 308].



\bibitem{GPT2}  	
V.~N.~Gribov, I.~Y.~.Pomeranchuk and K.~A.~Ter-Martirosian,
  Phys.\ Rev.\  {\bf 139} (1965) B184.
  \bibitem{S.Mandelstam63}    
 S.~Mandelstam,
  Nuovo Cim.\  {\bf 30} (1963) 1148.
	
	
\bibitem{PomeronCalculus}       
V.~N.~Gribov,
  Sov.\ Phys.\ JETP {\bf 26} (1968) 414
   [Zh.\ Eksp.\ Teor.\ Fiz.\  {\bf 53} (1967) 654].

 
\bibitem{GribovMigdal} 
 V.~N.~Gribov and A.~A.~Migdal,
  Zh.\ Eksp.\ Teor.\ Fiz.\  {\bf 55} (1968) 4.

\
  \bibitem{Gribovladder}V.~N.~Gribov,
  Sov.\ J.\ Nucl.\ Phys.\  {\bf 9} (1969) 246
\bibitem{Feynman}  R.~P.~Feynman,
  ``Photon-hadron interactions,'' W.A. Benjamin, Inc, 
  Reading, Messachusetts,  1972, 282p



 
  \bibitem{asymmetry}G.~Fidecaro, M.~Fidecaro, L.~Lanceri, S.~Nurushev, L.~Piemontese, V.~Solovyanov, A.~Vascotto and F.~Gasparini {\it et al.},
  Phys.\ Lett.\ B {\bf 105} (1981) 309.

\bibitem{AbrBat}V.~A.~Abramovsky and R.~G.~Betman,
  Sov.\ J.\ Nucl.\ Phys.\  {\bf 55} (1992) 912.

\bibitem{LandauN}   L.~D.~Landau, E.~M.~Lifshitz,  Quantum Mechanics Non-Relativistic Theory, Third Edition: Volume 3, Pergamon Press,1977. 

\bibitem{AMarten}  
A.~Martin,
  Phys.\ Rev.\  {\bf 129} (1963) 1432;

\bibitem{LPTM} L.~D.~ Landau and I.~Ya.~Pomeranchuk, Dokl.\ Akad.\  Nauk\  SSSR {\bf  92} (1953) 535, 735.

M.L. Ter-Mikaelian, Dokl. \ Akad.\ Nauk\ SSSR {\bf 94} (1954) 1033
  
A.~B.~Migdal,
  Phys.\ Rev.\  {\bf 103} (1956) 1811.
  
  M.L. Ter-Mikaelian, High Energy Electromagnetic Processes in Condensed Media, John Wiley \& Sons, NY, 1972.

\bibitem{GPI} V.~N.~Gribov, B.~L.~Ioffe and I.~Y.~.Pomeranchuk,
  Sov.\ J.\ Nucl.\ Phys.\  {\bf 2} (1966) 549
   [Yad.\ Fiz.\  {\bf 2} (1965) 768].
 
\bibitem{GribovST} 
V.N. Gribov.  In *Moscow 1 ITEP school, v.1 'Elementary particles'*, 65,1973, hep-ph/0006158.

\bibitem{Blok:2006ns}
  B.~Blok and L.~Frankfurt,
  Phys.\ Rev.\ D {\bf 75} (2007) 074001
  [hep-ph/0611062].
  
 \bibitem{Gribov-Glauber} . 	
 V.~N.~Gribov,
  Sov.\ Phys.\ JETP {\bf 29} (1969) 483
   [Zh.\ Eksp.\ Teor.\ Fiz.\  {\bf 56} (1969) 892].
 
\bibitem{Karmanov-Kondratuk}  V.~A.~Karmanov and L.~A.~Kondratyuk,
  Pisma Zh.\ Eksp.\ Teor.\ Fiz.\  {\bf 18} (1973) 451.
  
  
 \bibitem{CFS96}  
 J.~C.~Collins, L.~Frankfurt and M.~Strikman,
  Phys.\ Rev.\ D {\bf 56} (1997) 2982
  [hep-ph/9611433].
 
 

\bibitem{CT}   D.~Dutta and K.~Hafidi,
  Int.\ J.\ Mod.\ Phys.\ E {\bf 21} (2012) 1230004
  [arXiv:1209.5295 [nucl-ex]].
 
\bibitem{F.Low} F.~E.~Low,
  Phys.\ Rev.\ D {\bf 12} (1975) 163.


\bibitem{S.Nussinov} S.~Nussinov,
ÊÊPhys.\ Rev.\ Lett.\  {\bf 34} (1975) 1286.


\bibitem{GS:1976}  J.~F.~Gunion and D.~E.~Soper,
  Phys.\ Rev.\ D {\bf 15} (1977) 2617.


 \bibitem{Blaettel:1993rd}
  B.~Blaettel, G.~Baym, L.~L.~Frankfurt and M.~Strikman,
  Phys.\ Rev.\ Lett.\  {\bf 70} (1993) 896. 
  
  
\bibitem{Miller93}  L.~Frankfurt, G.~A.~Miller and M.~Strikman,
  Phys.\ Lett.\ B {\bf 304} (1993) 1
  [hep-ph/9305228].

\bibitem{Radushkin97}  L.~Frankfurt, A.~Radyushkin and M.~Strikman,
  Phys.\ Rev.\ D {\bf 55} (1997) 98
  [hep-ph/9610274].
  \bibitem{Muellereikonal} A.~H.~Mueller,
  Nucl.\ Phys.\ B {\bf 335} (1990) 115.
  
  
  
\bibitem{FGL}  
G.~V.~Frolov, V.~N.~Gribov and L.~N.~Lipatov,
  Phys.\ Lett.\ B {\bf 31} (1970) 34.

 V.~N.~Gribov, L.~N.~Lipatov and G.~V.~Frolov,
  Sov.\ J.\ Nucl.\ Phys.\  {\bf 12} (1971) 543
   [Yad.\ Fiz.\  {\bf 12} (1970) 994]. 	

\bibitem{BFKL} 
 E.~A.~Kuraev, L.~N.~Lipatov and V.~S.~Fadin,
  Sov.\ Phys.\ JETP {\bf 45} (1977) 199
   [Zh.\ Eksp.\ Teor.\ Fiz.\  {\bf 72} (1977) 377].

   
  I.~I.~Balitsky and L.~N.~Lipatov,
  Sov.\ J.\ Nucl.\ Phys.\  {\bf 28} (1978) 822
   [Yad.\ Fiz.\  {\bf 28} (1978) 1597].
 
\bibitem{MuellerBFKL}
A.~H.~Mueller,
  Nucl.\ Phys.\ B {\bf 415} (1994) 373.
  
Z.~Chen and A.~H.~Mueller,
  Nucl.\ Phys.\ B {\bf 451} (1995) 579.

\bibitem{Muellerrestriction} 
 A.~H.~Mueller,
  Phys.\ Lett.\ B {\bf 396} (1997) 251
  [hep-ph/9612251].
 

\bibitem{CTM}  
 M.~Ciafaloni, M.~Taiuti and A.~H.~Mueller,
  Nucl.\ Phys.\ B {\bf 616} (2001) 349
  [hep-ph/0107009].
\bibitem{Ciafaloni}  M.~Ciafaloni and D.~Colferai,
  Phys.\ Lett.\ B {\bf 452} (1999) 372
  [hep-ph/9812366].
  
  
  \bibitem{Lipatov-Fadin}  V.~S.~Fadin and L.~N.~Lipatov,
  Phys.\ Lett.\ B {\bf 429} (1998) 127
  [hep-ph/9802290].
  

\bibitem{KL} E.\ A. \ Kuraev and L.\ N.\ Lipatov, Yad.\ Fiz.\ {\bf 16} (1972)1060 [Sov. J. Nucl. Phys. {\bf 16} (1973) 584]

\bibitem{Altarelli} G.~Altarelli, R.~D.~Ball and S.~Forte,
  Nucl.\ Phys.\ B {\bf 575} (2000) 313
  [hep-ph/9911273].

\bibitem{Ciafaloni:2003rd}
  M.~Ciafaloni, D.~Colferai, G.~P.~Salam and A.~M.~Stasto,
  Phys.\ Rev.\ D {\bf 68} (2003) 114003.


  \bibitem{PM}  H.~I.~Miettinen and J.~Pumplin,
  Phys.\ Rev.\ D {\bf 18} (1978) 1696.
  \bibitem{Baym93} 
  B.~Blaettel, G.~Baym, L.~L.~Frankfurt, H.~Heiselberg and M.~Strikman,
  Phys.\ Rev.\ D {\bf 47} (1993) 2761.
  
  

 \bibitem{Good:1960ba}
  M.~L.~Good and W.~D.~Walker,
  Phys.\ Rev.\  {\bf 120} (1960) 1857.
  
  
\bibitem{Kopeliovich:1978qz}
  B.~Z.~Kopeliovich and L.~I.~Lapidus,
  Pisma Zh.\ Eksp.\ Teor.\ Fiz.\  {\bf 28} (1978) 664.

\bibitem{Frankfurt:1993qi}
  L.~Frankfurt, G.~A.~Miller and M.~Strikman,
  Phys.\ Rev.\ Lett.\  {\bf 71} (1993) 2859
  [hep-ph/9309285]. 
\bibitem{SG} M.~Strikman and V.~Guzey,
  Phys.\ Rev.\ C {\bf 52} (1995) 1189
  [nucl-th/9506010].
  \bibitem{LFGS}  L.~Frankfurt, V.~Guzey and M.~Strikman,
  J.\ Phys.\ G G {\bf 27} (2001) R23
  [hep-ph/0010248].

 \bibitem{Baym:1995cz}
  G.~Baym, B.~Blattel, L.~L.~Frankfurt, H.~Heiselberg and M.~Strikman,
  Phys.\ Rev.\ C {\bf 52} (1995) 1604
  [nucl-th/9502038].

\bibitem{HERA} F.~D.~Aaron {\it et al.}  [H1 and ZEUS Collaborations],
  arXiv:1207.4864 [hep-ex].
\bibitem{Collins} J.~C.~Collins,
  Phys.\ Rev.\ D {\bf 57} (1998) 3051
   [Erratum-ibid.\ D {\bf 61} (2000) 019902]
  [hep-ph/9709499].
\bibitem{Bj1}
   J.~D.~Bjorken,
  Conf.\ Proc.\ C {\bf 710823} (1971) 281.
  
  J.~D.~Bjorken and J.~B.~Kogut,
  Phys.\ Rev.\ D {\bf 8} (1973) 1341.

  \bibitem{Bj2} J.~D.~Bjorken,
  Lect.\ Notes Phys.\  {\bf 56} (1976) 93.
 
 
\bibitem{Frankfurt:1988nt} L. L.Frankfurt and M. Strikman,   Phys.\ Rep. \ {\bf 160} (1988) 235-427.
  

  \bibitem{Abramowicz}
  H.~Abramowicz, L.~Frankfurt and M.~Strikman,
  eConf C {\bf 940808} (1994) 033
   [Surveys High Energ.\ Phys.\  {\bf 11} (1997) 51]

 
 \bibitem{Kowalski:2008sa}
  H.~Kowalski, T.~Lappi, C.~Marquet and R.~Venugopalan,
  Phys.\ Rev.\ C {\bf 78} (2008) 045201
  [arXiv:0805.4071 [hep-ph]].
  

 

\bibitem{Bartels}  J.~Bartels and M.~Loewe,
  Z.\ Phys.\ C {\bf 12} (1982) 263.
  
\bibitem{BB}I.~I.~Balitsky and V.~M.~Braun,
  Nucl.\ Phys.\ B {\bf 311} (1989) 541.

\bibitem{Dittes}D.~Mueller, D.~Robaschik, B.~Geyer, F.~M.~Dittes and J.~Horejsi,
  Fortsch.\ Phys.\  {\bf 42} (1994) 101
  [hep-ph/9812448].
\bibitem{Ji1}  X.~-D.~Ji,
  Phys.\ Rev.\ Lett.\  {\bf 78} (1997) 610
  [hep-ph/9603249]; Phys.\ Rev.\ D {\bf 55} (1997) 7114
  [hep-ph/9609381].

\bibitem{Rad}  A.~V.~Radyushkin,
  Phys.\ Lett.\ B {\bf 380} (1996) 417
  [hep-ph/9604317]; Phys.\ Rev.\ D {\bf 56} (1997) 5524
  [hep-ph/9704207].
\bibitem{Jireview}
X.~-D.~Ji,
  J.\ Phys.\ G {\bf 24} (1998) 1181
  [hep-ph/9807358].
\bibitem{Freund}L.~L.~Frankfurt, A.~Freund and M.~Strikman,
  Phys.\ Rev.\ D {\bf 58} (1998) 114001
   [Erratum-ibid.\ D {\bf 59} (1999) 119901]
  [hep-ph/9710356].
  
   \bibitem{Diehl}
 M.~Diehl, T.~Gousset, B.~Pire and O.~Teryaev,
  Phys.\ Rev.\ Lett.\  {\bf 81} (1998) 1782
  [hep-ph/9805380].
 \bibitem{FPPS}   
 L.~L.~Frankfurt, P.~V.~Pobylitsa, M.~V.~Polyakov and M.~Strikman,
  Phys.\ Rev.\ D {\bf 60} (1999) 014010
  [hep-ph/9901429].
   \bibitem{Radyushkinrho}
A.~V.~Radyushkin,
  Phys.\ Lett.\ B {\bf 385} (1996) 333
  [hep-ph/9605431].
\bibitem{Frankfurt:1997ha}
  L.~Frankfurt, A.~Freund, V.~Guzey and M.~Strikman,
  Phys.\ Lett.\ B {\bf 418} (1998) 345
   [Erratum-ibid.\ B {\bf 429} (1998) 414]
  [hep-ph/9703449].
\bibitem{Shuvaev:1999ce}
  A.~G.~Shuvaev, K.~J.~Golec-Biernat, A.~D.~Martin and M.~G.~Ryskin,
 Phys.\ Rev.\ D {\bf 60} (1999) 014015
  [hep-ph/9902410].
\bibitem{Ryskin}   M.~G.~Ryskin,
  Z.\ Phys.\ C {\bf 57} (1993) 89.

\bibitem{Koepf} L.~Frankfurt, W.~Koepf and M.~Strikman,
  Phys.\ Rev.\ D {\bf 54} (1996) 3194
  [hep-ph/9509311]; Phys.\ Rev.\ D {\bf 57} (1998) 512
  [hep-ph/9702216]. 
 
\bibitem{Baltz:2007kq}
A.~J.~Baltz, G.~Baur, D.~d'Enterria, L.~Frankfurt, F.~Gelis, V.~Guzey, K.~Hencken, (ed.) and Y.~.Kharlov {\it et al.},
  Phys.\ Rept.\  {\bf 458} (2008) 1
  [arXiv:0706.3356 [nucl-ex]].
  
  \bibitem{Frankfurt:1998yf}
  L.~L.~Frankfurt, M.~F.~McDermott and M.~Strikman,
  JHEP {\bf 9902} (1999) 002
  [hep-ph/9812316].
  \bibitem{Martin:1999rn}
  A.~D.~Martin, M.~G.~Ryskin and T.~Teubner,
  Phys.\ Lett.\ B {\bf 454} (1999) 339
  [hep-ph/9901420].


  
  \bibitem{Chekanov:2004mw}
  S.~Chekanov {\it et al.}  [ZEUS Collaboration],
  Nucl.\ Phys.\ B {\bf 695} (2004) 3
  [hep-ex/0404008].
 
 \bibitem{Chekanov:2007zr}
S.~Chekanov {\it et al.}  [ZEUS Collaboration],
  PMC Phys.\ A {\bf 1} (2007) 6
  [arXiv:0708.1478 [hep-ex]].

\bibitem{Mankiewicz:1999tt}
  L.~Mankiewicz and G.~Piller,
  Phys.\ Rev.\ D {\bf 61} (2000) 074013
 

\bibitem{Goeke:2001tz}
  K.~Goeke, M.~V.~Polyakov and M.~Vanderhaeghen,
  Prog.\ Part.\ Nucl.\ Phys.\  {\bf 47} (2001) 401
 
  
  \bibitem{Freund:2001hm}
  A.~Freund and M.~F.~McDermott,
  Phys.\ Rev.\  D {\bf 65}  (2002) 091901
  
  
  \bibitem{Strikman:2003gz}
  M.~Strikman and C.~Weiss,
  Phys.\ Rev.\ D {\bf 69} (2004) 054012
  [hep-ph/0308191].
  
  
  \bibitem{Frankfurt:2010ea}
  L.~Frankfurt, M.~Strikman and C.~Weiss,
  Phys.\ Rev.\ D {\bf 83} (2011) 054012
  [arXiv:1009.2559 [hep-ph]].
  \bibitem{Frankfurt:2002ka} L.~Frankfurt and M.~Strikman,
  Phys.\ Rev.\ D {\bf 66} (2002) 031502
  [hep-ph/0205223].
  
  
  
\bibitem{Frankfurt:2008vi}
  L.~Frankfurt, M.~Strikman, D.~Treleani and C.~Weiss,
  Phys.\ Rev.\ Lett.\  {\bf 101} (2008) 202003 
  [arXiv:0808.0182 [hep-ph]].
  
  \bibitem{Frankfurt:1990nc}
  L.~Frankfurt and M.~Strikman,
  Phys.\ Rev.\ Lett.\  {\bf 63} (1989) 1914
  [Erratum-ibid.\  {\bf 64} (1990) 815].
  
  
  \bibitem{HERAgap}
S. Chekanov et al. [ZEUS Collaboration], Nucl.\ Phys.\  {\bf B 695} (2004) 3 [arXiv:hep-ex/0404008]; 

A. Aktas et al. [H1 Collaboration], Eur.\  Phys.\  J.\ C \ {\bf 46} (2006) 585 [arXiv:hep-ex/0510016].


    \bibitem{Frankfurt:2000jm}
  L.~Frankfurt, G.~A.~Miller and M.~Strikman,
  Phys.\ Rev.\  D {\bf 65}, 094015 (2002)
  [arXiv:hep-ph/0010297].
  
  
  
\bibitem{Aitala:2000hc}
  E.~M.~Aitala {\it et al.}  [E791 Collaboration],
  {\em Phys.\ Rev.\ Lett.\ }  {\bf 86}, 4768 (2001), ibid {\bf 86}, 4773 (2001).

\bibitem{Braun} V.~M.~Braun, D.~Y.~.Ivanov, A.~Schafer and L.~Szymanowski,
   Phys.\ Lett.\ B {\bf 509} (2001) 43; Nucl.\ Phys.\ B {\bf 638} (2002) 111
  

  \bibitem{Sokoloff:1986prl} M. D. Sokoloff {\it et al.} Phys.\ Rev.\ Lett.\ {\bf 57}, 3003 (1986).
  
  \bibitem{FPSS}
 L.~Frankfurt, G.~Piller, M.~Sargsian and M.~Strikman,
  Eur.\ Phys.\ J.\ A {\bf 2} (1998) 301
  [nucl-th/9801041].


  
 \bibitem{FFS} G.~R.~Farrar, H.~Liu, L.~L.~Frankfurt and M.~I.~Strikman,
  Phys.\ Rev.\ Lett.\  {\bf 61} (1988) 686.
  
  
   \bibitem{Adams:1995} M.~R.~Adams {\it et al.} (E665), Phys. Rev. Lett.~{\bf
74}, 1525 (1995).

\bibitem{Airape:2003} A.~Airapetian {\it et al.} (HERMES),
  Phys. Rev. Lett.~{\bf 90}, 052501 (2003).

\bibitem{Clasie:2007}
  B.~Clasie {\it et al.}, Phys.\ Rev.\ Lett.\ {\bf 99}, 242502 (2007).  

\bibitem{ElFassi:2012nr} 
  L.~El Fassi, L.~Zana, K.~Hafidi, M.~Holtrop, B.~Mustapha, W.~K.~Brooks, H.~Hakobyan and X.~Zheng {\it et al.},
  Phys. Lett. {\bf B 712}, 326 (2012).

\bibitem{MGFS}  L.~Frankfurt, V.~Guzey, M.~McDermott and M.~Strikman,
  Phys.\ Rev.\ Lett.\  {\bf 87} (2001) 192301
  [hep-ph/0104154].

\bibitem{BF1} 
B.~Blok and L.~Frankfurt,
  Phys.\ Rev.\ D {\bf 73} (2006) 054008
  [hep-ph/0508218].
  \bibitem{Watt:2007nr}
  G.~Watt and H.~Kowalski,
  Phys.\ Rev.\ D {\bf 78} (2008) 014016
  [arXiv:0712.2670 [hep-ph]].


\bibitem{MSM}
S.~Munier, A.~M.~Stasto and A.~H.~Mueller,
  Nucl.\ Phys.\ B {\bf 603} (2001) 427
  [hep-ph/0102291].

\bibitem{Rogers:2003vi}
  T.~Rogers, V.~Guzey, M.~Strikman and X.~Zu,
  Phys.\ Rev.\ D {\bf 69} (2004) 074011.
  [hep-ph/0309099].
 

\bibitem{UBFKL} 
A.~ H. Mueller,
Nucl.Phys.B437:107-126,1995.
\bibitem{Ted} L.~Frankfurt, T.~Rogers, M.~Strikman, in preparation.  
 \bibitem{GFS} 
  L.~Frankfurt, V.~Guzey and M.~Strikman,
  Phys.\ Rept.\  {\bf 512} (2012) 255
  [arXiv:1106.2091 [hep-ph]].

\bibitem{Gribovphoton} 
V.~N.~Gribov,
  Sov.\ Phys.\ JETP {\bf 30} (1970) 709
   [Zh.\ Eksp.\ Teor.\ Fiz.\  {\bf 57} (1969) 1306].



\bibitem{BK}   I.~Balitsky,
  Nucl.\ Phys.\ B {\bf 463} (1996) 99
  [hep-ph/9509348].
  
  
  Y.~V.~Kovchegov,
  Phys.\ Rev.\ D {\bf 61} (2000) 074018
  [hep-ph/9905214].
 
                    

\bibitem{amati1} D.~Amati, G.~Marchesini, M.~Ciafaloni and G.~Parisi,
  Nucl.\ Phys.\ B {\bf 114} (1976) 483.

\bibitem{amati2}  D.~Amati, \ M.~Le Bellac,\  G.~Marchesini and M.~Ciafaloni,
  Nucl.\ Phys.\ B {\bf 112} (1976) 107.

 \bibitem{Gribovcopies} 
 V.~N.~Gribov,
  Nucl.\ Phys.\ B {\bf 139} (1978) 1.
\end{thebibliography}
\end{document}